\documentclass[12pt]{article}
\usepackage{amsmath}
\usepackage{amssymb}
\usepackage{epsfig}
\usepackage{cite}
\usepackage{color,colordvi}

\begin{document}
\vspace{0.01cm}
\begin{center}
{\Large\bf   Entanglement, Observers and Cosmology: a view from von Neumann Algebras} 

\end{center}

\vspace{0.1cm}

\begin{center}

{\bf C\'esar G\'omez}\footnote{cesar.gomez@uam.es}

\vspace{.6truecm}

{\em 
Instituto de F\'{\i}sica Te\'orica UAM-CSIC\\
Universidad Aut\'onoma de Madrid,
Cantoblanco, 28049 Madrid, Spain}\\

\end{center}


\begin{abstract}
\noindent  
 
{\small

 Infinite entanglement fluctuations appear when a quantum field theory on a causally complete domain of space-time is a type $III$ factor. In the weak gravity limit $G_N=0$ this factor can be transformed into a crossed product type $II$ factor with finite entanglement fluctuations by adding a physical reference frame system (observer). The use of a physical reference frame to define a regularization of divergent entanglement is formally identical to the quantum information approach to superselection charges. In that case the added reference frame allows quantum superpositions between different superselection sectors. For the case of cosmological horizons we map the primordial inflationary slow rolling phase into a type $II$ modification of the pure de Sitter type $III$ factor and we use the so defined type $II$ finite entanglement fluctuations to predict the primordial power spectrum of scalar curvature fluctuations. For the Hagedorn high temperature phase of large $N$ Yang Mills, the type $II$ description of the Hagedorn phase accounts, in the large $N$ limit, for the quantum fluctuations of the interval of the corresponding matrix model eigenvalue distribution.}

\end{abstract}

\thispagestyle{empty}
\clearpage
\tableofcontents
\newpage

\section{Introduction and Summary}
Entanglement is the basic property of the quantum states describing composed physical systems \footnote{The first use of the word entanglement \cite{Schr} for states describing quantum composed systems appears in  \cite{Schr}, an insightful discussion of  the EPR paradox \cite{EPR}.} . Relative to a given partition into subsystems the amount of entanglement of a given quantum state can be measured by the von Neumann entropy \cite{vN} of the corresponding reduced density matrices. This entropy has not classical analog and measures the amount of quantum information encoded in the entangled state.

In local quantum field theory (QFT) we can extend the notion of quantum entanglement to a given partition of space itself \footnote{For an excellent recent review on entanglement and von Neumann algebras see \cite{wittenrev1} and \cite{wittenrev2} and references therein}. For a given fixed time hyper surface $\Sigma_t$ and a given partition of space into a bounded region $D$ and its complement $D^c$ we can define the algebras ${\cal{A}}(D)$ and ${\cal{A}}(D^c)$ of local observables with support in $D$ and $D^c$ respectively. In QFT these algebras are defined by their {\it representation} in the space ${\cal{B}}({\cal{H}})$ of bounded operators of the Hilbert space ${\cal{H}}$ describing the full QFT.
Under certain topological conditions \footnote{In this topology the convergency in the algebra of operators of $a_n$ is defined, relative to the Hilbert space representation, in terms of the convergency of the expectation values $\langle \phi|a_n|\phi\rangle$ for arbitrary $|\phi\rangle$. This topology is known as {\it weak topology} and the corresponding completion of the algebra as weak completion.} the so defined subalgebras of ${\cal{B}}({\cal{H}})$ are von Neumann algebras \footnote{von Neumann algebras can be defined either through the condition of weak completion or by the equivalent bi-commutant condition ${\cal{A}}^{''} = {\cal{A}}$.}. In such a case and assuming locality in the form of $[{\cal{A}}(D),{\cal{A}}(D^c)]=0$ \footnote{Note that in the present example $D$ and $D^c$ are space like separated.} we get the split of 
${\cal{B}}({\cal{H}})$ as
\begin{equation}
{\cal{B}}({\cal{H}}) = {\cal{A}}(D)\otimes {\cal{A}}(D^c)
\end{equation}
provided ${\cal{A}}(D)$ and ${\cal{A}}(D^c)$ have trivial center \footnote{ That means that the center is generated by the identity operator. von Neumann algebras with trivial center are known as {\it factors}.}.
This split property of ${\cal{B}}({\cal{H}})$ is {\it the algebraic definition of quantum localized subsystem in QFT}. 

The familiar approach to entanglement appears when associated with the split of ${\cal{B}}({\cal{H}})$ we can define the split of the Hilbert space 
\begin{equation}
{\cal{H}} = {\cal{H}}(D) \otimes {\cal{H}}(D^c)
\end{equation}
with ${\cal{A}}(D) = {\cal{B}}({\cal{H}}(D))$ and ${\cal{A}}(D^c) = {\cal{B}}({\cal{H}}(D^c))$ \footnote{This split implies the existence of quantum micro states describing the physics localized in regions $D$ and $D^c$.}. In this case generic states in ${\cal{H}}$ are quantum superpositions $\sum_{i,j} c_{i,j} \psi^{D}_i \otimes \psi^{D^c}_j$ with $\psi^{D}_i$ and $\psi^{D^c}_j$ defining basis of ${\cal{H}}(D)$ and ${\cal{H}}(D^c)$ respectively and the reduced density matrices for $D$ or $D^c$ as well as the corresponding entanglement entropy can be defined in the usual manner.

The crucial property of local QFT is that for the algebra ${\cal{A}}(D)$ associated with a bounded region of space the corresponding split property of the Hilbert space {\it does not exist}. The mathematical reason is that in this case ${\cal{A}}(D)$ is a type $III$ factor. This apparently exotic mathematical fact has, however,  important physical consequences specially when we consider QFT on a curved space-time classical background \footnote{These problems reflect the deep relation between quantum entanglement and space-time geometry \cite{Raam,Malda}, unveiled by the AdS/CFT correspondence \cite{Ads1,Ads2,Ads3}.}. 

To understand these consequences let us consider the algebra ${\cal{A}}({\cal{D}})$ of local observables associated to a {\it causally complete domain} ${\cal{D}}$ of space-time. Let $t$ be the global time on ${\cal{D}}$ and $\hat H$ a bounded Hamiltonian defining time evolution on ${\cal{A}}({\cal{D}})$ as $a(t)= e^{-i\hat H t}a e^{i\hat Ht}$ for $a$ any element in ${\cal{A}}({\cal{D}})$ \footnote{Note that for a causally complete domain of space time we can always define a global time $t$ foliating the domain.}.

For a curved space-time where ${\cal{D}}$ is covering a {\it subregion of the space-time} we can define the complementary region ${\cal{D}}^c$ and to identify as a {\it horizon} the common boundary of ${\cal{D}}$ and ${\cal{D}}^c$. In that case we can extend the former discussion to 
${\cal{A}}({\cal{D}})$ and ${\cal{A}}({\cal{D}}^c)$ and to ask for the corresponding split of the full Hilbert space ${\cal{H}}$ describing QFT in the full space-time including the causally complete domain ${\cal{D}}$ and its complement ${\cal{D}}^c$. In other words we can ask ourselves for {\it the quantum entanglement induced by the existence of a classical background horizon}. As examples of this situation we can consider the case where ${\cal{D}}$ is the static patch of an observer in de Sitter with ${\cal{D}}^c$ the complementary region in the planar patch\footnote{The region beyond the cosmological horizon of the observer.} or the case where ${\cal{D}}$ is identified with the exterior of a black hole \footnote{The causally complete domain of the asymptotic observer.} and ${\cal{D}}^c$ with the black hole interior. 

For ${\cal{A}}({\cal{D}})$ a type $III$ factor we cannot associate to ${\cal{D}}$ neither quantum micro states not finite density matrices. Intuitively the reason is that since in this case the Hilbert space ${\cal{H}}$ cannot be split, then we cannot define any reduced density matrix for describing ${\cal{D}}$ because no formal trace on ${\cal{H}}({\cal{D}}^c)$ is available given the lack of existence of ${\cal{H}}({\cal{D}}^c)$ itself.

Another related consequence is that $\hat H$ defining the time translations in ${\cal{D}}$ is {\it ill defined}. Intuitively we could have advanced that by simply thinking that if $\hat H$ is well defined we could have also a well defined density matrix, formally defined as $e^{-\hat H}$, which, as said, for type $III$ factors is not the case. The physics reason are the divergent quantum fluctuations $\Delta(\hat H^2)$ of $\hat H$ on any state $|\psi\rangle$ in ${\cal{H}}$. This divergence directly implies that $\hat H|\psi\rangle$ is not in ${\cal{H}}$ \footnote{Obviously in that case the state $\hat H|\psi\rangle$ is not normalizable.}.

This situation can look rather paradoxical mostly because, and based on the intuition provided by the well known Unruh effect \cite{unruh}, we are used to describe the physics either in the black hole exterior or in the dS static patch, two typical regions bounded by a horizon, in terms of a thermal density matrix and to use those density matrices to assign the Bekenstein-Hawking entropy (\cite{Bek} and \cite{H}) to the black hole or the Gibbons-Hawking entropy \cite{GH} to dS. Actually most of the mysteries that surround these classic results, derived using an Euclidean extension,  have their root in the, a priori unpleasant, type $III$ nature of the underlying algebras of local observables \footnote{One early use of von Neumann algebras in the context of black hole physics can be found in \cite{Papa1} and \cite{Papa2} where the Tomita Takesaki theory was used as the natural tool to define ( reconstruct ) the BH interior ( the definition of mirror operators ) in terms of the algebra of observables describing the BH exterior.}.

Once we localize the problem in the divergent quantum fluctuations of $\hat H$ we can try to improve the situation identifying some mathematically precise way to tame or to {\it regularize} these entanglement divergences. This can be done if we manage to modify the QFT type $III$ factor ${\cal{A}}({\cal{D}})$ into a more tractable type $II$ factor where we can describe the local physics in ${\cal{D}}$ in terms of density matrices. 

The most natural way to try to tame the entanglement infinities, induced by the type $III$ nature of the QFT algebras of observables, could be to include a natural UV cutoff to regulate the universal UV divergences of the QFT entanglement \cite{sorkin}. A reasonable approach will consist in using quantum gravity effects, organized as a perturbative series in powers of $G_N$, to do the job. Obviously the possibility that quantum gravity corrections could modify a QFT type $III$ factor into a more tractable type $II$ factor with at least well defined density matrices or, in the best case, into a type $I$ factor \footnote{Let us note, on passing, that some of the most recent interesting results in the understanding of the black hole information paradox, as for instance the derivation  of Page's curve \cite{Page} (see \cite{Almieri} for a recent review) are based on the so called {\it Central Dogma} \cite{Almieri} that, in essence, postulates a type $I$ description of the black hole exterior. How to develop the transition from the QFT type $III$ into a type $I$ is a deep and open question. A possibility is that the dynamics underlying this transition is somehow deeply related with the mechanism of {\it classicalization} \cite{class1,class2}.}. with well defined quantum micro-states is fascinating and at the very core of the 
holographic reconstruction problem \footnote{Indeed the general problem of reconstructing localized bulk operators in terms of boundary operators is, in essence, a UV problem where the bulk energy translates into the {\it operator complexity} of the boundary operator. In this sense any UV completion should host arbitrarily highly complex operators becoming, most likely,  a type $III$ factor.}.

 The preliminary question on how quantum gravity corrections could modify the nature of a QFT type $III$ factor into a type $II$ has been recently addressed by Witten in \cite{Witten1}.

This approach is however not fully satisfactory for explaining basic QFT results, in curved space-time with horizons, derived in the weak gravity limit. For example, the black hole Bekenstein-Hawking entropy and the Gibbons-Hawking de Sitter entropy or, in a different context, the power spectrum of primordial quantum fluctuations in Inflationary Cosmology.

Recently in \cite{Witten2,Witten3} a new approach to modify, {\it in the weak gravity limit},
 a QFT type $III$ factor into a type $II$ factor has been suggested on the basis of {\it crossed products}. In this approach you extend the algebra of observables of the QFT adding a quantum algebra representing a reference frame observer. The importance of this approach is to provide a way to regularize the QFT type $III$ divergences associated with $\Delta(\hat H^2)$ {\it without invoking quantum gravity order $G_N$ effects or in the case of large $N$ without invoking $\frac{1}{N}$ effects} \footnote{To regularize the type $III$ infinite entanglement without adding any fundamental quantum gravity UV cutoff is, in spirit, analogous to the standard renormalization program in QFT. In such a case renormalizability, in the Wilsonian sense, sets the formal {\it renormalization group} conditions under which a well defined theory can be defined in the limit of infinite UV cutoff. In the holographic setup the analog of the Wilsonian approach could be designed in terms of quantum error correcting transformations ( see \cite{QEC} for a recent attempt).}

 In a nutshell the crossed product approach is based on the following steps \footnote{Along the paper these steps will be discussed in detail. Here we simply try to stress the logic flow of the construction.}.
 
 {\bf Step 1} For a given causally complete domain ${\cal{D}}$ with associated type $III$ factor algebra 
 ${\cal{A}}({\cal{D}})$ define a {\it reference frame} Heisenberg algebra
 \begin{equation}\label{RFalg}
 [\hat H_{RF},\hat t] = i {\hbar}
 \end{equation}
 with $[\hat H_{RF},{\cal{A}}({\cal{D}})]=0$ and with $\hat H_{RF}$ and $\hat t$ having continuous real spectrum. As usual the operator $\hat H_{RF}$ can be represented as $-i{ \hbar}\frac{d}{dt}$ for $t$ representing the spectrum of $\hat t$. Although we will normally use the standard convention $\hbar =1$ we will keep $\hbar$ in this case to track the quantum effects associated to the added reference frame algebra.
 
 {\bf Step2} For a given GNS representation of ${\cal{A}}({\cal{D}})$\footnote{See definitions in \cite{wittenrev1} and in section 5.} we identify the spectrum of $\hat t$ with the "time" parameter of the outer automorphisms $a(t) = e^{-i\hat ht} a e^{i\hat h t}$ defined by the Tomita Takesaki modular Hamiltonian $\hat h$ associated with ${\cal{A}}({\cal{D}})$ for a given GNS reference state $|\Psi\rangle$. Using this identification we define the {\it regularized} and state dependent type $II$ Hamiltonian as
 \begin{equation}\label{defhamiltonian}
 \hat h^{II}= \hat h + \hat H_{RF}
 \end{equation}
 Intuitively the definition (\ref{defhamiltonian}), replacing the non existent hamiltonian $\hat H$ by {\it the regularized} $\hat h^{II}$, is equivalent to define a type $II$ modular hamiltonian $\hat h$ as $\hat h^{II} -\hat H_{RF}$ with $\hat H_{RF}$ in the commutant ${\cal{A}}'({\cal{D}})$.
 
 {\bf Step 3} We define an extended Hilbert space ${\cal{H}}^{ext}$ as ${\cal{H}}^{GNS} \otimes {\cal{H}}^{RF}$ with 
${\cal{H}}^{RF}$ an irrep of the reference frame Heisenberg algebra (\ref{RFalg}) i.e. $L^2(R)$ for $R$ the real spectrum of $\hat t$. Product states in ${\cal{H}}^{ext}$ are of the type
\begin{equation}
|\Psi\rangle \otimes |\psi\rangle_{f}
\end{equation}
with $|\psi\rangle_{f}= \int d\epsilon f(\epsilon) |\epsilon\rangle$ for $\hat H_{RF}|\epsilon\rangle = \epsilon|\epsilon\rangle$ and for the quantum wave function $f$ satisfying $\int d\epsilon|f|^2=1$. These states are characterized by the quantum uncertainties $\delta(H_{RF})$ and $\delta(t)$ satisfying Heisenberg relation 
\begin{equation}
\delta(H_{RF}) \delta(t) \geq \hbar
\end{equation}

{\bf Step 4} In this extended Hilbert space we get states with the desired regularization of quantum entanglement fluctuations in the form of finite $\Delta(\hat h^{II})^2$.

The crucial property of this construction lies in the {\it non vanishing value} of $\hbar$ in (\ref{RFalg}). Indeed, in the limit $\hbar=0$, we could get an extended algebra generated by ${\cal{A}}({\cal{D}})$ and $\hat H_{RF}$ that is still type $III$ although {\it not } a factor \footnote{Since in this extended algebra $\hat H_{RF}$ is central.}. For this extended algebra we can define {\it classical statistical ensembles} on the spectrum of $\hat H_{RF}$ with an statistical finite value of $\Delta(\hat H_{RF})^2$ set by the probability distribution defining the ensemble. However only for $\hbar\neq 0$ we get the crossed product and the regularized $\hat h^{II}$ Hamiltonian. Indeed only for $\hbar\neq 0$ $\hat H_{RF}$ becomes a generator of "time" translations. \footnote{Note that is the $\hbar$ in (\ref{RFalg}) what effectively changes the statistical ensemble, formally defined on the spectrum of a central element, into a real quantum state in the Hilbert space representation of (\ref{RFalg}).}.

An obvious question at this point is to try to relate $\hbar$ defining the reference frame algebra with a potential modification of the type $III$ into type $II$ based on quantum gravity corrections or $\frac{1}{N}$ corrections. This is equivalent to identify $\hbar \sim G_N$ or $\hbar \sim \frac{1}{N}$ and consequently to erase type $II$ $\hbar$ quantum effects in the weak gravity $G_N=0$ limit or in the $N=\infty$ limit. In that case finite quantum entanglement fluctuations in the $G_N=0$ limit or in the $N=\infty$ limit can be only defined using an statistical ensemble. Note that if we define formally $\hat H_{RF} = \hat H^0_{RF} + \hbar \hat H_{RF}^1$ with $\hat H^0_{RF}$ a {\it central element}, then $\hat h^{II} = (\hat h + \hat H^0_{RF}) + \hbar \hat H_{RF}^1$. Thus if we think the quantum corrections defining the regularized $\hat h^{II}$ as quantum gravity or $\frac{1}{N}$ corrections we get the identification $\hbar \sim G_N$ or $\hbar \sim \frac{1}{N}$.

In this paper we are interested in taking the $G_N=0$ and $N=\infty$ limits but working in a type $II$ factor with non vanishing $\hbar$ for the corresponding reference frame algebra. In summary we try to address the question: 

{\it What is the physical meaning of quantum $\hbar$ type $II$ effects in the $G_N=0$ limit or in the $N=\infty$ limit ?}

Since by construction they are neither quantum gravity effects nor $\frac{1}{N}$ effects, then: What they represent ?

In the type $II$ factor obtained after adding the reference frame algebra (\ref{RFalg}) we have a well defined density matrix description of the physics in the causally complete domain ${\cal{D}}$. As shown in \cite{Witten1,Witten2,Witten3} when we focus our attention on the associated entropy we naturally get, for any product quantum state in the extended Hilbert space, a {\it generalized entropy}\cite{Bek2} that we can formally think as composed of two pieces $S=S_{{\cal{D}}} + S_{RF}$ with $S_{{\cal{D}}}$ depending on the modular properties of ${\cal{A}}({\cal{D}})$ and with $S_{RF}$ depending on the {\it added} reference frame algebra. The cancellation of entanglement divergences  in $S=S_{{\cal{D}}} + S_{RF}$, in the weak gravity limit,  is, in part, a sort of {\it normal ordering renormalization} and is based on the underlying "symmetry" of the type $II$ factor under transformations $\hat H_{RF} \rightarrow \hat H_{RF} + cte$ defining the {\it fundamental group}\cite{Connes} of the type $II$ factor \footnote{The relation between the type $II$ "regularization" and generalized entropies nicely agrees with the original intuition about cancellations of divergences originally discussed in \cite{Unglum}.Very interesting formulas for generalized entropy in the AdS/CFT holographic context has been recently derived using Euclidean quantum field theory and the replica trick. In particular the generalizations of the Ryu-Takayanagi \cite{RT, HRT} prescription, the Lewkowycz- Maldacena prescription for bulk quantum corrections \cite{LM} and the QES prescription \cite{QES1,QES2,QES3}.}.

An interesting place where type $II$ quantum effects, surviving in the weak gravity limit, can have important and {\it measurable} physical consequences is in the context of {\it Inflationary Cosmology} \cite{Gomezcross} ( see also \cite{corean} for a different approach). Obviously the relevant quantum features of Inflation, that we measure in the CMB spectrum are properties of a QFT -- defined in a curved background describing the primordial slowly rolling phase -- derived in the weak gravity limit (see \cite{mukhanov0,mukhanov1} and reference therein). Thus a natural question is: 

{\it What is the physical meaning of the type $II$ quantum fluctuations $\Delta(\hat h^{II})^2$ in the type $II$ modification of pure de Sitter ?} 

And, more specifically: {\it How these weak gravity quantum effects, evaluated on different states in the extended Hilbert space, are related with the standard definition of the power spectrum in the slow rolling models of Inflation ?}

An heuristic hint can be designed along the following simple qualitative argument. After adding the reference frame algebra (\ref{RFalg}) and keeping $\hbar$ finite and non vanishing we get unavoidably quantum fluctuations $\delta(\hat t)$ on the state representing, in the extended Hilbert space, the ground state. It is pretty natural to interpret the reference frame algebra as defining a {\it quantum physical clock}. In that sense the reference frame algebra, used to define the type $II$ factor, plays the role of the inflaton that, naturally, defines {\it a clock}. Thus you can try to link the type $II$ quantum uncertainty $\delta(\hat t)$ with the typical time delay at horizon crossing and to think of $\Delta(\hat h^{II})^2$ as representing the finite power spectrum. More precisely you could expect to get, from the type $II$ quantum effects in the weak gravity limit, direct type $II$ information about inflationary parameters. In more concrete terms we interpret the pure dS divergences of the power spectrum for gauge invariant fluctuations as a type $III$ effect and we identify the type $II$ crossed product version as the quantum definition of the inflationary slow rolling phase. 

Note that in this approach {\it slow roll appears as a type $II$ "regularization" of the pure de Sitter type $III$ factor}. Hence Inflation is fully encoded {\it in the quantum effects of a type $II$ crossed product version of pure de Sitter at weak gravity limit}. 
In more intuitive words slow roll parameters as $\epsilon$,  play the role of the non vanishing $\hbar$ in the reference frame algebra (\ref{RFalg}) that you need to add in order to transform, in the weak gravity limit, the type $III$ factor of pure de Sitter into a type $II$ factor. We will develop this discussion in section VII.

In \cite{LL1,LL2} it has been suggested that the high temperature CFT algebra of single trace operators in the large $N$ limit of pure Yang Mills in finite volume is a type $III_1$ factor.
Thus, following the same logic, we can consider the non trivial quantum type $II$ effects  in the description, in the large $N$ limit, of the high temperature Hagedorn phase of large $N$ gauge theories in finite volume\cite{GW,Wadia,Wittenconf,Sudorg,Minwalla1, Minwalla2,Minwalla3,gomezliu}. 

In this case the addition of the reference frame algebra provides a type $II$ definition of the order parameter which we identify again with $\Delta(\hat h^{II})^2$. In the matrix model representation we can identify the type $II$ quantum effects ( although not $\frac{1}{N}$ effects ) with intrinsic {\it quantum fluctuations} in the {\it size} of the interval where eigenvalues are confined in the high temperature phase. The natural definition of generalized entropies in the type $II$ frame allows us to associate the added reference frame algebra with the quantum dynamics of the interval of eigenvalues. These fluctuations are a consequence of adding the full fledged reference frame algebra with a non trivial and finite $\hat t$ conjugated operator. Since, in the holographic setup, the distribution of eigenvalues represents a bulk geometry the corresponding type $II$ quantum fluctuations could explain the dynamics underlying the extremality of the celebrated Ryu-Takayanagi (RT) \cite{RT, HRT}  prescription. In other words the extremal surfaces defining, in the RT prescription, the entanglement wedges have, even in the $N=\infty$ limit, quantum fluctuations reflecting the quantumness of (\ref{RFalg}). As we will see, in this approach, the first moment of the eigenvalue distribution is determined by the entanglement capacity of the ground state
in the type $II$ extended Hilbert space.

The outline of the paper is as follows. In section 2 we review some of the typical problems associated with localization of information in QFT, stressing the potential meaning of the {\it nuclearity} condition in the case of Hagedorn phase. In section 3 we try to provide a simple physical intuition to answer the question: Why we need a reference frame algebra and what is its meaning ? To do that we use the quantum information approach to superselection rules \cite{RF} \cite{kitaev}. In fact the use of {\it reference frame} terminology along the paper comes from this analysis. In section 4 we summarize the mathematical structure of crossed products in an intuitive manner stressing the difference between crossed products and statistical ensembles. In section 5 we focus on de Sitter and we review the recent results in \cite{Witten2}. In section 6 we discuss the idea of purifications in the particular context of the de Sitter planar patch. In section 7 we present the Type $II$ version of Inflationary Cosmology and finally in section 8 we discuss the type $II$ quantum effects for the large $N$ Hagedorn phase \footnote{Unfortunately the paper becomes longer than needed probably reflecting that the dust is not settle down yet, at least in what concerns the author understanding.}.

\section{Localization of Information and Factors}
In the context of quantum gravity the general problem of {\it localization of information} in QFT has recently attracted some attention. For a recent discussion and relevant references see \cite{Papa3}. In this section we will review what we consider the key aspects of this discussion from the point of view of algebraic quantum field theory. Although most of what follows in this section is well known it will be useful to set notation and to frame the problem. 

In a nutshell the problem of localizing information in QFT can be presented as follows. Consider a spacelike hyper surface $\Sigma$ divided into two regions, a bounded region $D$ and its complement $D'$. For simplicity let us imagine we work in Minkowski space-time. Let us now define the algebra ${\cal{A}}(D)$ of local operators with support in the bounded region of space $D$. For a generic quantum state $|\phi\rangle$ define the state $|\phi'\rangle \equiv a|\phi\rangle$ with $a$ a {\it unitary} operator in the algebra ${\cal{A}}(D)$. Then you will say that the information associated with the unitary operator $a$, is, for the state $|\phi\rangle$, localized in the region $D$ if the states $|\phi\rangle$ and $|\phi'\rangle$ {\it canont be distinguished by performing any local measurement in $D'$}. This means that for any local operator $O$ with support in $D'$ you will get that
\begin{equation}\label{local}
\langle \phi'|O|\phi'\rangle = \langle \phi |O|\phi\rangle
\end{equation}
This equality follows trivially if we assume $[O,a]=0$ and that $a$ is a unitary operator. Algebraically this means that the algebra ${\cal{A}}(D')$ of local operators with support in the complement of $D$ are in the commutant of ${\cal{A}}(D)$. 

Until this point we have not considered any time evolution. Obviously to give a stronger sense to the idea of localization we should consider the time dependent expectation values 
$\langle \phi'_t|O|\phi'_t\rangle$ and $\langle \phi_t |O|\phi_t\rangle$ defined by some Hamiltonian $H$ describing the 
full system with $|\phi_t\rangle =e^{iHt}|\phi\rangle$. Thus for given $a$ representing the information we try to localize in $D$ we can define the function
\begin{equation}\label{distin}
f(t) = \max_{O\in {\cal{A}}(D')}|\langle \phi'_t|O|\phi'_t\rangle - \langle \phi_t |O|\phi_t\rangle|
\end{equation}
Locality implies that the information can be localized if the function $f(t)$ is zero  for $t$ smaller than $t_c=\frac{d}{c}$ with $d$ representing the distance on $\Sigma$ between the support of $a$ and the complementary region $D'$. However for times larger than $t_c$ we can expect that some signal can go from the localization of $a$ into $D'$ leading to $f(t)\neq 0$ for $t>\frac{d}{c}$.

In order to understand the meaning of (\ref{distin}) note that in QFT we could define the probability to detect in $D'$ the information located in $D$ after a time $t$ using a local projection operator $P\in {\cal{A}}(D')$ as
\begin{equation}
 g(t)= \langle \phi'_t|P|\phi'_t\rangle
\end{equation}
From locality we know that $g(t)=0$ for $t<t_c$. Since $P$ is a projection i.e. $P^2=P$ we can write $g(t)= \langle \phi'_t|P^2|\phi'_t\rangle$ which is equal to
\begin{equation}
|| P |\phi'_t\rangle ||^2
\end{equation}
and therefore we get that $P|\phi'_t\rangle =0$ for $t<t_c$.
However in QFT is well known that if the Hamiltonian $H$ is bounded the vector valued function $P|\phi'_t\rangle$
can be analytically extended in time and consequently if it is zero on some finite interval it is {\it identically zero}. Thus we conclude that $P|\phi'_t\rangle$ is identically zero and therefore $g(t)$ is also identically zero. That means that although the causal domains of $D$ and $D'$ will  overlap after a time $t>t_c$ no signal carrying information from $D$ can be detected in $D'$ {\it using a local projection} $P\in {\cal{A}}(D')$.  

This situation is sometimes known as the Fermi paradox ( see \cite{Fermi1,Fermi2,Fermi3,Fermi4} ). In that presentation you imagine an excited atom $A$ in $D$ and you ask yourself about the probability to detect in $D'$ the decay of atom $A$. You can imagine a localized atom $B$ in $D'$ at certain distance $d$ from the atom $A$ and you ask about the probability of atom $B$ to absorb the energy emitted in the decay of atom $A$. If the corresponding 
$g(t)$, is defined for an arbitrary {\it bounded} Hamiltonian describing the two atoms ( including the radiation ), then you will reach the paradoxical conclusion that the probability to detect the decay of atom $A$ localized in $D$ using a projection operator localized in $D'$ is simply zero.

The solution to this puzzle is that ${\cal{A}}(D')$ is a type $III$ factor and consequently the projection $P$ simply {\it does not exist}. However we can have for ${\cal{A}}(D)$ and ${\cal{A}}(D')$ type $III$ factors $f(t)$ satisfying the conditions of information localization namely vanishing for $t<t_c$ but non vanishing for $t>t_c$\footnote{Note that for $f$ you cannot use the former analyticity argument.}.

A similar situation in the context of holography can be easily defined \cite{Papa3}. Imagine now the CFT algebra of local observables localized in a time band. Let us ignore for the time being the formal problem of how to define this algebra in such a way that is different i.e a sub algebra, of the total algebra ${\cal{A}}_{CFT}$ of the CFT. We will come back to this discussion in future sections. For the time being you can think the algebra ${\cal{A}}_{\delta}$, associated with a time band of size $\delta$ around $t=0$, as fully characterized by sufficiently simple operators, for instance finite products of single trace operators, unable to describe the time evolved operators for some time larger than $\delta$ \footnote{To define this time band algebra we need to give sense to some product, defined for instance using some OPE, naturally truncated to the so defined space of simple operators \cite{LL3}.}. Let us define the bulk {\it causal wedge} of the time band. Thus in the bulk we can identify an interior causal diamond that is space like with respect to the causal wedge of the time band. In this case you can define $D$ as the intersection of the interior causal diamond and the bulk hyper surface at time $t=0$ and $D'$ as the corresponding intersection of the causal wedges of the time band. Now our former discussion on information localization can be posed as the question of weather we can add some information in $D$, the interior causal diamond, such that cannot be detected using the boundary algebra ${\cal{A}}_{\delta}$. As in the previous discussion this information could be localized in $D$ if the corresponding unitary operator $a$ commutes with ${\cal{A}}_{\delta}$. In such a case a given state $|\psi\rangle$ describing the bulk and the state $a|\psi\rangle$ will be indistinguishable with respect to the algebra ${\cal{A}}_{\delta}$. However the state $a|\psi\rangle$ could be distinguished if we use operators in ${\cal{A}}_{CFT}$ located outside the time band. In this example the size of the time band plays the role of $t_c$ in the former simpler example. In \cite{Papa3} it is argued that the corresponding function $f(t)$ for $|\phi\rangle$ a CFT state dual to a semiclassical AdS geometry satisfies the conditions of information localization introduced above. 

The typical problem of this sort of approach is related with the Gauss law. In fact for $a$ to commute with ${\cal{A}}_{\delta}$ we need to impose that all the features associated with $a$ that can be measured using asymptotic charges i.e. using the associated Gauss law, cannot be practically resolved using ${\cal{A}}_{\delta}$.

\subsection{Split property and species}
In the former discussion of information localization is implicitly assumed that we can prepare the state in $D$ where we want to localize the information, independently of what is in the complement $D'$. In order to do that we need to eliminate the entanglement between $D$ and $D'$.  This can be done, for two commuting von Neumann algebras acting on the same Hilbert space, if they satisfy the condition known as {\it causal independence}. Generically for two commuting algebras ${\cal{A}}(D_1)$ and ${\cal{A}}(D_2)$ causal independence is defined by means of the split property as the existence of a type $I$ factor ${\cal{N}}$ such that
\begin{equation}
{\cal{A}}(D_1)\subset {\cal{N}} \subset {\cal{A}}'(D_2)
\end{equation}
where by ${\cal{A}}'(D_2)$ we mean the commutant of ${\cal{A}}(D_2)$. In this case the Hilbert space ${\cal{H}}$ can be decomposed as ${\cal{H}}_1 \otimes {\cal{H}}_2$ with ${\cal{N}} = B({\cal{H}}_1)$ and ${\cal{N}}' = B({\cal{H}}_2)$. Thus we can define {\it split states} in the Hilbert space ${\cal{H}}$ as those states for which for any $a\in{\cal{A}}(D_1)$ and $b\in{\cal{A}}(D_2)$ we have that
\begin{equation}
\langle \phi|ab|\phi\rangle = \langle \phi_1|a|\phi_1\rangle \langle \phi_2|b|\phi_2\rangle
\end{equation}
for two states $|\phi_1\rangle \in {\cal{H}}_1$  and $|\phi_2\rangle  \in {\cal{H}}_2 $. 

In QFT is expected to have this split property provided some condition on the growth with energy of the local density of states is satisfied. More precisely if we have a bounded region $D$ of space we can define in the full Hilbert space, the set of states ${\cal{L}}(D)$ representing quantum excitations of the vacuum localized in $D$. Now you can define the subset of ${\cal{L}}(D)$ of states with some energy smaller or equal to some given $E$ and to check how this subset increases size when you increase the energy $E$. In essence you check {\it how the number of particle species increases with energy}. It is important to keep in mind that in the Hagedorn phase when the increase of the number of species with energy leads to a maximal temperature {\it the split property is expected to fail} \cite{split1,split2,split3,split4}. This is potentially important in the holographic setup where we expect that the boundary CFT in the large $N$ limit can be in a Hagedorn high temperature phase. In such a phase we should not expect the existence of split states. We will briefly discuss this question in section 8.

Note that if we work on a split state we can in principle {\it localize species information} i.e. we can create a species excitation in a certain local region $D_1$ that cannot be detected performing experiments in $D_2$. This is due to the lack of entanglement between $D_1$ and $D_2$ in the split state. However if the theory enters into a Hagedorn phase at high temperatures this protocol for hiding species information can only work if the energy needed to create the split state is smaller than the typical Hagedorn temperature.

\subsection{Localization, Dressing and Commutant}
Generically Gauss's law constraints are defined for some asymptotic charge $Q$. Gauge invariant observables ${\cal{O}}$ should satisfy the Gauss law condition $[Q,{\cal{O}}]=0$. In the case of GR physical observables should be Diff invariant. In the simpler case of gauge theories they should be gauge invariant which implies $[Q,{\cal{O}}]=0$ for $Q$ the corresponding {\it charge}. Let us simply think in a $U(1)$ gauge theory with $Q$ representing the electric charge. In this case we can define two types of operators satisfying $[Q,{\cal{O}}]=0$, namely local operators that are {\it neutral}, but also {\it dressed charged} operators. 

For instance imagine ${\cal{O}}(x)$ a local charged operator satisfying 
$[Q,{\cal{O}}(x)]={\cal{O}}(x)$. The corresponding dressed operator can be defined as $\hat {\cal{O}}(x) = {\cal{O}}(x)
e^{ie \int_x^{\infty} A}$. Then if we reduce the discussion to {\it small gauge transformations} i.e. those gauge transformations vanishing at infinity, then we get $[Q,\hat {\cal{O}}(x) ]=0$. As it is obvious from the former typical dressing, the so dressed operator, although charged and gauge invariant, is {\it non local}. This creates an obvious problem for localization. Namely even if $x$ is in the domain $D$ of space the dressed operator $\hat {\cal{O}}(x)$ will not be in the commutant of ${\cal{A}}(D')$ and therefore condition (\ref{local}) will not be satisfied.

The former discussion can be easily posed in more abstract algebraic terms. The notion of localization is defined relative to an algebra ${\cal{A}}(D)$ representing local observables that can be measured performing local experiments in a bounded region of space-time $D$. In such a case localization requires to find a {\it gauge invariant physical observable} $\hat {\cal{O}}$ satisfying $[Q,\hat {\cal{O}}]=0$ and such that 
\begin{equation}\label{commutant}
[\hat {\cal{O}},{\cal{A}}(D)]=0
\end{equation}
i.e. such that $\hat {\cal{O}}$ is in the commutant ${\cal{A}}^{'}(D)$ of ${\cal{A}}(D)$. This reflects the intuition that the presence of the information associated with $\hat {\cal{O}}$ cannot be detected using any experiment performed in the region $D$ defining the algebra ${\cal{A}}(D)$ and therefore we can say that such information is {\it localized with respect
to the algebra ${\cal{A}}(D)$}. 

In the holographic setup we can think of ${\cal{A}}(D)$ as the boundary algebra associated with a  time band and the operator $\hat {\cal{O}}$ satisfying (\ref{commutant}) as an operator representing a {\it localized} ( relative to ${\cal{A}}(D)$) bulk excitation. In these conditions the corresponding function $f(t)$ defined in (\ref{distin}) will be zero for $t$ in the time band used to define ${\cal{A}}(D)$.

From the previous discussion we learn something interesting, recently stressed in \cite{Papa3}. Namely, in order to localize gauge invariant information relative to a Gauss law generator $Q$ for a given algebra ${\cal{A}}(D)$ the {\it necessary condition} is that the algebra of gauge invariant observables  ${{\cal{A}}(D)}^{Q}$ should have a {\it non trivial commutant}.

In case the generator $Q$ ( where in GR you can think instead of in $Q$ in the Hamiltonian generator $H$ of time translations ) has divergent quantum fluctuations $\Delta(Q^2) = \infty$ you should use a regularized type $II$ representation of the algebra of gauge invariant ( dressed) observables. This defines the type $II$ crossed product algebra, briefly described in the introduction, with non trivial commutant and well defined localization of gauge invariant observables.

An alternative {\it code subspace} approach can be defined in  the holographic setup using $H$ as the Gauss law operator $Q$. In this case the algebra ${\cal{A}}(D)$ associated with  a boundary time band will be a well defined subalgebra if we reduce the size $T$ of the time band in such a way that $a(T)= e^{-iTH} a e^{iTH}$ should admit a representation in terms of the simple operators used to define ${\cal{A}}(D)$ or equivalently if $a(T)$ can be represented in the corresponding {\it code subspace} associated with ${\cal{A}}(D)$. Then, heuristically, any {\it dressing} with respect to $H$ that involves integrating over a time period much larger than $T$ will make the corresponding dressed operator $\hat {\cal{O}}$ to effectively commute with 
${\cal{A}}(D)$  on the code subspace (i.e. $\langle \psi|[\hat {\cal{O}},{\cal{A}}(D)]|\phi_0\rangle = e^{-{\Delta(H^2)}}$ for $|\phi_0\rangle$ the state used to generate the code subspace as ${\cal{A}}(D)|\phi_0\rangle$ and $|\psi\rangle$ any state in the code subspace) and consequently to be in the {\it effective} commutant of ${\cal{A}}(D)$ \cite{Papa3}.

The connection between both pictures is an interesting issue that will not be discuss in this paper.

\subsection{A comment on global charges and the species scale}
At the light of the previous discussion let us say few words on the existence of black holes with localized conserved global charge. It is widely accepted that global charges are incompatible with quantum gravity \cite{Banks0}. The argument relies on semiclassical black hole dynamics and on the general Bekenstein \cite{Bekenstein} limits on the amount of information we can enclose in a finite region of space. Indeed if we have a global charge $Q$ we can define large black holes with this charge localized inside. The semiclassical Hawking evaporation of this black hole is not changing the value of $Q$ so we can always manufacture a final state of the semiclassical evaporation process that will violate Bekenstein's bound. 

Once we enclose the global charge $Q$ in the interior of the black hole, the type $III_1$ factor ${\cal{A}}$ describing the interior should be replaced, if we assume that the global charge is conserved, by ${\cal{A}}^{\hat Q}$ for $\hat Q$ the global charge generator. This simply means that {\it we cannot have the black hole in a quantum superposition of states with different values of the global charge $\hat Q$ \footnote{Equivalently no quantum localized fluctuation in the black hole interior changing the value of the global charge $Q$.}}. The former argument based on Bekenstein limits is essentially telling us that this algebra cannot be the correct description. So it looks that gravity forces us to modify this algebra into the crossed product algebra or in other words forces us to add the reference frame algebra. The added reference frame "Hamiltonian", that we have denoted as $\hat H_{RF}$ in the introduction, accounts for the existence of {\it quantum superpositions} of states, describing the black hole interior, with different charge and consequently makes the original charge $Q$ effectively not conserved. Thus the conjecture, basic in the "swampland program" \cite{vafa2}, on the non existence of global conserved charges seems to be equivalent to postulate that in presence of gravity, the global and central ( in case it is conserved) charge that you will like to assign to a black hole should be embedded into the Heisenberg sub algebra of a crossed product type $II$ factor where finite quantum fluctuations of the supposed global charge are unsuppressed as well as unavoidable \footnote{Note that this type $II$ argument can be developed in the weak gravity limit. In that sense a stronger version of the conjecture on the non existence of global charges in presence of gravity could be that {\it in QFT on a space-time classical background with horizons conserved global charges are not allowed in the type $II$ version}.}

A different way to address the problem can be presented using {\it particle species}. In those conditions you can say that the state representing a black hole cannot be a {\it split state}. The reason is that if we localize a certain number of different species inside i.e. a certain amount of a global conserved charge, then

 {\it The  nuclearity condition needed for statistical independence and existence of split states is necessarily violated at some point in the evaporation process}. 
 
 In general we can define the {\it species scale} \cite{species1,species2} as the scale at which the nuclearity condition is violated. At that scale species i.e. {\it flavors} cannot be localized inside the black hole.

\section{ Superselection rules and Reference frames}
In this section we will use superselection charges {\it to motivate} a natural way to introduce the notion of {\it reference frames} We will present this discussion in standard quantum mechanical terms i.e. using type $I$ factors to describe the algebra of observables. Thus some of the results presented in this section cannot be directly exported to those cases where the relevant algebra of observables is a type $III$ factor. 
\subsection{Definitions}
In Quantum Mechanics, symmetries with respect to a group $G$, are implemented by means of a representation
\begin{equation}\label{aut1}
\alpha: G \rightarrow Aut({\cal{A}})
\end{equation}
from the symmetry group $G$ into the automorphisms of the  algebra ${\cal{A}}$ of the physical observables. For the simple case of $G=U(1)$ we have 
\begin{equation}\label{auto}
\alpha(\phi) a = e^{i\phi \hat X}ae^{-i\phi \hat X}
\end{equation} 
with $\phi$ parametrizing the group element, $a$ any element in ${\cal{A}}$ and $\hat X$ the unitary operator generating the group action. We will refer to $\hat X$ as the {\it associated charge}. The charge $\hat X$ will be part of the algebra ${\cal{A}}$ if the corresponding automorphism (\ref{auto}) is {\it inner}. 

We will say that $\hat X$ is a {\it superselection charge} ((SS)-charge) if we declare that the algebra of physical observables should satisfy
\begin{equation}\label{comm}
[{\cal{A}},\hat X]=0
\end{equation}
Hence condition (\ref{comm}) is equivalent, in the case of dealing with type $I$ factors, to say that the algebra of {\it physical obsevables} is identical  to the invariant subalgebra $B({\cal{H}})^G$ of the full algebra $B({\cal{H}})$ of bounded operators
\begin{equation}\label{invariant}
{\cal{A}}=B({\cal{H}})^G
\end{equation}
with $B({\cal{H}})^G=\{a \in B({\cal{H}});\alpha(g)a=a\}$. Note
that the deep meaning of (\ref{invariant}) is to postulate that not all self adjoint operators represent physical observables. Once we identify the physical Hilbert space with an irrep of ${\cal{A}}$ the existence of a superselection charge $\hat X$ implies the Hilbert space decomposition
\begin{equation}
{\cal{H}} = \oplus_q {\cal{H}}_q
\end{equation}
with ${\cal{H}}_q$ the Hilbert space of states with eigenvalue $q$ of the SS-charge. 
Thus imposing that the algebra of physical observables ${\cal{A}}$ is equal to the invariant subalgebra $B({\cal{H}})^G$ implies that {\it quantum superpositions of pure states with different value of the SS-charge cannot be physically realized} \footnote{The axiomatic approach to SS-charges was initiated by Wigner ( see \cite{Wigner}). From this point of view condition (\ref{comm}) can be considered as a way to identify the algebra of physical observables that not necessarily should coincide with the whole algebra of self adjoint operators. The case of the electric charge was studied in \cite{Wigner2} and \cite{Wigner3}. The first discussion on the possibility to define coherent quantum superpositions of states with different eigenvalues of a SS-charge was done in \cite{AS}.}.

\subsection{SS-charges and reference frames}
In many cases the physical characterization of an observable requires to fix a {\it reference frame}. For instance if we are describing the photons of some optical system we need to set {\it the reference frame of polarizations}. In particular, in order to put in correspondence the measurements of two different observers, we need to know the explicit transformation relating the different reference frames used by the observers. Once we have such knowledge we can require {\it covariance} with respect to changes of reference frame. The interesting problem appears when we should work with {\it total absence of knowledge about the reference frame}.

An interesting observation appearing in the context of quantum information theory \cite{RF} is that {\it absence of knowledge about the reference frame leads to SS-charges}. Let us briefly explain this phenomena. 

Imagine a physical system ${\cal{S}}$ described by the algebra of observables ${\cal{A}}_{\cal{S}}$. Imagine now that in order to characterize the observables ( or at least part of them) you need to use some reference frame (RF) as for instance a polarization basis. Assume that it exists a group $G$ of transformations relating the different RF's. For simplicity we will consider this group to be compact and isomorphic to $U(1)$.

As we did in the previous section let us define the representation of $G$ into the automorphisms of ${\cal{A}}_{\cal{S}}$ and let us denote $\hat X_{\cal{S}}$ the corresponding generator. Obviously {\it if you don't have any information about the RF} the best you can do, in order to describe the system ${\cal{S}}$, is to use the $G$-invariant part of ${\cal{A}}_{\cal{S}}$ i.e. ${\cal{A}}_{\cal{S}}^G$. 

Now you don't impose ${\cal{A}}_{\cal{S}} ={\cal{A}}_{\cal{S}}^G$ for ${\cal{A}}_{\cal{S}}$ the algebra of physical observables of the system ${\cal{S}}$. You simply define the invariant subalgebra ${\cal{A}}_{\cal{S}}^G$ as the algebra {\it representing the maximal information you can have about the system ${\cal{S}}$ if you don't have any information about the reference frame needed to give observable meaning to the whole set of observables in the algebra ${\cal{A}}_{\cal{S}}$}.

Thus, for the observer that lacks information about the RF, the algebra of physical observables is ${\cal{A}}_{\cal{S}}^G$ and $\hat X_{\cal{S}}$ works as the SS-charge. What about the Hilbert space of states ? Lack of information about the RF forces us to assign to ${\cal{S}}$, as Hilbert space of quantum states, the space of states with {\it fixed eigenvalue} of the SS-charge $\hat X_{\cal{S}}$. Hence, coherent superpositions of states with different eigenvalues of $\hat X_{\cal{S}}$ are {\it physically prohibited}. 

However, this observer can assign to ${\cal{S}}$ {\it mixed states} of the type
\begin{equation}\label{one1}
\rho = \int d\phi U(\phi) |\psi\rangle\langle\psi|U^{\dagger}(\phi)
\end{equation}
for $|\psi\rangle$ any pure state of ${\cal{S}}$ and for $U(\phi)=e^{i\phi \hat X_{\cal{S}}}$ with the integral in (\ref{one1}) defined with respect to the Haar measure of the group $G$. This mixed state reflects, as usual, a lack of knowledge, in this case the lack of knowledge of the observer about the RF. 

In order to fix ideas let us consider as a simple example an optical system ${\cal{S}}$ where photons can be in $K$ different polarization modes. The Fock space is generated by states
\begin{equation}
|n_1,n_2,...n_K\rangle
\end{equation}
with $n_i$ the number of photons in mode with polarization vector $e_i$. In this case changing RF is defined by a $U(1)$ rotation $U(\phi)$ acting on the polarization basis as
\begin{equation}
e_i' = U(\phi)_{i,j} e_j
\end{equation}
This RF transformation can be implemented on the algebra ${\cal{A}}_{\cal{S}}$, generated by the photon creation annihilation operators, using, as generator, the {\it total photon number operator} $\hat N_{tot}$. So, in this simple case, the total number of photons is, when we lack information about the RF of polarizations, playing the role of a SS-charge. The algebra describing the system in these conditions is
\begin{equation}
{\cal{A}}_{\cal{S}}^G = \{ a\in {\cal{A}}_{\cal{S}} ; [a,\hat N_{tot}] =0\}
\end{equation}
Moreover quantum superpositions of the type
\begin{equation}\label{two}
a|N_1\rangle +b|N_2\rangle
\end{equation}
with different total numbers of photons will not be physically realizable. 

Note that an obvious consequence of this fact is that the {\it quantum variance} of $\hat X_{\cal{S}}$ (that in the former example is $\hat N_{tot}$ ) is vanishing on all physically realizable states. Indeed this variance is only non vanishing for coherent superpositions of quantum states with different eigenvalues of $\hat X_{\cal{S}}$. 

This last comment can be nicely described in terms of quantum information theory. Indeed, imagine an observer Alice that prepares a state of ${\cal{S}}$ using some RF. Let us denote this state $|\psi;\phi\rangle$ where $\phi$ is a parameter that carries the information about the concrete RF used by Alice to prepare her state. Let us now consider other observer Bob that lacks information about what reference frame has been used by Alice. The quantum information of Bob about the value of the RF parameter $\phi$ is encoded in the {\it quantum Fisher information}, namely in the variance of $\hat X_{\cal{S}}$ for the state $|\psi;\phi\rangle$. For Bob, that lacks complete information about Alice's RF, this quantum information is simply vanishing \footnote{Bob's information about Alice's reference frame requires that Bob can {\it distinguish} between states prepared by Alice with different values of the RF parameter $\phi$. This information is the quantum Fisher information that Bob can have about the RF used by Alice. Bob's total ignorance about Alice's RF implies that this quantum information is vanishing. Thus for Bob the state prepared by Alice is, always, an eigenvector with a well defined eigenvalue of the generator $\hat X_{\cal{S}}$.}. This in particular means that Bob will never associate to ${\cal{S}}$ coherent quantum superpositions of states with different eigenvalue of $\hat X_{\cal{S}}$ and will interpret $\hat X_{\cal{S}}$ as a SS-charge.

\subsection{Adding a RF (observer)}
A very interesting observation by Aharonov and Susskind \cite{AS} was that by adding a quantum mechanical system representing the RF we can effectively construct -- in the {\it extended Hilbert space} representing the system ${\cal{S}}$ and the quantum system ${\cal{R}}$ defining the RF -- states with {\it non vanishing variance of $\hat X_{\cal{S}}$}.

Let us see how this construction is working. The physical system ${\cal{R}}$ must represent a physical way to set the reference frame. For instance in the case the reference frame sets the basis of polarizations we can characterize quantum mechanically this system ${\cal{R}}$ using an angle operator $\hat \phi$ and its conjugated operator $\hat X_{\cal{R}}$ with \footnote{Generically we can use as $\hat \phi$ the {\it position} operator on the group $G$ that we will consider one dimensional and $\hat X_{\cal{R}}$ its conjugated momentum.}
\begin{equation}\label{rfalgebra}
[\hat \phi,\hat X_{\cal{R}}] = -i
\end{equation}
Let us denote the algebra generated by $\hat \phi$ and $\hat X_{\cal{R}}$ as ${\cal{A}}_{\cal{R}}$. The natural Hilbert space associated with the reference frame algebra (\ref{rfalgebra}) will be ${\cal{H}}_R = L^2(G)$ i.e. square integrable, relative to the Haar measure of the group, wave functions on the group $G$ of RF transformations. The algebra ${\cal{A}}_R$ can be identified with the algebra of bounded operators in $L^2(G)$.  Thus the algebra associated with the combined system ${\cal{S}}$ and ${\cal{R}}$ will be
\begin{equation}
{\cal{A}}_{\cal{S}} \otimes {\cal{A}}_{\cal{R}} 
\end{equation}
The generator defining changes of RF for the {\it whole} system is
\begin{equation}
\hat X = \hat X_{\cal{S}} + \hat X_{\cal{R}}
\end{equation}
Thus we can now impose that $\hat X$ is a SS-charge for the whole system that includes the quantum mechanical system representing the RF. In that case the algebra of physical observables of the combined system will be
\begin{equation}\label{alg}
({\cal{A}}_{\cal{S}} \otimes {\cal{A}}_{\cal{R}})^G
\end{equation}
with $G$ being now generated by $\hat X = \hat X_{\cal{S}} + \hat X_{\cal{R}}$. Now comes the interesting connection with {\it crossed products} \cite{crossed} \cite{Witten1}. Namely the algebra (\ref{alg}) is generated by
\begin{equation}\label{elements}
\{ e^{i\hat \phi\hat X_{\cal{S}}} a e^{-i\hat \phi\hat X_{\cal{S}}}, \hat X_{\cal{R}} \}
\end{equation}
which is {\it the crossed product} of ${\cal{A}}_{\cal{S}}$ with the group $G$ of changes of reference frame acting as automorphisms of ${\cal{A}}_{\cal{S}}$ with generator $\hat X_{\cal{S}}$ \footnote{ The elements $\hat a$ in the algebra (\ref{elements}) satisfy $[\hat X_{\cal{S}}+\hat X_{\cal{R}},\hat a]=0$ as a consequence of the commutation relation (\ref{rfalgebra}) defining the reference frame algebra ${\cal{A}}_{\cal{R}}$.}. We will denote this algebra as
\begin{equation}
{\cal{A}}_{\cal{S}}^{cr}   
\end{equation}
In the extended Hilbert space representing the algebra (\ref{alg}) we will have the desired states with non vanishing variance of $\hat X_{\cal{S}}$. Moreover in this extended Hilbert space the SS-charge $\hat X$ will reduce the allowed pure states to those with a well defined eigenvalue of $\hat X$ \footnote{Coming back to the previous example of Alice and Bob we see that what Bob is trying to do is to {\it estimate} the RF  used by Alice. In quantum estimation theory \cite{Paris} the information available to Bob is the quantum Fisher information about the reference frame used by Alice. If this information is zero Bob will have infinite uncertainty i.e. total lack of knowledge about Alice reference frame. From quantum estimation theory this will means that the {\it quantum estimator} operator is ill defined. Adding a reference frame system is equivalent to add a well defined operator representing this quantum estimator.} . 

\subsection{A toy model example}
To put in more concrete terms the former abstract construction of the crossed product associated with a SS-charge we will use the simple optical model defined above. In this case the group $G$ are the transformations relating different polarization basis. The algebra ${\cal{A}}_{\cal{S}}$ is generated by creation annihilation operators $a_i$ and $a^{\dagger}_i$ of the photons with polarization $i$. The SS-charge is the total number operator $\hat N_{tot}$ and the invariant subalgebra ${\cal{A}}_{\cal{S}}^{\hat N_{tot}}$ the set of elements in ${\cal{A}}_{\cal{S}}$ commuting with $\hat N_{tot}$. Irreducible Hilbert space representations of the invariant subalgebra are the subspaces ${\cal{H}}_{N}$ representing the space of eigenstates of $\hat N_{tot}$ with eigenvalue $N$. Obviously $a_i$ and $a^{\dagger}_i$ are not part of ${\cal{A}}_{\cal{S}}^{\hat N_{tot}}$. 

Let us now define the RF algebra. This is defined by the operator $\hat X_R$ and $\hat \phi$ canonically conjugated to $\hat X_R$. Let us assume $\hat X_R$ has as spectrum the set of integer numbers and denote $|\epsilon\rangle_R$ the corresponding eigenvectors i.e. $\hat X_R|\epsilon\rangle_R =\epsilon|\epsilon\rangle_R$. In this case we can formally define the operator $\hat \phi$ conjugated to $\hat X_R$ as defining translations of $\epsilon$ i.e.
\begin{equation}
e^{i\alpha \hat \phi}|\epsilon\rangle = |\epsilon+\alpha\rangle
\end{equation}
for $\alpha$ an integer number. 

In these conditions we can map the algebra ${\cal{A}}_{\cal{S}}$ into the crossed product algebra ${\cal{A}}_{\cal{S}}^{cr}$ as follows
\begin{equation}
a_i\rightarrow \hat a_i = a_i\otimes e^{i\hat \phi}
\end{equation}
and
\begin{equation}
a^{\dagger}_i \rightarrow \hat a^{\dagger}_i = a^{\dagger}_i\otimes e^{-i\hat \phi}
\end{equation}
In the extended Hilbert space ${\cal{H}}_{\cal{S}}\otimes {\cal{H}}_R$ the subspace {\it representing the crossed product algebra} are formally states of the type \footnote{This statement will not be true in case the algebra ${\cal{A}}_{\cal{S}}$ is not a type $I$ factor. In this section, that is intended to motivate physically the notion of reference frame, we will assume ${\cal{A}}_{\cal{S}}$ to be type $I$. Thus in this discussion we are not assuming any form of space localization of our system.}
\begin{equation}
|\tilde N\rangle_f = \sum_{\epsilon} f(\epsilon)|\tilde N -\epsilon\rangle_S|\epsilon\rangle_R
\end{equation}
for a generic $\tilde N$ and with support of $f$ in the interval $[0,\tilde N]$ and such that $\sum_{\epsilon}|f(\epsilon)|^2 =1$. 

Note that these states are invariant under the action of $e^{i\alpha\hat X}$ for $\hat X=\hat X_S+\hat X_R$ \footnote{By invariant we mean that $e^{i\alpha\hat X}$ is acting as the identity on the projective space of rays.}. The action of the elements in the crossed product algebra is well defined on these states, namely
\begin{equation}
\hat a_i |\tilde N\rangle_f = \sum_{\epsilon} f(\epsilon)|\tilde N -\epsilon-1\rangle_S e^{i\hat \phi}|\epsilon\rangle_R 
\end{equation}
i.e. $\sum_{\epsilon} f(\epsilon)|\tilde N -\epsilon-1\rangle_S |\epsilon +1\rangle_R$
and similarly for $a_i^{\dagger}$. 

Generically these states describe entanglement between the system ${\cal{S}}$ and the reference frame ${\cal{R}}$. Moreover the variance of $\hat X_{\cal{S}}$ and $\hat X_{\cal{R}}$ on these states
\begin{equation}
\Delta(\hat X_{({\cal{S,R}})}^2) = \langle \tilde N|\hat X_{({\cal{S,R}})}^2|\tilde N\rangle - (\langle \tilde N|\hat X_{({\cal{S,R}})}|\tilde N\rangle)^2
\end{equation}
will be non vanishing.
Note that the variance of $\hat X_R$ fixes the quantum uncertainty in the variable $\phi$ characterizing the reference frame. This uncertainty is given by
\begin{equation}
\Delta(\phi) = \frac{1}{\sqrt{\Delta(\hat X_{{\cal{R}}}^2)}}
\end{equation}

\subsection{Reference frames and topology: the case of $\theta$ vacua}
Next we can ask ourselves about SS-rules associated with topological charges ( see \cite{Gomez-s} for a previous discussion ). Let us consider as an example pure $SU(N)$ Yang Mills in four dimensions. In the temporal gauge $A_0=0$ and after compactifying the space to $S^3$ we can classify pure vacua by the homotopy group $\Pi_3(SU(N))$. These are the well known $|n\rangle $ vacua \cite{Jakiw1,Jakiw2}. 

Can the topological charge $n$ work as a superselection charge ? In principle we know the answer. Indeed if $n$ would be associated with a SS-charge we will not have quantum superpositions of states with different values of $n$. We know that this is not the case and that such quantum superpositions are the ones that define the physical $\theta$ vacua. Moreover quantum superpositions of states with different values of $n$ can be motivated, for instance, invoking instanton effects \cite{tHooft}. Let us assume we don't know all these results and that we try to apply the former reference frame formalism to create, without invoking instantons, quantum superpositions of states with different value of the topological charge $n$. The way we will proceed will be defining first the topological charge $\hat T$ such that $\hat T|n\rangle =n|n\rangle$.

Let us introduce the algebra ${\cal{A}}_{YM}$ of gauge invariant physical observables and let us define the automorphism in $Aut({\cal{A}}_{YM})$ by
\begin{equation}\label{aut}
e^{i\alpha \hat T} a e^{-i\alpha \hat T}
\end{equation}
for $a\in {\cal{A}}_{YM}$. Note that at this level $\alpha$ is just a  variable formally defining the action of transformations on ${\cal{A}}_{YM}$ with generator  $\hat T$. The automorphism (\ref{aut}) is the one playing the formal role of changes of reference frame that we parametrize by $\alpha$. Let us then formally interpret $\alpha$ as {\it a reference frame parameter}. Recalling the discussion in the previous section we will associate {\it total lack of information about $\alpha$ } with the invariant algebra
\begin{equation}\label{YM}
{\cal{A}}_{YM}^{\hat T}
\end{equation}
The physics described by this algebra corresponds to take $\hat T$ as a SS-charge which will prevent the existence of coherent quantum superpositions of states with different topological charge. Thus lacking total information about the $\alpha$ reference frame parameter is equivalent to promote the topological charge $\hat T$ into a superselection charge. We know that generically instanton effects create superpositions of states with different value of the topological charge $\hat T$ so the algebra (\ref{YM}) cannot be the right candidate for the physical algebra of observables of Yang Mills. In other words to work with the algebra (\ref{YM}) is equivalent to effectively eliminate any instanton tunneling effect.

However, we can follow the same conceptual path described in the previous section, namely {\it we can add a reference frame system}. This reference frame system will be defined by an operator $\hat \alpha$ and its conjugated operator that we will denote, as we did in the former examples, $\hat H_R$ with $[\hat \alpha,\hat H_R]=-i$. Let us denote the reference frame algebra generated by $\hat \alpha$ and $\hat H_R$ {\it the axion algebra} ${\cal{A}}_{axion}$ and the algebra of the combined system of YM and the reference frame
\begin{equation}
{\cal{A}}_{YM}\otimes {\cal{A}}_{axion}
\end{equation}
As before the {\it crossed product} algebra could be defined by
\begin{equation}\label{ax}
({\cal{A}}_{YM}\otimes {\cal{A}}_{axion})^{\hat T+\hat H_R}
\end{equation}

Let us now denote $U(g;1)$ a topologically non trivial gauge transformation acting on the $|n\rangle$ vacua as $U(g;1)|n\rangle =|n+1\rangle$. Obviously $U(g;1)$ is not part of the algebra ${\cal{A}}_{YM}^{\hat T}$ \footnote{This is the analog problem we have for creation annihilation operators for photons in our former example where the charge that would be playing the role of $\hat T$ is the total photon number $\hat N_{tot}$.}. However we can try to define $U(g;1)$ as an element of the extended crossed product algebra (\ref{ax}). In essence we re trying to embed the standard instanton effects as part of (\ref{ax}). Let us see how this can be formally done.

First of all let us figure out the typical states in the extended Hilbert space representation of the crossed product algebra (\ref{ax}). To do that let us fix an integer number $\tilde N$ {\it much bigger than one} and let us define the state
\begin{equation}
|\tilde N, \theta\rangle = \sum_{\epsilon =0}^{\tilde N} f_{\theta}(\epsilon)|\tilde N -\epsilon\rangle|\epsilon\rangle
\end{equation}
Here $\theta$ is just a way to parametrize the function $f_{\theta}$ used in the definition of the state. 

Now we can extend the action of $U(g;1)$ on this state as we did in the previous section in order to define the creation annihilation operators, namely
\begin{equation}
\hat U(g;1)|\tilde N, \theta\rangle = \sum_0^{\tilde N} f_{\theta}(\epsilon)|\tilde N -\epsilon +1\rangle|\epsilon-1\rangle
\end{equation}
In the formal limit of $\tilde N=\infty$ and defining $f_{\theta}(\epsilon)= C e^{i\theta \epsilon}$ for $C$ some constant \footnote{This constant should be fixed by the normalization condition of $f_{\theta}$.} we get
\begin{equation}
\hat U(g;1)|\tilde N, \theta\rangle = e^{i\theta}|\tilde N, \theta\rangle
\end{equation}
that is the typical transformation of the standard $\theta$ vacua under topologically non trivial gauge transformations. 

This leads us to suggest, as a way to represent the $\theta$ vacua, to use the large $\tilde N$ limit of the states $|\tilde N, \theta\rangle$ in the extended Hilbert space representing the crossed product algebra (\ref{ax}). Note also that the reference frame Hamiltonian $\hat H_R$ induces translations of $\theta$, namely
\begin{equation}
e^{i\alpha \hat H_R}|\tilde N, \theta\rangle = \sum_0^{\tilde N} f_{\theta}(\epsilon) e^{i\alpha\epsilon}|\tilde N -\epsilon\rangle|\epsilon\rangle = |\tilde N, (\theta+\alpha)\rangle
\end{equation}
where in the last equality we used $f_{\theta}(\epsilon)=e^{i\theta \epsilon}$. This translation of the $\theta$ angle is the reason we have denoted the reference frame algebra as ${\cal{A}}_{axion}$.

The non vanishing variance $\Delta(\hat T^2)$ on these states will be the natural definition of {\it the topological susceptibility} \cite{Wittentop1,Wittentop2}. From this point of view {\it the axion reflects the crossed product of the Yang Mills algebra by the action (\ref{aut})}.

\section{The mathematical picture}
In this section we will review, following \cite{Witten1,Witten2,Witten3}, the general mathematical picture used to transform a type $III_{1}$ factor into a type $II$ factor. The physics setup is QFT in a classical space-time background with horizons. The two main examples we will consider along the paper will be a black hole horizon and a cosmological horizon. We will work in the weak gravity limit $G_N=0$. In this limit we will ignore any back reaction of the background metric. We can, however, consider the quantum gravitational fluctuations described by the linearized gravity on the classical background. Hence quantum gravity effects $O(G_N)=O(\frac{1}{M_P})$ will not be taken into account in this construction of type $II$ factors.

The typical type $III_1$ factors that we will use will be the von Neumann algebra ${\cal{A}}$ of local observables associated with the {\it causal diamond} of an observer\footnote{By that we mean a causally complete domain of space-time.}. In the case of de Sitter geometry this region will be the static patch of the observer while in the case of the black hole geometry we will consider the {\it causal diamond} of the asymptotic observer that agrees with the exterior region of the black hole. We will assume that these algebras are, in both cases, type $III_1$ Murray von Neumann factors. From a physics point of view this assumption implies several things. First of all, the condition of being a von Neumann factor implies:

i) The existence of a Hilbert space ${\cal{H}}$ and a representation $\pi:{\cal{A}} \rightarrow B({\cal{H}})$ for $B({\cal{H}})$ the space of bounded operators such that:

i-1) $\pi({\cal{A}}) = \pi({\cal{A}})^{''}$ for $\pi({\cal{A}})^{'}$ the {\it commutant} of $\pi({\cal{A}})$ i.e. the set of operators in $B({\cal{H}})$ commuting with all elements in $\pi({\cal{A}})$.

i-2) The center of $\pi({\cal{A}})$ is trivial

From now on we will use ${\cal{A}}$ instead of $\pi({\cal{A}})$ for notational simplicity. Physically we can think of ${\cal{H}}$ as the Hilbert space describing QFT on the full space-time background.

If ${\cal{A}}$ is a {\it factor} we automatically get the {\it split property} for $B({\cal{H}})$ as
\begin{equation}
B({\cal{H}}) = {\cal{A}}\otimes {\cal{A}}'
\end{equation}

From the algebraic point of view {\it states} associated with ${\cal{A}}$ are defined as linear forms $f:{\cal{A}}\rightarrow R$ satisfying $f(a^*a)>0$ for $a\neq 0$ and $f(1)=1$. The physics meaning of {\it states} is to assign to any observable $a\in {\cal{A}}$ its {\it observable value} i.e. the value the observer will assign to this observable performing {\it local measurements in her causal diamond}. Now comes the first subtlety. In the cases we are interested the causal diamond of the observer is not covering the full space-time. This situation leads to three possible options:

{\it Option 1} The causal diamond is {\it completely disentangled} from the rest of space-time. In this case we can define a linear form $f$ on ${\cal{A}}$ that agrees with the expectation value $\langle \psi|a|\psi\rangle$ for $|\psi\rangle$ a {\it pure state} in the Hilbert space associated to the causal diamond. This possibility implies the split property for the full Hilbert space ${\cal{H}}$ into two pieces one describing the states in the causal diamond and the other the states in its complement.

{\it Option 2} The causal diamond is intrinsically entangled with the rest of space-time but this entanglement, as normally measured by the von Neumann entropy, can be {\it any real value}. In this case linear forms on ${\cal{A}}$ can define {\it density matrices} $\rho$ representing the lack of knowledge of the observer on the unobservable region i.e. the complement of her causal diamond.

{\it Option 3} The observer causal diamond is {\it infinitely entangled} with the rest of space-time. In this case does not exist any finite linear form $f$ on ${\cal{A}}$ satisfying the trace property or, equivalently, no density matrix description of the quantum physics taking place in the observer causal diamond.

Option 1 means that ${\cal{A}}$ is a type $I$ factor. Option 2 that ${\cal{A}}$ is a type $II$ factor and Option 3 that ${\cal{A}}$ is a type $III$ factor \footnote{This is a very qualitative description intended to provide some physical intuition. The classification of factors is properly defined in terms of the different types of {\it projection operators} in the algebra with type $I$ having minimal projections that we associate with pure states, the type $II$ having finite projections but not minimal and the type $III$ case with neither minimal nor finite projections.}. Thus the assumption that ${\cal{A}}$ is a type $III_1$ factor implies that the causal diamond of the observer is infinitely entangled with the rest of the space-time.

Moreover assuming that ${\cal{A}}$ is a type $III$ factor means that the Hamiltonian $H$ defining {\it local time translations in the causal diamond is ill defined} with divergent quantum fluctuations and consequently no Heisenberg like equations are available to define the time evolution of the observables in ${\cal{A}}$ \footnote{Recall that by construction ${\cal{A}}$ is associated with a causally complete domain so you can foliate the domain using equal time hyper surfaces $\Sigma_t$. The non existence of a well defined local Hamiltonian in the domain means that we cannot define the local observables on $\Sigma_t$ using the Heisenberg equations and the local observables on $\Sigma_{t_0}$ for $t_0 <t$.}.

Hence the question is:

How to tame these entanglement divergences ? 

\subsection{The crossed product recipe}
For ${\cal{A}}$ a type $III_1$ factor we know the existence of a {\it state dependent} outer automorphism generated by the Tomita Takesaki modular Hamiltonian $\hat h$\footnote{This Hamiltonian depends on a selected state $|\psi_0\rangle$ that satisfies $\hat h |\psi_0\rangle=0$ and should be denoted $\hat h_{ |\psi_0\rangle}$. To simplify the notation we will use $\hat h$ and we will specify the state if needed.}. The automorphism is defined by
\begin{equation}\label{modular}
e^{it\hat h}ae^{-it\hat h}
\end{equation}
The crossed product recipe consists in replacing the algebra ${\cal{A}}$ by the algebra ${\cal{A}}^{cr}$ generated by 
\begin{equation}
\{ e^{i\hat t \hat h}a e^{-i\hat t \hat h} , \hat h_{RF} \}
\end{equation}
with the added operators $\hat t$ and $\hat h_{RF}$ defining a standard position/momentum Heisenberg algebra 
\begin{equation}\label{reference}
[\hat t, \hat h_{RF}] =-i\hbar
\end{equation}
The algebra (\ref{reference}) is the reference frame algebra introduced in section 1. As we stressed there we keep in the definition the $\hbar$ defining the quantumness of this algebra. In what follows we will, for notational simplicity use the standard convention $\hbar=1$.
The new algebra ${\cal{A}}^{cr}$ is the desired type $II$ factor\footnote{Note that this is exactly the qualitative construction we were presenting for the case of superselection charges in the previous section.}. We will assign to $\hat h_{RF}$ units of {\it energy} and therefore we will introduce a {\it unit of length} $\beta$ such that $\beta \hat h_{RF}$ is dimensionless as it is the modular Hamiltonian $\hat h$.

As advanced in the introduction we will define the type $II$ {\it "renormalized Hamiltonian"} $\hat h^{II}$ as
\begin{equation}
\beta \hat  h^{II}= \hat h + \beta\hat h_{RF}
\end{equation}

Elements  $\hat a \in {\cal{A}}^{cr}$ are defined by
\begin{equation}\label{operator}
\hat a = \int dt a(t) e^{it \beta h^{II}}
\end{equation}
for $a(t)$ a square integrable function from $R$ ( the real line) into ${\cal{A}}$. Note that the key ingredient in the construction of the crossed product algebra lies in adding the Heisenberg algebra (\ref{reference}). Based on the discussion in the previous section we will denote this algebra as the {\it reference frame} RF algebra. Intuitively you can think this algebra as associated with a {\it physical clock} with $\hat h_{RF}$ the clock Hamiltonian and with the spectrum of $\hat t$ representing the values of the time measured by such a clock.  We call this algebra a reference frame algebra since it plays the role of setting a reference for measuring physical time. In summary the crossed product construction is based on i) changing the c-number $t$ in (\ref{modular}) by the {\it operator} $\hat t$ and ii) to add the conjugated reference frame Hamiltonian $\hat h_{RF}$. 

The factor $e^{it \beta h^{II}}$ in (\ref{operator}) can be interpreted as a {\it dressing factor} (in the sense given to dressing in section 2) needed to define the elements in the crossed product algebra\footnote{Note that this dressing becomes ill defined in the $\hbar=0$ limit of the reference frame algebra (\ref{reference}).}. Note that by construction elements in ${\cal{A}}^{cr}$ commute with $\hat h + \beta \hat h_{RF}$ in the same way as the elements in the crossed  product algebra in the case of superselection charges commute with $\hat X+\hat X_R$ with the SS charge $\hat X$ being here replaced by the Tomita Takesaki Hamiltonian $\hat h$. In this sense the constraint $[H,a]=0$ on physical and local observables is replaced, in the crossed product extension, by $[\hat h + \beta \hat h_{RF}, \hat a]=0$ with the reference frame dependent factor in the definition of $\hat a$ playing the role of the {\it dressing factor}.

One of the consequences of the algebra ${\cal{A}}$ being a type $III$ factor is that the overlap \cite{Papa3}
\begin{equation}
R(T) \equiv \langle \Psi|e^{iTH}|\Psi\rangle
\end{equation}
for any state $|\Psi\rangle$ in the Hilbert space representation of ${\cal{A}}$ is zero. More specifically $R(T) \sim e^{-\Delta(H^2) T}$ for $\Delta(H^2)$ the variance of $H$ on the state $|\psi\rangle$. Thus the divergent variance of $H$ leads to vanishing $R(T)$. This quantity is very close to the well known {\it spectral form factor}. An obvious question at this point, to be discussed later, is what is the meaning of the modified {\it spectral form factor} $R^{II}(T)$ defined replacing $H$ by $\beta \hat h^{II}$ and using states in the extended Hilbert space.

\subsection{The Hilbert space}
The Hilbert space representation of (\ref{reference}) is simply $L^2(R)$ assuming the real line $R$ is the spectrum of $\hat t$ and $\hat h_{RF}$. In the {\it energy representation}, states in this Hilbert space are represented by probability amplitudes $f(\epsilon)$ with $\int |f|^2 =1$ and with $\epsilon$ representing the energy eigenvalues of $\hat h_{RF}$. Based on our intuition on the discussion on SS charges we can {\it formally} think states in the extended Hilbert space representation of ${\cal{A}}^{cr}$ as
\begin{equation}\label{entangled}
|\hat \Phi\rangle (E) = \int d \epsilon f(\epsilon) |\Phi_{E-\epsilon}\rangle |\epsilon\rangle
\end{equation}
with $\hat h |\Phi_{E-\epsilon}\rangle = (E-\epsilon)|\Phi_{E-\epsilon}\rangle$ and $|\Phi_{E-\epsilon}\rangle$ in the GNS representation of ${\cal{A}}$, with $\hat h_{RF}|\epsilon\rangle = \epsilon|\epsilon\rangle$ and with $\int d\epsilon |f|^2 =1$. We can think of these states as representing in the extended Hilbert space quantum excitations with $\beta \hat h^{II}$ "energy" equal $E$.  

States like the ones defined in (\ref{entangled}) describe entanglement between the reference frame system and the QFT modes. To evaluate the corresponding entanglement entropy requires to associate with these states a density matrix $\rho_{\hat \Phi_E}$ and to evaluate the corresponding von Neumann entropy. In what follows we will focus on states in the extended Hilbert space that are un-entengled in the former sense i.e. they can be written as $| \Phi\rangle |RF\rangle$ with $|RF\rangle = \int d\epsilon f(\epsilon)|\epsilon\rangle$ i.e. states of the type
\begin{equation}\label{classical}
|\hat \Phi\rangle = \int d\epsilon f(\epsilon) |\Phi\rangle |\epsilon\rangle
\end{equation}
with $|\Phi\rangle$ in the GNS representation of ${\cal{A}}$. For these states both the vev $\langle \hat h^{II} \rangle$ as well as the variance $\langle \Delta(\hat h^{II})^2 \rangle$ is fully determined "classically" by the probability distribution $|f|^2$. States like the ones defined in (\ref{classical}) are denoted in \cite{Witten1} classical-quantum states. The {\it spectral form factor $R^{II}(T)$} on these states is fully determined by the classical variance of the function $f$ used to define the state.

It is important tot distinguish the quantum state (\ref{classical}) from a generic {\it statistical ensemble} defined on the spectrum of $\hat h_{RF}$ by
\begin{equation}\label{classical2}
|\hat \Phi\rangle = \int d\epsilon p(\epsilon) |\Phi\rangle |\epsilon\rangle
\end{equation} 
even if we choose $p(\epsilon)= |f|^2$. While state (\ref{classical}) with $f$ a quantum wave function requires to add the  algebra (\ref{reference}) the {\it statistical ensemble} (\ref{classical2}) can be defined without adding any operator conjugated to $\epsilon$ and consequently without transforming the type $III$ factor into a type $II$ factor. As already stressed in the Introduction this fact is crucial to extract the quantum ( by contrast to statistical ) consequences of the type $II$ construction. 
 
\subsection{The {\it ground state} and the trace}
Since ${\cal{A}}^{cr}$ is a type $II$ factor we should be able to define a trace. Depending on the value of the trace of the identity this type $II$ factor will be a type $II_1$ ( finite trace of the identity) or a type $II_{\infty}$. Let us define the {\it ground state} $|\psi_0\rangle$ in ${\cal{H}}$ the state satisfying $\hat h |\psi_0\rangle=0$\footnote{This state can be thought as the one associated with the identity in a GNS representation. The state is not unique reflecting the fact that the GNS representation is far from being irreducible.}. Then for any $\hat a = \int dt a(t) e^{it h^{II}}$ we define \cite{Witten1} a trace by:
\begin{equation}\label{tracedef}
tr(\hat a) = \int d{\epsilon} e^{\beta \epsilon} {\cal{F}}_{\hat a,|\psi_0\rangle}(\epsilon)
\end{equation}
for ${\cal{F}}_{\hat a,|\psi_0\rangle}(\epsilon)$ the Fourier transform of
\begin{equation}
\langle \psi_0|a(t)|\psi_0\rangle
\end{equation}
Note that different choices of $|\psi_0\rangle$ i.e. different GNS representations of the algebra will lead to different definitions of the trace.
\subsubsection{Density matrices and purification}
Once we have defined a trace on ${\cal{A}}^{cr}$ we can proceed to define the density matrix associated to states $|\hat \Phi\rangle$ in the extended Hilbert space.
Given the trace on the algebra ${\cal{A}}^{cr}$ we associate to this state a density matrix operator $\rho_{\hat \Phi}$ by the equation
\begin{equation}\label{matrixequ}
tr(\rho_{\hat \Phi} \hat a) = \langle \hat \Phi|\hat a|\hat \Phi \rangle
\end{equation}
This equation can be solved \cite{Witten1,Witten2,Witten3} for $|\hat \Phi\rangle$ a classical-quantum state defined by a probability distribution $|f|^2$. In this case we have
\begin{equation}
\langle \hat \Phi|\hat a|\hat \Phi \rangle = \int d\epsilon |f|^2(\epsilon){\cal{F}}_{\hat a,\Phi}(\epsilon)
\end{equation}
with ${\cal{F}}_{\hat a,\Phi}(\epsilon)$ the Fourier transform of $\langle \Phi|a(t) e^{it \hat h}|\Phi\rangle$. In these conditions equation (\ref{matrixequ})  can be easily solved using Araki's relative modular operator if we assume that the uncertainty in energy of the formal state in $L^2(R)$ defined by the amplitude $f$ i.e. the state $\int d\epsilon f(\epsilon) |\epsilon\rangle$ is very large or equivalently that the corresponding uncertainty in time is very small. The state $|\hat \Phi\rangle$ in the extended Hilbert space representation of ${\cal{A}}^{cr}$ solving (\ref{matrixequ}) defines the {\it purification} of the density matrix $\rho_{\hat \Phi}$.

Note that the former definition of the trace is {\it quantum} and not purely statistical. Indeed the approximated solution of (\ref{matrixequ}) uses the {\it time uncertainty} for the quantum state $|\hat \Phi\rangle$ that cannot be defined for the statistical ensemble (\ref{classical2}).

Now in the case it exists a normalizable state such that $tr(\hat a) = \langle \hat \Phi|\hat a|\hat \Phi\rangle$ i.e. with $\rho_{\hat \Phi} =1$ the factor will be type $II_1$ otherwise the factor is type $II_{\infty}$.

\subsection{Generalized entropy}
Once we have solved (\ref{matrixequ}) we can proceed to define the entanglement entropy for any state $|\hat \Phi\rangle$ in the extended Hilbert space as
\begin{equation}
S(\hat \Phi) = -tr(\rho_{\hat \Phi} \ln \rho_{\hat \Phi})\end{equation}
In particular for a state like (\ref{entangled}) this entropy will gives us the {\it quantum} entanglement entropy of such state. For a classical-quantum state we get, up to small errors associated with the time uncertainty for the state defined by the defined by the quantum wave function  $f$ \cite{Witten2,Witten3}
\begin{equation}\label{entropy1}
S(\hat \Phi) = \langle \hat \Phi|\beta \hat h_{RF}|\hat \Phi \rangle - \langle \hat \Phi|\beta \hat h_{\psi_0|\Phi}|\hat \Phi \rangle -\langle \hat \Phi|\log |f|^2|\hat \Phi \rangle
\end{equation}

This entanglement entropy depends on two basic quantities:

i) The {\it distinguishability} in ${\cal{H}}$ between the state $|\Phi\rangle$ and the {\it ground state} $|\psi_0\rangle$. By that we mean how much we can distinguish these states using local measurements in ${\cal{A}}$. This is a close relative to the function defined in (\ref{distin}). 

ii) The expectation value $\langle \hat \Phi|\beta \hat h_{RF}|\hat \Phi\rangle$ of the {\it reference frame} Hamiltonian $\hat h_{RF}$ on the state $|\hat \Phi\rangle$ and the term $\langle \hat \Phi|\log |f|^2|\hat \Phi \rangle$. 

The first contribution informs us about the QFT state in the causal diamond of the observer. The second contribution that depends on the quantum state of the added reference frame i.e. on the amplitude $f$, informs us about the expectation value of the reference frame Hamiltonian that, by construction, commutes with the whole algebra ${\cal{A}}$.  

The deep connection between (\ref{entropy1}) and Bekenstein generalized entropy \cite{Witten2,Witten3} unveils a new meaning of the reference frame. Indeed while the contribution i) in (\ref{entropy1}) is telling us about the quantum fluctuations in the causally complete domain where we are defining ${\cal{A}}$, the contribution ii) is giving us information about the other piece of the generalized entropy namely, the piece associated with the horizon itself. This makes natural to associate the reference frame algebra with a sort of {\it quantum horizon algebra} generated by an area operator $\hat A$ and its conjugated "time" \footnote{In principle the probability distribution $|f|^2$ accounts for the statistical fluctuations of the horizon \cite{Das}.}. Thus in this context the reference frame has two main roles. On one side to set the dressing factor and on the other side to define {\it a quantum horizon algebra} \footnote{In the Introduction we have denoted the two pieces contributing to the generalized entropy as $S_{{\cal{D}}}$ and $S_{RF}$. Using Bekenstein formula for the generalized entropy the part associated with the reference frame algebra is the one we put in correspondence with the contribution of the horizon to the entropy. It is in this sense that the reference frame algebra acquires its meaning as a quantum horizon algebra.}.

\subsection{Entanglement Capacity and Quantum Fisher Information}
For future use it will be important to define the entanglement capacity of a generic state $|\hat \Phi\rangle$ in the extended Hilbert space representing the crossed product type $II$ factor. Given $|\hat \Phi\rangle$ you define the corresponding density matrix $\rho_{\hat \Phi}$ solving equation (\ref{matrixequ}). Now you formally define
\begin{equation}
\hat h_{\hat \Phi}^{II} = -\log \rho_{\hat \Phi}
\end{equation}
The {\it entanglement capacity} is now defined as \cite{boer}
\begin{equation}\label{capacity}
{\cal{C}}(\hat \Phi) = \Delta(\hat h_{\hat \Phi}^{II})^2
\end{equation}
i.e. as the variance of $\hat h_{\hat \Phi}^{II}$ on the state $|\hat \Phi\rangle$. The corresponding quantum Fisher information is defined by $I_F(\hat \Phi) = \frac{{\cal{C}}(\hat \Phi)}{4}$.

\subsection{The fundamental group}
Finally note that changes of the RF Hamiltonian $\hat h_{RF}$ of the type
\begin{equation}
\hat h_{RF} \rightarrow \hat h_{RF} + cte
\end{equation} 
modify $h^{II}$ into $h^{II} + cte$ and consequently {\it rescale} the trace defined by (\ref{tracedef}) as well as the corresponding density matrix, leading to a shift of the entropy by the constant. For a type $II$ factor the group of outer automorphisms rescaling the trace is known as {\it the fundamental group}. To evaluate this group is a hard problem originally solved by Connes \cite{Connes}.  We will reduce ourselves to some very qualitative physics comments. In the case of type $II_1$ factor we can fix the normalization for the maximal entropy state as $\rho_{\hat \Phi}=1$. However we can think in some other "trace" states different from the maximal entropy state and to compare for these states the corresponding value of the trace of the identity. The set of these rescaled values will define the fundamental group. In type $II_1$ since there exist a maximal entropy state these rescaled values should lead to smaller entropies i.e. in the interval $[-\infty,0]$. Thus we can change $\hat h_{RF}$ only by arbitrary negative constants. By contrast in the type $II_{\infty}$ case we could change $\hat h_{RF}$ adding an arbitrary real constant making the allowed rescalings isomorphic to the whole real line $R$. 

Note that while the generalized entropies in (\ref{entropy1}) are ambiguous, in the type $II_{\infty}$ case, with respect to an arbitrary constant corresponding to an arbitrary fundamental group rescaling, this is not the case with respect to the entanglement capacity or the quantum Fisher information.

\section{Gravity and the clock as reference frame}
In this section and following \cite{Witten2} we will consider the algebra of QFT local observables that can be measured by a dS observer. Using the previous section we will describe the type $II$ description of the algebra of local observables with support in the static patch of the observer. Let us denote ${\cal{A}}_{dS}$ this algebra. The Hilbert space representation of this algebra, in terms of bounded operators, is a von Neumann factor of type $III_1$. As usual a Hilbert space representation of this algebra can be defined using the GNS construction\footnote{For the benefit of the reader we will summarize the main ingredients of the GNS construction.}. 

Recall (see \cite{wittenrev1} and \cite{wittenrev2} and references therein) that for a $C^*$ algebra ${\cal{A}}$ the GNS Hilbert space is defined associating to each element $a\in {\cal{A}}$ a vector state $|a\rangle$ and defining, on this set of states, a scalar product $\langle a|b\rangle = f(a^*b)$ in terms of some linear form $f$ on ${\cal{A}}$, satisfying $f(a^*a)>0$ for $a\neq 0$ and $f(1)=1$ \footnote{The Hilbert space is defined as the completion of the so defined pre Hilbert space.}. Thus we can write $f(a^*b) = \langle\Phi|a^*b|\Phi\rangle$ with $|\Phi\rangle$ the state associated, in the GNS construction, with the identity in ${\cal{A}}$. Technically this construction requires to mod by the ideal $I_{f}$ of elements $x$ in ${\cal{A}}$ such that $f(x^*x)=0$, so, in practice, we are representing the space of equivalence classes ${\cal{A}}/I_{f}$. 

The GNS Hilbert space $H_{GNS}$ defines a representation, generically non irreducible, of ${\cal{A}}$ in the space of bounded operators $B(H_{GNS})$ and therefore, it also defines the commutant ${\cal{A}}'$ as those elements in $B(H_{GNS})$ commuting with ${\cal{A}}$. The representation of ${\cal{A}}$ in $H_{GNS}$ is a von Neumann algebra if ${\cal{A}} = {\cal{A}}^{''}$. Moreover, this von Neumann algebra will be a {\it factor} if the center is trivial. The state $|\Phi\rangle$ associated with the identity will be cyclic and separating both with respect to ${\cal{A}}$ and ${\cal{A}}'$. When this representation of ${\cal{A}}$ is a type $III$ factor we don't have any state, in the sense of a finite linear form $f$ on ${\cal{A}}$, satisfying the trace property, namely $tr(ab)=tr(ba)$. Thus in this case we cannot represent correlators $\langle \Phi|a|\Phi\rangle$, defined  for a generic $a\in {\cal{A}}$, as $tr(\hat \rho_{\Phi} a)$ for some finite density matrix representing the GNS state $|\Phi \rangle$. 

This creates an obvious problem for de Sitter where we expect that correlators for operators in the algebra ${\cal{A}}_{dS}$ associated with a static patch should correspond to thermal correlators for the Gibbons-Hawking density matrix \cite{GH}\footnote{The Gibbons-Hawking expectation values for operators with support in the static patch are defined, in the $G_N=0$ limit i.e. without taking into account gravitational back reaction on the metric, by analytic continuation to Euclidean signature and using as vacua the de Sitter invariant Bunch-Davis vacuum. These expectation values can be interpreted as thermal, for a temperature $\beta_{dS}$ set by the de Sitter radius, and with canonical thermal density matrix $\rho= e^{ -\beta_{dS}H}$ with $H$ defining energy in the static patch i.e. the generator of time translations in the static patch. As already mentioned in section II, the main problem with this construction is that the Hamiltonian $H$, defining energy in the static patch in this QFT limit, is divergent. Thus, in order to give some rigorous meaning to the Gibbons-Hawking analytic continuation, we would need to introduce some "regulator" for the divergences of the quantum fluctuations of $H$.}. This leads to the problem of {\it how to modify the algebra ${\cal{A}}_{dS}$ into a new algebra, let us say $\tilde{\cal{A}}_{dS}$, such that we can define on it states of trace type}. This implies that the new algebra $\tilde{\cal{A}}_{dS}$ should be at least a type $II$ factor.  The construction of $\tilde{\cal{A}}_{dS}$ as a type $II_1$ factor has been developed in \cite{Witten2} under the general lines described in section 4. In what follows we will use the former discussion of section 3 on SS-charges to motivate a natural way to define this extended algebra.

First of all for the type $III_1$ algebra ${\cal{A}}_{dS}$ we can define using Tomita-Takesaki construction a {\it state dependent} modular Hamiltonian. Let us denote $|\Psi_{dS}\rangle$ the state associated in the GNS construction with the identity. The Tomita-Takesaki construction defines the associated modular Hamiltonian $\hat K_{|\Psi_{dS}\rangle}$ that implements, as an outer automorphism, the group of time translations. The state $|\Psi_{dS}\rangle$ satisfies $\hat K_{|\Psi_{dS}\rangle}|\Psi_{dS}\rangle =0$. We can formally define a "Hamiltonian" $\hat H_{dS}$ as
\begin{equation}\label{beta}
\hat K_{|\Psi_{dS}\rangle} = \beta_{|\Psi_{dS}\rangle} \hat H_{dS}
\end{equation}
for some $\beta$ \footnote{Note that this $\beta$ plays the role of the unit length discussed in the section 4 in order to make $\beta_{|\Psi_{dS}\rangle} \hat H_{dS}$ dimensionless as it is the modular Tomita Takesaki Hamiltonian $\hat K$.}. On the basis of Gibbons-Hawking results we can use $\beta_{dS}$, the de Sitter GH temperature, to define $\hat H_{dS}$. This means that we identify the state $|\Psi_{dS}\rangle$ used in the GNS construction with the Bunch-Davis de Sitter invariant vacuum  \footnote{ For the definition of de Sitter invariant vacua see \cite{BD1,BD2,BD3,BD4,BD5}.}.

Geometrically $\hat H_{dS}$ is associated to the Killing of full dS acting as opposite time translations on both static patches of the full Penrose diagram. Note that $\hat H_{dS}$  defines a good outer automorphism of ${\cal{A}}_{dS}$ i.e. $\hat H_{dS}$ is not in ${\cal{A}}_{dS}$ \footnote{Note that $\hat H_{dS}$ is not the natural Hamiltonian $H$ that the observer will use in her static patch to define energy. As we have already mentioned such Hamiltonian is ill defined due to quantum fluctuations on the cosmological horizon boundary of the static patch.}. 

For the special case of dS, and due to the fact that spatial sections are compact, general covariance implies that $\hat H_{dS}$ acts as a {\it constraint}. At the level of the algebra we can implement this constraint declaring as {\it the algebra of physical 
observables} the  subalgebra invariant under the action of the group of time translations generated by $\beta_{dS}\hat H_{dS}$ i.e.
\begin{equation}\label{time}
{\cal{A}}_{dS}^{\beta_{dS}\hat H_{dS}}
\end{equation}
Unfortunately this invariant subalgebra is trivial due  to the fact that the modular Hamiltonian $\hat K$ ( and consequently also $\hat H_{dS}$ ) is acting ergodically. A different way to see the problem is observing that the obvious time invariant operator in ${\cal{A}}_{dS}^{\hat H_{dS}}$ defined as $\int dt e^{i\hat H_{dS}t}ae^{-i\hat H_{dS}t}$ for any $a\in {\cal{A}}_{dS}$ has divergent expectation value on any state {\it satisfying the constraint}.

At the level of the Hilbert space the constraint could be used to identify the physical subspace  ${\cal{H}}_{GNS}^{phys}$ of the GNS Hilbert space. Recall that by construction the state $|\Psi_{dS}\rangle$ is cyclic and therefore ${\cal{H}}_{GNS} \sim {\cal{A}}_{dS} |\Psi_{dS}\rangle$. Naively we could try to define ${\cal{H}}_{GNS}^{phys}$ as ${\cal{A}}_{dS}^{\hat H_{dS}} |\Psi_{dS}\rangle$ i.e. states obtained acting on $|\Psi_{dS}\rangle$ with the invariant subalgebra. This leads to a trivial Hilbert space. 

At this point it could be worth to interpret the general covariance constraint from the point of view of our former discussion on SS-charges and reference frames. Condition (\ref{time}) can be read, in this spirit, as reflecting complete lack of knowledge about how the clock, measuring the time conjugated to $\hat H_{dS}$ is practically defined i.e. about {\it the clock reference frame}. Alternatively we can interpret (\ref{time}) as defining physics by the condition of {\it invariance} under arbitrary changes of this clock reference frame. 

Hence, and inspired by the discussion on SS-charges, let us think of $\hat H_{dS}$ as the SS-charge associated with {\it the complete lack of knowledge about the clock reference frame}.  Once we think in this way the former discussion in section 3 for SS-charges indicates the natural steps to be taken, namely 

{\it i) to add a quantum system representing the clock reference frame and 

ii) to impose the constraint (\ref{time}) on the extended system including the reference frame.}

This is actually the rule suggested in \cite{Witten2} to define the algebra $\tilde{\cal{A}}_{dS}$. In \cite{Witten2} what we call a RF is introduced as an {\it observer} equipped with a clock.

As we did in the previous section  the quantum system representing the {\it clock RF} should be associated with a reference frame algebra ${\cal{A}}_R$ generated by an operator $\hat t$ formally representing the "clock time" and its conjugated that we will denote, as usual, $\hat H_{RF}$ and that we can interpret as the "clock Hamiltonian". Now once we have introduced the RF algebra ${\cal{A}}_R$ we will implement the constraint defining as the physical algebra the analog of (\ref{alg}) i.e.
\begin{equation}\label{dSalg}
({\cal{A}}_{dS}\otimes {\cal{A}}_R)^{(\hat H_{dS} + \hat H_{RF})}
\end{equation}
instead of (\ref{time}). In what follows we will denote this algebra, as we did in section 4, 
${\cal{A}}_{dS}^{cr}$.

Note that (\ref{dSalg}) can be interpreted as an interesting modification of general covariance where we extend the principle to the combined system, that in this case is, the quantum field theory, including linearized gravitational fluctuations, defined on a dS space-time background, plus the quantum representation of the clock RF.

\subsection{The clock reference frame} 

Let us start describing quantum mechanically the RF defined by the clock. The simplest possibility is to define the Hilbert space ${\cal{H}}_R$ as an irrep of the clock algebra defined by $[\hat H_{RF},\hat t]=-i$ where we think in this algebra as the algebra in quantum mechanics defined by the position and momentum operator. Thus we can assume that both $\hat H_{RF}$ and $\hat t$ have a continuous real spectrum and therefore the clock Hilbert space can be identified as $L^2(R)$. In this representation $\hat t$ is simply the {\it position} on $R$ and $\hat H_{RF}$ the generator of translations on $R$ \footnote{Obviously we can use to represent the algebra ${\cal{A}}_R$ either the "position" or the "momentum" representation. If we identify $\hat t$ as position then $\hat H_{RF}= -i\frac{d}{dt}$ for $t$ representing the spectrum of $\hat t$.}.

Let us denote $|\epsilon\rangle$ the spectrum of $\hat H_{RF}$. Thus typical pure state of the clock RF will be
\begin{equation}
|RF(f)\rangle = \int d\epsilon f(\epsilon)|\epsilon\rangle
\end{equation}
for some wave function $f\in L^2(R)$ with $\int d\epsilon|f|^2=1$ \footnote{The function $f$ defines the probability amplitude and therefore can be complex with non trivial phases.}. This pure state has a typical uncertainty $\delta (\hat H_{RF})$ for $\hat H_{RF}$ that we can derive from the variance
\begin{equation}
\Delta (\hat H_{RF}^2)= \langle RF(f)|\hat H_{RF}^2|RF(f)\rangle - (\langle RF(f)|\hat H_{RF}|RF(f)\rangle)^2
\end{equation}
as $\delta(\hat H_{RF}) = \sqrt {\Delta (\hat H_{RF}^2)}$. Using now the canonical commutation relation $[\hat H_{RF},\hat t]=-i$ we get the corresponding minimal uncertainty bound on $\delta(t)$.

We can also associate with the clock RF mixed states of the type
\begin{equation}
\hat \rho_R = \int d\epsilon p(\epsilon)|\epsilon\rangle\langle\epsilon|
\end{equation}
associated with a probability distribution $p(\epsilon)$. For this density matrix we can evaluate $tr(\hat \rho_R H_{RF})= \langle \hat H_{RF}\rangle$ as well as the fluctuations $tr(\hat \rho_R (\hat H_{RF} - \langle \hat H_{RF}\rangle)^2)$ for any probability distribution $p(\epsilon)$.

Finally note that $\hat t$ as well as $\hat H_{RF}$ define observables with units of length and energy respectively. Thus we need to use a {\it unit} to measure these quantities. The natural unit of length working in de Sitter should be, as already mentioned, the de Sitter radius i.e. $\beta_{dS}=2\pi r_{dS}$.

\subsection{The observables in ${\cal{A}}_{dS}^{cr}$}
Let us now identify the elements $\hat a$ in ${\cal{A}}_{dS}^{cr}$. By construction these observables must satisfy $[\hat a,\hat H_{dS}+\hat H_R]=0$. Thus the generators of ${\cal{A}}_{dS}^{cr}$ will be, in addition to $\hat H_R$, elements of the type
\begin{equation}\label{crossed}
\hat a = e^{i\hat t \hat H_{dS}}ae^{-i\hat t \hat H_{dS}}
\end{equation}
for any $a\in {\cal{A}}_{dS}$ and where we use the canonical commutation relation $[\hat t,\hat H_{RF}]=-i$. Generic elements can be associated with functions $a(t) :R \rightarrow {\cal{A}}_{dS}$ that are square integrable:
\begin{equation}
\hat a= \int dt a(t) e^{it(\hat H_{dS}+\beta_{dS}\hat H_{RF})}
\end{equation}
Note ( see section 4) that to promote the physical c-number time into the clock operator $\hat t$ conjugated to $\hat H_{RF}$ is {\it crucial} for the definition of the crossed product algebra. 

How to identify the physical Hilbert space is a subtle issue discussed in \cite{Witten2}. Formally we could identify this space, extending the GNS construction to ${\cal{A}}_{dS}^{cr}$,  as the appropriated completion of ${\cal{A}}_{dS}^{cr} |\hat\Psi_{dS}\rangle$ for some state $|\hat \Psi_{dS}\rangle$ formally representing the identity.  However, as it is standard in the GNS construction, we first need, in order to promote  ${\cal{A}}_{dS}^{cr} |\hat\Psi_{dS}\rangle$ into a Hilbert space, to define the associated linear form on ${\cal{A}}_{dS}^{cr}$.

If ${\cal{A}}_{dS}^{cr}$ is type $II$ we could require this linear form to satisfy the trace property i.e.
\begin{equation}\label{trace}
\langle \hat \Psi_{dS}|\hat a|\hat \Psi_{dS}\rangle = tr (\hat a)
\end{equation}
How to define the state $|\hat \Psi_{dS}\rangle$ and the corresponding trace? A solution based on the results in \cite{Witten1} is presented in \cite{Witten2}. 

Let us start defining the state $|\tilde \Psi_{dS}\rangle$ in the extended Hilbert space $H_{ext}= H_{GNS}\otimes L^2(R)$ 
\begin{equation}\label{tilde}
|\tilde \Psi_{dS}\rangle = \int d{\epsilon} f(\epsilon)|\Psi_{dS}\rangle |\epsilon\rangle
\end{equation}
for $\int d{\epsilon} |f|^2 =1$. Let us now consider a generic element $\hat a= \int dt a(t) e^{it(\hat H_{dS} +\beta_{dS}\hat H_{RF})}$ and let us define the linear form $f_{|\tilde \Psi_{dS}\rangle}(\hat a)$ as
\begin{equation}
f_{|\tilde \Psi_{dS}\rangle}(\hat a)= \langle \tilde \Psi_{dS}| \hat a |\tilde \Psi_{dS}\rangle
\end{equation}
Using the former definitions we get
\begin{equation}\label{form}
f_{|\tilde \Psi_{dS}\rangle}(\hat a) =\int d{\epsilon}|f|^2(\epsilon) {\cal{F}}_a(\epsilon)
\end{equation}
with ${\cal{F}}_a(\epsilon)$ the Fourier transform of $\langle \Psi_{dS}|a(t)|\Psi_{dS}\rangle$ \footnote{See discussion in section 4.}. 

Now we want to transform the linear form $f_{|\tilde \Psi_{dS}\rangle}$ into a trace. In \cite{Witten1} it was shown , using the KMS condition, that a trace form can be defined as
\begin{equation}
tr(\hat a) = \langle \tilde \Psi_{dS}|
\frac{\hat a e^{-\beta_{dS} \hat H_{RF}}}{|f|^2} |\tilde \Psi_{dS}\rangle
\end{equation}
that leads to
\begin{equation}
tr(\hat a) = \int d{\epsilon}\beta_{dS}e^{-\beta_{dS}\epsilon} {\cal{F}}_a(\epsilon)
\end{equation}
This trace is independent of the function $f$ used in the definition of $|\tilde \Psi_{dS}\rangle$ and only depends on the state $|\Psi_{dS}\rangle$ i.e. on the cyclic state used in the GNS representation of the de Sitter algebra ${\cal{A}}_{dS}$ \footnote{In principle this state can be associated with the standard Bunch-Davis de Sitter vacuum.}.

Using this definition of trace we can easily find that the state in the extended Hilbert space satisfying (\ref{trace}) is 
\begin{equation}\label{state}
|\hat \Psi_{dS}\rangle = \int d{\epsilon}\sqrt{\beta_{dS}} e^{-\frac{\beta_{dS}\epsilon}{2}} |\Psi_{dS}\rangle |\epsilon\rangle
\end{equation}

For $\hat a$ the identity we get $tr1= \int d{\epsilon}\beta_{dS} e^{-\beta_{dS}\epsilon}$. Now as discussed in section 4 is the value of $tr(1)$ what will make the crossed product algebra either type $II_1$, in case is finite, or type $II_{\infty}$. From $tr1= \int d{\epsilon}\beta_{dS} e^{-\beta_{dS}\epsilon}$ we observe that this quantity will be finite if we restrict the spectrum of $\hat H_{RF}$ to be in the interval $[-a,\infty]$ for some finite $a$. This means that in the fundamental group of transformations $\hat H_{RF} \rightarrow \hat H_{RF} + cte$ we need to restrict the cte to be in $R^+$. If we impose this restriction the state (\ref{state}) will be normalizable.

Now note that relative to $|\hat\Psi_{dS}\rangle$ we could define the associated density matrix by
\begin{equation}
tr(\hat a) =tr(\hat \rho_{|\hat\Psi_{dS}\rangle} \hat a) 
\end{equation}
that by construction leads to
\begin{equation}
\hat \rho_{|\hat\Psi_{dS}\rangle} = 1
\end{equation}
i.e. to {\it flat entanglement}. This makes the state $|\hat\Psi_{dS}\rangle$ the {\it maximal entropy state.}\cite{Torroba}.

The maximal entropy state (\ref{state}) admits a nice physical interpretation \cite{Witten2}. Indeed we can interpret the extra factor
$e^{-\frac{\beta_{dS}\epsilon}{2}}$ defining this state, as accounting for the {\it RF effective action} and the normalization factor  $\int d{\epsilon}\beta_{dS} e^{-\beta_{dS}\epsilon}$ as effectively defining the RF generating functional.

We can define different candidates of {\it physical states} in the extended Hilbert space. The simplest example could be a {\it classical-quantum} state
$|\hat \Phi\rangle = \int d{\epsilon} g(\epsilon) |\Phi \rangle |\epsilon\rangle$ where we replace the dS GNS ground state $|\Psi_{dS}\rangle$ representing the Bunch-Davis vacuum by some other state $|\Phi\rangle$ in $H_{GNS}$. The associated density matrix $\hat \rho_{|\hat \Phi\rangle}$ is defined by
\begin{equation}\label{identity}
\langle \hat \Phi |\hat a|\hat \Phi \rangle = tr(\hat \rho_{|\hat \Phi\rangle} \hat a)
\end{equation}
where we can use the former definition of $tr$ i.e.
\begin{equation}
tr(\hat \rho_{|\hat \Phi\rangle} \hat a) = \langle \hat \Psi_{dS}|\frac{\hat \rho_{|\hat \Phi\rangle} \hat a e^{\beta_{dS}\hat H_{RF}}}{e^{-\beta_{dS}\epsilon}}|\hat \Psi_{dS}\rangle
\end{equation}
leading to
\begin{equation}
\langle\hat \Phi| \hat a|\hat \Phi\rangle = \int d{\epsilon} |g|^2 {\cal{F}}(\epsilon,\Phi)
\end{equation}
where ${\cal{F}}(\epsilon,\Phi)$ is now the Fourier transform of $\langle \Phi |a(t) e^{it\hat H_{dS}} |\Phi\rangle$. If now we assume that $|\hat \Phi\rangle$ is {\it semiclassical} in the sense of \cite{Witten2} we can choose the function $g$ in such a way that the {\it time uncertainty} of the clock RF is much smaller than one. In this case we can replace, up to small errors, the Fourier transform ${\cal{F}}(\epsilon,\Phi)$ by the one of $\langle \Phi |a(t) |\Phi\rangle$. We can visualize this condition thinking in a RF clock pure state $\int d\epsilon g(\epsilon)|\epsilon\rangle$ with large energy uncertainty $\Delta(\epsilon)$ and consequently,  due to the canonical commutation relation $[\hat t,\hat H_{RF}]=-i$, small time uncertainty.

Once this condition is implemented the relation $\langle \Psi_{dS}|\Delta_{\Psi_{dS}|\Phi}a|\Psi_{dS}\rangle = \langle \Phi| a |\Phi\rangle$ with  $\Delta_{\Psi_{dS}|\Phi}$ Araki's relative modular operator, leads to
\begin{equation}\label{density}
\hat \rho_{|\hat \Phi\rangle}= e^{h_{(\Psi_{dS}|\Phi)}}|g(\beta_{dS}\hat H_{RF})|^2e^{\beta_{dS}\hat H_{RF}}
\end{equation}
with $h_{(\Psi_{dS}|\Phi)} = -\log \Delta_{\Psi_{dS}|\Phi}$. 
The quantum {\it distinguishability} between the representative of the maximal entropy dS ground state and a generic state $|\hat \Phi\rangle$ can be directly measured by the von Neumann entropy of $\hat \rho_{|\hat \Phi\rangle}$. For notational simplicity we will refer to these two types of contributions to the entropy as $S(\hat \Phi, RF)$ for the piece accounting for the RF contribution and $S(\hat \Phi|\Psi_{dS})$ for the one measuring the quantum distinguishability between the state $\hat \Phi$ and the "reference state" $\Psi_{dS}$ used to define the $tr$.

 Note that, as already discussed in section 4, this entropy, for {\it classical-quantum} states contains three pieces. One is the relative entropy 
$\langle \Phi|h_{(\Psi_{dS}|\Phi)}|\Phi\rangle$ between the state $|\Phi\rangle$ and $|\Psi_{dS}\rangle$. This piece accounts for the distinguishability, using operators in ${\cal{A}}_{dS}$, between the Bunch-Davis state $|\Psi_{dS}\rangle$ and the state $|\Phi\rangle$. The other piece is associated with the RF clock energy $\langle \hat \Phi|\beta_{dS}\hat H_{RF}|\hat \Phi\rangle$ and finally we have a piece that accounts for the quantum fluctuations of the RF as they are encoded in the function $g$ defining the state $|\hat\Phi\rangle$. These last two pieces, for classical-quantum states, only depend on the probability distribution $|g|^2$.

As discussed in section 4 the former construction allows us to define for a generic state ( not necessarily of the type classical-quantum) $|\hat\Phi\rangle$ in the extend Hilbert space the corresponding type $II$ Hamiltonian $h^{II}_{|\hat \Phi\rangle}$ as
\begin{equation}
h^{II}_{|\hat \Phi\rangle} = - \log \hat \rho_{|\hat \Phi\rangle}
\end{equation}
for $\hat \rho_{|\hat \Phi\rangle}$ the density matrix purified by the state $|\hat\Phi\rangle$ i.e. the matrix solving equation (\ref{matrixequ}). 

\subsection{Some comments on classical-quantum states in dS}
First of all recall that states $\int d\epsilon f(\epsilon) |\Phi\rangle |\epsilon\rangle$ with the integral over the spectrum of $\hat H_{RF}$ are characterized by a quantum wave function $f(\epsilon)$ on which the operator $\hat t$ in the reference frame algebra acts as the translation operator $i\frac{d}{d\epsilon}$. Thus as stressed in several places before these states are characterized by a quantum uncertainty $\delta(\hat t)$ as well as by a quantum uncertainty in $\hat H_{RF}$. 

For fixed $|\Phi\rangle$ in the GNS Hilbert space we can have different types of these states depending on the form of the wave function $f$. We can have states where $|f|^2$ is a gaussian centered around some $\epsilon_0$ in the spectrum of $\hat H_{RF}$. This corresponds to a {\it coherent state} for the reference frame. We can also have {\it thermal wave functions} with $|f|^2 = e^{-\beta_{DS} \epsilon}$ in case the spectrum of $\hat H_{RF}$ is projected to positive values. As discussed the maximal entropy state is a thermal state in the former sense with $|\Phi\rangle$ the dS Bunch Davis vacuum $|\Psi_{dS}\rangle$. 

States like $\int d\epsilon g(\epsilon) |\Psi_{dS}\rangle |\epsilon\rangle$ with $|g|^2$ a gaussian are a sort of {\it thermal coherent states} with density matrix $|g|^2 e^{\beta_{dS} \epsilon}$. All these states differ from {\it statistical ensembles} defined on the spectrum of $\hat H_{RF}$. Indeed for those {\it statistical ensembles} we cannot define {\it quantum fluctuations} of $\hat H_{RF}$ but just statistical variance. 

If we consider dS in the planar patch, as we will discuss in the next section, we can define states in the extended Hilbert space corresponding to {\it squeezed states} for the reference frame system. These states will be important in the discussion of Inflation as a type $II$ factor. 

An aspect of the maximal entropy state associated with a thermal wave function is that it leads to a quantum uncertainty in $\hat t$ of the order $\beta_{dS}$. A question we will consider is section 7 is the physical meaning of this large time uncertainty of the maximal entropy state.

\subsection{Entropy deficit}

Let us now introduce the notion of {\it entropy deficit}. We have started with the GNS representation of ${\cal{A}}_{dS}$ that defines the Hilbert space ${\cal{H}}_{GNS}$ and has as cyclic state $|\Psi_{dS}\rangle$ that we normally identify with the de Sitter invariant Bunch Davis vacuum. In this Hilbert space we have the split of the algebra $B({\cal{H}}_{GNS})$ of bounded operators in ${\cal{H}}_{GNS}$ as
$B({\cal{H}}_{GNS})={\cal{A}}\otimes {\cal{A}}'$
with ${\cal{A}}'$ the commutant. We can formally extend this decomposition to the extended Hilbert space and to the algebra ${\cal{A}}_{dS}^{cr}$ defining the corresponding crossed product commutant. This commutant could be interpreted as the crossed product  $({\cal{A}'}_{dS}\otimes {\cal{A}}_{R'})^{\hat H_{dS}+\beta_{dS}\hat H_{{RF}'}}$ where ${\cal{A}'}_{dS}$ is the algebra of observables with support in the complementary static patch\footnote{Note that since ${\cal{A}}_{dS}^{cr}$ is a von Neumann algebra its representation in the extended Hilbert space should automatically define its commutant. The simplest guess is to represent the commutant as $({\cal{A}'}_{dS}\otimes {\cal{A}}_{R'})^{\hat H_{dS}+\beta_{dS}\hat H_{{RF}'}}$ i.e. by simply adding a "mirror" reference frame system. Although this is a very natural guess we don't know how to justify the necessity of the mirror reference frame ( see \cite{Witten 2} for a more elaborated discussion).}. The algebra ${\cal{A}}_{R'}$ and the corresponding Hamiltonian $\hat H_{{RF}'}$ represent a {\it mirror RF clock} located in the other static patch.  Thus, the constraint defining the physical Hilbert space is $(\hat H_{dS}+\beta_{dS}\hat H_{RF} + \beta_{dS}\hat H_{RF'})=0$. Using the notion of {\it coinvariance} it was shown in \cite{Witten2} that the physical Hilbert subspace of $H_{GNS}\otimes L^2(R)_{R}\otimes L^2(R)_{R'}$ satisfying the constraint is precisely the extended Hilbert space $H_{ext}$ used above.

Imagine, for a moment, that the former decomposition of $B({\cal{H}}_{GNS})$ reflects a decomposition of the extended Hilbert space into two pieces one associated with ${\cal{A}}_{dS}^{cr}$ and the other with the commutant that we will denote ${\cal{H}}_A$ and ${\cal{H}}_{A'}$ respectively for simplicity. We know that this is actually not the case even if ${\cal{A}}_{dS}^{cr}$ is a type $II$ factor, but let us use that assumption as an heuristic tool to introduce the notion of {\it entropy deficit}. Assuming this split of the Hilbert space we will think that any density matrix $\hat \rho$ on ${\cal{A}}_{dS}^{cr}$ could be represented as $Tr_{{\cal{H}}_{A'}} |\Phi\rangle\langle\Phi|$ for some state $|\Phi\rangle$ in the extended Hilbert space and where by $Tr$ we mean the standard trace over the Hilbert space of the commutant.

Using this intuition we will conclude that the maximal von Neumann entropy of $\rho$ will be determined by the logarithm of the {\it dimension} of the smaller Hilbert space i.e. $\log \dim {\cal{H}}_A$ if $\dim {\cal{H}}_A<\dim {\cal{H}}_{A'}$ and the other way around in case $\dim {\cal{H}}_{A'}<\dim {\cal{H}}_{A}$. Thus we can associate, with any type $II$ von Neumann algebra, an entropy deficit as the analog of $\log \frac{\dim {\cal{H}}_A} { \dim{\cal{H}_{A'}}}$. Of course this is purely formal because we don't have the split property for the Hilbert space. However Murray and von Neumann extended the former heuristic notion to a classification of representations of type $II$ factors. Recall that each representation defines the von Neumann algebra as well as its commutant so, intuitively, each representation defines the analog of a "deficit" parameter $d=\frac{\dim {\cal{H}}_A} { \dim{\cal{H}_{A'}}}$. 

For instance when $A$ and $A'$ are identical as it is the case in the TFD representation we get $d=1$. When one algebra is "infinitely bigger" than the other we get either $d=0$ or $d=\infty$ and the algebra that is "infinitely bigger" becomes a type $II_{\infty}$ factor.

How we can change the "size" of the two ( although non existent ) Hilbert spaces ${\cal{H}}_A$ and ${\cal{H}}_{A'}$ ? Physically since elements in ${\cal{A}}'$ commute with all elements in ${\cal{A}}$  we can reduce the size of the representation of ${\cal{A}}$ using a projector in ${\cal{A}}'$. In standard physical terms this is equivalent to say that we consider states with a given fixed "eigenvalue" of some observable in ${\cal{A}}'$. This can be understood as a reduction of the size induced by imposing some symmetry as a "constraint" on Hilbert space. Thus, in some sense, reducing or increasing $d$ ( depends if we look from the point of view of $A$ or $A'$) is equivalent to increase symmetry constraints or to break symmetries. 

At this point a natural question is how the entropy deficit informs us about the difference between the reference frame used to make ${\cal{A}}_{dS}$ a type $II$ factor and the "mirror" reference frame  used to define the crossed product commutant \cite{Witten3}. In order to gain some intuition we need to discuss the geometry underlying a given purification.

\section{Purification and Geometry}

\subsection{One side black hole purification}
Let us consider the classical background defined by the Schwarzschild black hole metric. 
In abstract terms, and for a given asymptotic observer, we can try to define the algebra ${\cal{A}}_{out}$ of local operators representing all the physical observables the asymptotic observer can actually measure performing local observations. We expect that this algebra 
${\cal{A}}_{out}$ is contained in the larger algebra $B({\cal{H}})$ of bounded operators acting on the full Hilbert space ${\cal{H}}$. The {\it information paradox} as well  as the {\it purity} of the quantum state describing the full system can be translated into the preliminary algebraic problem on how to define the algebra of observables describing the black hole interior. 

 Let us formally define an algebra ${\cal{A}}_{in}$ of local observables describing {\it the black hole interior}. Obviously this algebra represents observations that the asymptotic observer cannot perform locally. However we know {\it a priori} some desired properties of this algebra. 
In fact we should require that $[{\cal{A}}_{out},{\cal{A}}_{in}]=0$ i.e. that ${\cal{A}}_{in}$ should be in the {\it commutant} of ${\cal{A}}_{out}$. By that we mean that, relative to the Hilbert space ${\cal{H}}$ where we are representing ${\cal{A}}_{out}$, the elements in ${\cal{A}}_{in}$ should be operators in $B({\cal{H}})$ commuting with all the bounded operators representing ${\cal{A}}_{out}$. Moreover we should also expect that the whole algebra $B({\cal{H}})$ should factorize in the form 
\begin{equation}\label{split}
B({\cal{H}}) = {\cal{A}}_{out}\otimes {\cal{A}}_{in}
\end{equation}
This split at the level of the algebra of observables is not equivalent to the most familiar split property at the level of the Hilbert space into a product of two Hilbert spaces one associated with the exterior and the other with the interior of the black hole. Actually such split of the Hilbert space does not exist. 

Let us concentrate on (\ref{split}). This is achieved if ${\cal{A}}_{out}$ \footnote{Defined by the corresponding representation in $B({\cal{H}})$.} is a {\it factor} i.e. a von Neumann algebra with trivial center and if ${\cal{A}}_{in}$ is equal to the {\it commutant} i.e. ${\cal{A}}_{out} = {\cal{A}}_{in}^{'}$. This already implies that in order to achieve (\ref{split}) we must consider the algebra of observables of the asymptotic observer {\it without including any central term} \footnote{The central terms for the asymptotic observer are the global properties of the black hole, as the ADM mass or the total charge. Those global properties will not be included in the von Neumann algebra ${\cal{A}}_{out}$.} . In addition since we are considering ${\cal{A}}_{out}$ a von Neumann algebra we must require {\it completeness} relative to the topology defined by the scalar product of our Hilbert space \footnote{This means that the limit of a sequence $a_n$ of elements in ${\cal{A}}_{out}$ is defined by the element $a$ such that $\langle \psi|a|\psi\rangle =\lim_{n=\infty}\langle \psi|a_n|\psi\rangle$ for any $|\psi\rangle$ in ${\cal{H}}$. This condition is equivalent to the relation ${\cal{A}}_{out} = {\cal{A}}_{in}^{''}$}. In that sense the algebraic data leading to the desired split property (\ref{split}) are dependent on how we identify the Hilbert space ${\cal{H}}$.

Our next question is of course what type of factor we expect  ${\cal{A}}_{out}$
should be. On general grounds we expect that ${\cal{A}}_{out}$ should be a type $III$ factor. If this is the case we know that there exist a GNS representation on the space of bounded operators of a Hilbert space ${\cal{H}}_{GNS}$ characterized by a cyclic state $|\Phi_{GNS}\rangle$. This means that ${\cal{H}}_{GNS}$ is the completion of the set of states obtained acting with ${\cal{A}}_{out}$ on the state $|\Phi_{GNS}\rangle$. This is an extremely interesting property that says that the Hilbert space can be represented either as the completion of the set of states obtained by acting with ${\cal{A}}_{out}$ on $|\Phi_{GNS}\rangle$ or as the completion of the set defined by the commutant ${\cal{A}}_{in}$. This is reminiscent of the popular {\it black hole complementarity} \cite{complementarity}. Thus, if we assume that ${\cal{H}}$ is ${\cal{H}}_{GNS}$ and that our "vacuum" state $|\Phi_0\rangle$ is the GNS cyclic state we will achieve (\ref{split}) in a way dependent on the selected state $|\Phi_0\rangle$. 

At this level we have learned that the black hole interior is described by the {\it commutant} of the algebra describing the observations of the asymptotic observer and that the definition of this commutant is {\it state dependent} \cite{Papa1},\cite{Papa2}.

The type $III$ nature of ${\cal{A}}_{out}$ provides, thanks to Tomita Takesaki theory, a {\it state dependent map} to relate the interior and the exterior algebras of the black hole. This is actually the map used in \cite{Papa1} and \cite{Papa2} to define the so called {\it mirror operators}. Briefly the Tomita operator is defined by
\begin{equation}\label{TT}
S_{{\cal{A}}_{out}} a |\Phi_0\rangle = a^{\dagger}|\Phi_0\rangle
\end{equation}
where we assume $|\Phi_0\rangle$ to be the cyclic state of the GNS representation we are using and where $a$ in (\ref{TT}) is any element in ${\cal{A}}_{out}$. The important property is that $S_{{\cal{A}}_{out}}$ can be written as ${\cal{J}} \Delta^{\frac{1}{2}}$ with $\Delta$ the modular operator defined by the modular Hamiltonian $\hat h_{\Phi_0}$ and with ${\cal J}{\cal{A}}_{out} = {\cal{A}}_{in}$ assuming, as already discussed, that ${\cal{A}}_{in}$ is the commutant of ${\cal{A}}_{out}$. Thus, we see that (\ref{TT}) maps, thanks to the existence of the map ${\cal{J}}$, elements in the exterior algebra ${\cal{A}}_{out}$ into elements in the interior algebra ${\cal{A}}_{in}$. This for a concrete state $|\Phi_0\rangle$ of the type of the TFD state is the definition of {\it mirror operators} in \cite{Papa2}.

After these preliminaries we can try following the former general discussion in section 4 to define a type $II$ factor associated with ${\cal{A}}_{out}$ by adding a RF and to use some "mirror" reference frame to promote the interior algebra ${\cal{A}}_{in}$ into a type $II$ factor. If we succeed we could define density matrices describing the region outside the black hole and the corresponding purification in the extended Hilbert space. The key point is that this purification defines a pure quantum state that describes the exterior as well as the black hole interior. In the case we use a TFD approach the corresponding type $II$ factors are defined for the the left and right "external" sides of the two sided black hole. However if we work with von Neumann algebras satisfying ${\cal{A}}_{in} = {\cal{A}}_{out}^{'}$ and we develop the type $II$ crossed product construction in this frame we get purifications describing both the interior as well as the exterior of the black hole. We can denote these purifications {\it one sided purifications}. 

If now we define in the extended type $II$ Hilbert space a state $|\hat \Psi_t\rangle$ describing the time evolution of the black hole evaporation process the corresponding type $II$ entropy associated with $\rho_{\hat \Psi_t}$ could be written in two different ways depending if we consider ${\cal{A}}_{in}$ or its commutant ${\cal{A}}_{out}$, namely as \footnote{For notation see discussion around equation (\ref{density}).}
\begin{equation}
S(\hat \Psi_t, RF) + S(\hat \Psi_t|\Phi_0;{\cal{A}}_{out})
\end{equation}
where formally $S(\hat \Psi_t, RF)$ is the RF contribution that, for classical-quantum states, will be $\langle \hat \Psi_t|\hat h_{\Phi_0}+\hat h_{RF}|\hat \Psi_t\rangle$ and where $S(\hat \Psi_t|\Phi_0;{\cal{A}}_{in})$ measures the distinguishability between the state $|\Phi_0\rangle$ and the state $|\Psi_t\rangle$ relative to the "exterior" algebra ${\cal{A}}_{out}$. Alternatively we can write
\begin{equation}
S(\hat \Psi_t, RF') + S(\hat \Psi_t|\Phi_0;{\cal{A}}_{in})
\end{equation}
where we use the mirror reference frame RF' and where we define distinguishability relative to the "interior" algebra ${\cal{A}}_{in}$.

As discussed in \cite{Witten3} and in section 4 there exist a natural connection between the reference frame 
hamiltonian $\hat h_{RF}$ and the {\it dressing} of the operators defining the crossed product algebra. Recall from section 4 that this dressing is associated with the factor $e^{it\beta h^{II}}$ in the definition of the elements in the crossed product algebra. This dressing is clearly {\it asymmetric} with respect to the role of the algebra and its commutant. In the two sided case this asymmetry manifest as {\it a time shift} \cite{Witten3} that you can associate with the RF hamiltonian. In the case of the one sided black hole with the algebras ${\cal{A}}_{out}$ and ${\cal{A}}_{in}$ playing the role of the algebra and its commutant the dressing factor, now defined in terms of the Tomita Takesaki modular Hamiltonian for the representation consistent with ${\cal{A}}_{out}^{'} ={\cal{A}}_{in}$, creates a clear asymmetry between the interior and the exterior of the black hole that we can interpret as defining a {\it time shift at horizon crossing}.

\subsection{de Sitter planar purification}
Let us now consider the case of de Sitter but instead of considering, as we did in the TFD construction, the two static patches let us take the {\it planar patch} of one observer either the north pole or the south pole observer. On the planar patch we have two well defined regions separated by a horizon. One region is the static patch of the observer and the other piece, in the planar patch, is representing what is outside the static patch i.e. what is beyond the cosmological horizon of the observer. 

We can define two algebras namely ${\cal{A}}_{dS}$ associated with the static patch and an algebra ${\cal{A}}_h$ associated with the planar patch complementary region i.e. with observables located beyond the cosmological horizon ( see next section for more details). Now assuming they are von Neumann algebras and requiring $[{\cal{A}}_{dS},{\cal{A}}_h]=0$, as it should be the case, we can impose that ${\cal{A}}_h$ is just the commutant of the algebra associated with the static patch. 

In essence what we are defining is what we can call a "{\it planar representation}" where instead of using, as in the TFD construction, the two static patches to define 
the algebra and its commutant we use the two pieces of the planar patch. 

In the most familiar TFD construction the {\it purifications}, of the density matrices defining the physics on one static patch i.e. the density matrices in the type $II$ factor ${\cal{A}}_{dS}^{cr}$, are pure states, as the TFD state, covering the union of both static patches. In this {\it planar representation} the purifications of the density matrices of the type $II$ factor should be instead pure states {\it covering the whole planar patch}.

\section{Inflation, Observers and the Early Universe Algebra}
\subsection{The type $II$ proposal}
In this section we will discuss the crossed product type $II$ algebraic version of inflation \cite{Gomezcross},\cite{corean}.
Normally Inflation is defined in terms of weakly coupled QFT in a space-time classical background solving the Einstein equations for a slow rolling primordial "dark energy". This geometry is characterized by the equation of state $\epsilon=\frac{3(\rho+p)}{2\rho}$ with $\epsilon=0$ the pure de Sitter limit. In this approximation we ignore, during the primordial and slow rolling period, {\it quantum gravity} effects.  Thus the primordial inflationary period can be described in the weak gravity limit. 

Obviously we could work out quantum gravity $O(\frac{1}{M_P})$ corrections and to try to see how these quantum gravity corrections modify the classical background. In particular we could be interested in quantum gravity corrections to the inflationary potential or equivalently to the classical equation of state. In this section we will not discuss these important effects and we will keep ourselves in the weak gravity limit \footnote{To rule out eternal de Sitter on the basis of {\it quantum gravity} corrections to $\epsilon$ was already discussed in \cite{Gia1,Giainfla1,Giainfla2} where a quantum gravity lower bound on the minimal value of $\epsilon$ was stablished as well as the associated maximal time of stability identified as a quantum breaking time. For a different approach see \cite{Vafa}. }. 

Let us focus on the {\it double limit} $G_N=0$ and $\epsilon=0$ corresponding to QFT in pure de Sitter. The first thing to be noticed is that in this limit the QFT algebra ${\cal{A}}_{dS}$ of local observables describing the static patch i.e. the causally complete domain of a generic observer,  is a type $III_1$ factor. This means, as extensively discussed, that the Hamiltonian $\hat H$ generating time translations in this domain has divergent quantum fluctuations.

The first observation defining our proposal for an algebraic description of Inflation consists in:

\vspace{7mm}
{\bf 1.-}  {\it Comparing these divergent fluctuations, induced, in the $G_N=0$ limit, by the type $III_1$ nature of ${\cal{A}}_{dS}$, with the divergent power spectrum of scalar curvature fluctuations in the pure dS limit defined by $\epsilon=0$.}
\vspace{7mm}

{\it Quantum mechanically} the power spectrum of scalar curvature fluctuations (see for a review \cite{Bauman}) is defined, up to numerical normalization factors, for each mode of comoving momentum $k$  by 
\begin{equation}\label{powerk}
{\cal{P}} \equiv k^3|\zeta_k(\eta)|^2
\end{equation}
with $\zeta_k(\eta)= \frac{v_k(\eta)}{z M_P}$ for $v_k$ the Mukhanov Sasaki variable. The {\it wave function} $v_k(\eta)$ satisfies the Chibisov-Mukhanov equation \cite{Chibisovphonon}
\begin{equation}\label{Chibisov}
v^{''}_k + ( k^2 - \frac{z^{''}}{z}) v_k =0
\end{equation}
with $z=a\sqrt{\epsilon}$ for $a$ the conformal factor of the metric and $\epsilon$ determined by the equation of state \footnote{In this formulation $\eta$ is the conformal time defined on the full planar patch. Derivatives $'$ in (\ref{Chibisov}) are defined with respect to the conformal time $\eta$.} Note that  for $\epsilon = cte$ we get 
\begin{equation}\label{dS}
\frac{z^{''}}{z} = 2H^2a^2 + \epsilon H^2 a^2
\end{equation} 
At horizon crossing $k\sim a H$ we get for the power spectrum
\begin{equation}
 {\cal{P}} \sim \frac{H^4}{\dot \phi^2} = \frac{H^2}{M_P^2 \epsilon}
\end{equation}  
which is divergent in the pure de Sitter limit $\epsilon=0$. The variable $\zeta_k$ represents the {\it scalar curvature} of hyper surfaces of uniform density. This is a {\it gauge invariant} quantity whose meaning can be easily understood. Indeed for $\rho$ the density we have
\begin{equation}
-\zeta = \Psi + \frac{H}{\dot {\bar \rho}} \delta \rho
\end{equation}
with $\Psi$ defined by the metric fluctuation $\delta g_{ii} = a^2(1-2\Psi)$ and with $\bar \rho$ the homogeneous ground state density. Obviously this quantity is invariant under time reparametrizations $t\rightarrow t+ \alpha$ under which $\delta \rho$ goes into $\delta \rho + \dot {\bar \rho} \alpha$. 

Moreover the fluctuation $\delta \rho$ contributes to 
$T_{00}$ and therefore we can think \footnote{In the gauge defined by spatially flat hyper surfaces.}  of $\zeta = \frac{\delta \rho}{2\epsilon\rho}$ as contributing to the Hamiltonian.

Alternatively we could think the power spectrum in terms of a classical {\it gaussian statistical ensemble} defined by a gaussian probability distribution $p(\zeta_k)$ on possible values $\zeta_k$ defining the Fourier modes of the scalar curvature on hyper surfaces of uniform density i.e.
\begin{equation}\label{statistical}
\langle \zeta_k \zeta_{-k} \rangle_{st} \sim \frac{{\cal{P}}}{k^3}
\end{equation}
where now $\langle..\rangle_{st}$ means average with respect to the statistical ensemble. In this statistical ensemble approach the power spectrum defines the variance of the Gaussian probability distribution. Thus we have two possible descriptions of the power spectrum. On one side the quantum mechanical (\ref{powerk}) where we use the quantum wave functions solving equation (\ref{Chibisov}) and on the other hand the statistical ensemble description (\ref{statistical}) where in first approximation we use a gaussian probability distribution \footnote{A priori there is not any reason for gaussianity that should be checked considering quantum three point correlators of $\zeta$. A Gaussian description will be natural if the variables $\zeta$ define central elements.}  A classical problem in Cosmology is how to distinguish, without invoking non gaussian effects, the quantum mechanical description (\ref{powerk}) from the purely statistical (\ref{statistical}) determined by a statistical ensemble. The natural way to proceed  is by  designing some form of cosmological Bell experiment \cite{Bell}. This has been discussed in \cite{MaldaB}. In this section we will not touch this problem.

The former comment on the quantum mechanical description versus the statistical ensemble description of the primordial power spectrum of scalar curvature fluctuations provides the first hint to develop the correspondent algebraic approach. 

Let us define the local and gauge invariant {\it operator} $\hat \zeta^{III}(x,t)$ measuring the scalar curvature quantum fluctuations in the pure de Sitter limit and let us denote $\hat \zeta_k^{III}$ the corresponding Fourier components. We introduce the upper label $III$ to stress that this operator is defined in the type $III$ factor describing pure de Sitter where, as already stressed, we include quantum fluctuations of the metric as well as of the matter on the classical background. What we know is that the quantum fluctuations in the pure dS limit, are divergent: 
\begin{equation}
\langle (\hat \zeta_k^{III})(\hat \zeta_{-k}^{III})\rangle \sim O(\frac{1}{\epsilon})
\end{equation}

 Thus we can substantiate {\it the statement} {\bf 1.-} above defining the map relating the pure de Sitter divergence of the power spectrum for scalar curvature fluctuations, at horizon crossing, with the divergent type $III_1$ quantum fluctuations of the generator $\hat H$ of time translations in the static patch, as the correspondence 
 \begin{equation}
 \hat H \Leftrightarrow \hat \zeta^{III}
\end{equation} 
 between the dS Hamiltonian $\hat H$ and the type $III$ version of $\zeta$. In summary

\vspace{7mm}
{\bf 2.-} {\it Type $III$ divergent quantum fluctuations $\Delta(\hat H^2)$ $\Leftrightarrow$ Divergent power spectrum $\langle (\hat \zeta_k^{III})^2\rangle $ in the dS limit.}
\vspace{7mm}

In this correspondence $\Delta(\hat H^2)$ is divergent for any state in the GNS Hilbert space representation of ${\cal{A}}_{dS}$. From this algebraic point of view the divergence of $\langle (\hat \zeta_k^{III})(\hat \zeta_{-k}^{III})\rangle$ in the pure dS limit means that the "modes" created by $\hat \zeta_k$ are {\it not} in the Hilbert space representation of ${\cal{A}}_{dS}$.

At this point we can define a {\it rescaled} $\tilde \zeta = \hat \zeta^{III} \sqrt{\epsilon}$ \footnote{This is the analog of the rescaled variable $\frac{\hat H-\langle\hat H\rangle}{N}$ in the most familiar case where the divergence of $\hat H$ goes as $N^2$ \cite{Witten3}. In that approach this rescaled variable becomes central in the $N=\infty$ limit.}. We can think this rescaled variable as a {\it tensor operator} by analogy with the standard definition of tensor fluctuations. Note that formally $\tilde \zeta$ is {\it central} in the dS limit $\epsilon=0$ and consequently we will have, in this limit, purely {\it Gaussian} correlators for $\tilde \zeta$ with variance $O(\frac{H^2}{M_P^2})$. Adding the rescaled $\tilde \zeta$ is equivalent, at the algebraic level, as adding a central term with Gaussian correlators. In the Cosmological context we can think of this extra central term as defining {\it tensor fluctuations}. Note that by adding this central term we keep ourselves in type $III$ although not anymore in a factor. 

Based on our former experience if we think $\tilde \zeta$
as an extra central term added to the pure dS algebra and we identify this mode with tensor perturbations we could think the gaussian spectrum of tensor modes in terms of the statistical ensemble controlling $\tilde \zeta$.

Now comes the key ingredient of our proposal namely {\it to move into a type $II$ version of de Sitter}. In order to define this type $II$ version we need to add the quantum reference frame algebra $[\hat h_{RF},\hat t]= -i\hbar$ and to work in an extended Hilbert space with generic product states characterized by quantum wave functions $\psi_{RF}(\epsilon)$ on the spectrum of $\hat h_{RF}$ and some state in the GNS representation of ${\cal{A}}_{dS}$. In this extended type $II$ algebra instead of using the rescaled and {\it central} variable $\tilde \zeta$ we need to introduce the type $II$ version $\hat \zeta^{II}$. If we think $\hat \zeta^{III}$ in parallel to $\hat H$ then we need to think $\hat \zeta^{II}$ in parallel to the regularized Hamiltonian $\hat h^{II}$ with
\begin{equation}
\hat h^{II} = \hat h + \beta_{dS}\hat h_{RF}
\end{equation}
with $\hat h$ the {\it state dependent} Tomita Takesaki modular Hamiltonian naturally defined by the type $III_1$ algebra ${\cal{A}}_{dS}$. 

In this case the type $II$ quantum mechanical version of the power spectrum becomes 
\begin{equation}
\langle (\hat \zeta^{II})^2 \rangle \sim \Delta(\hat h^{II})^2 
\end{equation}
on some state in the extended Hilbert space characterized by some wave function $\psi_{RF}(\epsilon)$.

We can summarize this step of the proposal as:

\vspace{7mm}
{\bf 3.-} {\it To use the type $II$ quantum regularization of $\Delta(\hat H^2)$, defined in the weak gravity limit as $\Delta(h^{II})^2$  for the appropriated type $II$ Hamiltonian $h^{II}$, to define a regularized type $II$ power spectrum.}
\vspace{7mm}

In essence what we are doing until this point is to {\it regularize} the pure dS divergent power spectrum i.e. the divergent quantum fluctuations of $\hat \zeta^{III}$ by adding the reference frame algebra in the form of a crossed product and looking for a quantum state in the extended Hilbert space with finite quantum fluctuations of $\hat \zeta^{II}$. Thus the proposal is to define {\it an emergent} effective $\epsilon$ by the formal correspondence
\begin{equation}
{\cal{P}} \sim \frac{H^2}{M_P^2 \epsilon} \Leftrightarrow \Delta(h^{II})^2
\end{equation}
with $\Delta(h^{II})^2$ defined on some particular state in the extended type $II$ Hilbert space.
Note already on the basis of (\ref{capacity}) that this correspondence relates, as we will see in a moment, the power spectrum with an {\it entanglement capacity}.

Before going on let us recap the main aspects of the former discussion. In the standard approach to Inflation in the slow roll approximation, we need to distinguish two aspects. On one side the {\it amplitude} of the power spectrum we observe at the CMB and the {\it scaling} of this power spectrum. Let us first focus on the amplitude. The amplitude, as explained above, is defined by $|\zeta_k(\eta)|^2$ at horizon crossing that goes like $\frac{|v_k(\eta)|^2}{a^2M_P^2 \epsilon}$. Independently of how precisely we solve equation (\ref{Chibisov}) to evaluate $v_k(\eta)$, the $\epsilon$ in the denominator appears due to impose gauge invariance of physical fluctuations \footnote{We can use Bessell approximation at horizon crossing or even the exact solution of (\ref{Chibisov}) for the no decaying mode but in any case the leading dependence of the amplitude on $\epsilon$ as $\frac{1}{\epsilon}$ is determined by {\it gauge invariance}.}.

 The first step of our algebraic approach has been i) to interpret this divergent value of $\zeta$ in the pure dS case $\epsilon=0$ as the typical type $III$ divergence of quantum energy fluctuations and ii) to interpret the finite value associated with a quasi de Sitter equation of state with non vanishing $\epsilon$ as a type $II$ regularization. 
 
 In this sense {\it instead of modifying dS ( $\epsilon=0$ ) into a quasi dS ($\epsilon\neq 0$) we modify the type $III$ algebra of dS into the type $II$ crossed product algebra}. It is under this claim that we map the power spectrum into the entanglement capacity for some state in the extended Hilbert space.

Until this point we have simply taming the pure dS divergence of the amplitude using a type $II$ crossed product version of the type $III$ pure de Sitter algebra. However this cannot be the end of the story. As stressed above, in standard inflation we don't simply correct the divergent pure de Sitter amplitude including some non vanishing slow roll, but we also include this information on slow roll in the key equation (\ref{Chibisov}) modifying in this way the dependence of $v_k$ on $\eta$. It is this modification what explains the scaling of $|v_k(\eta)|^2$ and determines the known spectral index $(1-n_s)$. In other words, regularizing the amplitude also implies a non trivial scaling. This is physically very clear since a regularized amplitude, with a non vanishing $\epsilon$, immediately implies a running of the power spectrum defined at horizon crossing. The reason is simply because, classically, for non vanishing $\epsilon$ the data at horizon crossing depend on the scale. 

After this comment we can try to continue with our algebraic approach. What we expect is that the type $II$ version of the power spectrum {\it not only accounts for a regularization of the amplitude in the form of some effective non vanishing $\epsilon$ but also for the non trivial scaling leading to the correct value of $(1-n_s)$}. More precisely to want to extract the effects of the semiclassical slow roll, both on the amplitude as well as on the scaling, from the type $II$ entanglement capacity of some state in the type $II$ extended Hilbert space.

In order to gain some preliminary intuition let us consider some concrete states in the extended Hilbert space. For the maximally entropic state with density matrix equal the identity the RF wave function is thermal and normalizable after projecting on positive eigenvalues of $\hat h_{RF}$. In this case we get a non vanishing entanglement capacity $\Delta(\hat h^{II})^2$ scaling like $H^2$. This is the naive power spectrum we get if we think in tensor fluctuations. However we don't get any specific slow roll information. When we consider instead a reference frame wave function typical of a coherent state with a gaussian wave function of some variance $\sigma$ we naturally get a finite power spectrum for $\hat \zeta^{II}$ characterized by $\sigma$ and therefore we can map the slow roll parameters, entering into the amplitude, into this reference frame wave function.

However to end the discusion at this point will be highly unsatisfactory. In essence the question is: Why among the plethora of states in the extended Hilbert space with finite quantum fluctuations for $\hat \zeta^{II}$ {\it Nature} selects a particular one with the observed value ? Equivalently: Can we imagine a way to select the reference frame wave function accounting for the observed power spectrum in a model independent way ?

\subsection{Toward a predictive approach to Inflation}

This question leads us to the last step of our proposal. In the type $II$ version we have generic states $|\psi\rangle$ in the extended Hilbert space with non vanishing entanglement capacity defined as
\begin{equation}
{\cal{C}}(|\psi\rangle) = \Delta (\hat h^{II})^2
\end{equation}
with $\Delta$ evaluated on the state $|\psi\rangle$. This definition of entanglement capacity is state dependent even at the level of the definition of $\hat h^{II}$. Now these states in the extended Hilbert space of the type $II$ factor define, in principle, {\it purifications} of well defined type $II$ density matrices. Moreover these states account for the physics described by the algebra ${\cal{A}}_{dS}$ of local observables defined in the static patch as well as for the physics, not accessible to the observer, described by the operators in the commutant ${\cal{A}}'_{dS}$. 

As discussed in the previous section we can formally define a one side or {\it planar} purification. Recall that this planar purification, defined relative to the planar patch corresponding to the causal future of north pole observer, can be used to describe de Sitter expanding Cosmology. In {\it static coordinates} this patch contains two regions separated by the cosmological {\it horizon}. One region is the static patch and the other is the region beyond the cosmological horizon of the observer \footnote{In these {\it static coordinates} the time translation  $\frac{\partial}{\partial t}$ is timelike in the static patch, null on the horizon and spacelike in the region outside the horizon.}.
In order to cover the planar patch we can use {\it planar coordinates} where we can define the conformal time $\eta$ and the associated physical time $t_p$ by $t_p=\int ad\eta$ for $a=e^{Ht_p}$. Hypersurfaces of fixed $\eta$ "foliate" the planar patch. 

Algebraically we can assign to the two regions defining the planar patch two algebras, namely ${\cal{A}}_{dS}$, defined above, for the static patch and ${\cal{A}}_{h}$ representing the local observables on {\it the region outside the cosmological horizon}. We will define a {\it planar representation} as a representation of ${\cal{A}}_{dS}$ such that its commutant is the algebra ${\cal{A}}_{h}$ describing {\it the exterior region}. Under these conditions the states $|\psi\rangle$ in the extended Hilbert space are defined on the whole planar patch provided the type $II$ factor is defined for the planar purification with 
${\cal{A}}_{h}= {\cal{A}}'_{dS}$. 

States with non vanishing quantum entanglement transform non trivially under the transformation generated by $\hat h^{II}$, namely $|\psi(t)\rangle = e^{it\hat h^{II}} |\psi\rangle$. Thus we can interpret these "time" dependent states as defining, {\it quantum mechanically}, a foliation of the full planar patch with each formal hyper surface associated with a common value of $t$ and with the state $|\psi(t)\rangle$ describing the physics on this hyper surface. Note that this is only possible for the type $II$ factor.  

Now comes the last and more speculative element of our proposal. A natural time foliating the planar patch is the conformal time $\eta$ so we could try to relate the time conjugated to $\hat h^{II}$ with the planar patch conformal time. It is well known that the quantum mechanical definition of the conformal time on the planar patch implies the use of a non trivial Bogolyubov transformation relating the quantam modes on hyper surfaces corresponding to different values of the conformal time $\eta$. In this sense we could try to associate the type $II$ operator $\hat h^{II}$ defining quantum mechanicaly a natural foliation of the planar patch with the Bogolyubov transformation needed to define free QFT on the planar patch i.e.

\vspace{7mm}
{\bf{C}} {\it (planar) Bogolyubov transformation $\Leftrightarrow$ type $II$ definition of $\hat h^{II}$}
\vspace{7mm}

On the basis of the qualitative correspondence {\bf C} we can look for the simplest and model independent Bogolyubov transformation, namely the one used to define quantum mechanically the conformal time foliation of the planar patch, for free fields.

Now recall we are looking for a state $|\psi\rangle$ in the type $II$ extended Hilbert space such that the corresponding entanglement capacity ${\cal{C}}(|\psi\rangle)$ accounts for the primordial power spectrum of scalar curvature fluctuations. The simplest although quite bold conjecture is to identify this state with one quantum pure state defined over the whole planar patch and fully determined by the {\it pure de Sitter}  Bogolyubov transformation. 

Natural quantum states on the planar patch can be defined for each comoving momentum $k$ using the standard solutions $a_k(\eta), a^{\dagger}_k(\eta)$ to the Bogolyubov transformation defining, in the planar patch, the conformal time dependence of creation annihilation operators ( see\cite{Martin} and references therein). Once we know $a_k(\eta), a^{\dagger}_k(\eta)$ we can define the state $|k,\eta\rangle$ as the solution to $a_k(\eta)|k,\eta\rangle =0$. This state is well known and is given by
\begin{equation}
|k,\eta \rangle = \sum_n c((k\eta),n) e^{in\Phi((k\eta))}|n_k,n_{-k}\rangle
\end{equation}
where we sum over all integers and where
$c((k\eta),n)=C \tanh({r(k\eta)})^{n}$
with $C$ a normalization constant, 
$r(k \eta) = - \sinh^{-1}(\frac{1}{2k\eta})$
and with the phase given by
$\Phi(k\eta) = -\frac{\pi}{4} -\frac{1}{2} \tan^{-1}(\frac{1}{2k \eta})$.

Let us notice several interesting properties of this state. For fixed momentum $k$ this state clearly depends on the conformal time $\eta$. The dependence appears on the probability distribution defined by $c((k\eta),n)$ as well as on the {\it quantum phases} defined by $\Phi(k\eta)$. Now our "bold conjecture" is that $|k,\eta\rangle$ for fixed $k$ defines a type $II$ {\it quantum foliation} of the planar patch. Equivalently we assume it exists a type $II$ Hamiltonian $\hat h^{II}$ satisfying locally
\begin{equation}\label{matching}
\frac{d|k,\eta\rangle}{d\eta} = (\hat h^{II}) |k,\eta\rangle
\end{equation}
with as usual $\hat h^{II} = \hat h_{planar} +\beta_{dS}\hat h_{RF}$ for some added RF algebra.

On the basis of this qualitative argument we identify the state $|k,\eta\rangle$ with the desired state in the type $II$ extended Hilbert space. 
Hence we can define the type $II$ {\it entanglement capacity} ${\cal{C}}(k,\eta)$ associated with this state as the {\it quantum Fisher information} relative to $\eta$ dependence of the purification state $|k,\eta\rangle$. This {\it quantum information}, by contrast with a classical information is determined by the quantum phases of the state $|k,\eta\rangle$ and is represented by

\begin{equation}
{\cal{C}}(k,\eta)= ( \sum_n c(k\eta,n)^2 (\frac{\partial \Phi(k\eta,n)}{\partial (k\eta)})^2 - (\sum_n c(k\eta,n)^2 (\frac{\partial \Phi(k\eta,n)}{\partial (k\eta)}))^2)
\end{equation}

This quantity has been evaluated numerically in \cite{GJ}. After reaching this point the final step of our proposal is to define the type $II$ regularized mode dependent power spectrum in terms of 
${\cal{C}}(k,\eta)$ i.e.
\begin{equation}
{\cal{P}}(k,\eta) = \langle \hat \zeta_k^{II}\hat \zeta_{-k}^{II}\rangle \Leftrightarrow {\cal{C}}(k,\eta)
\end{equation}
and to extract from this identification the effective inflationary parameters determined by the type $II$ construction.

The obvious question that the skeptical reader can pose at this point is: 

{\it How can we expect to get some relevant information about inflationary parameters if the whole input we are introducing is the definition of the state $|k,\eta\rangle$ that only depends on pure de Sitter information as the standard Bogolyubov transformations in pure dS ?}

The answer we can offer is the following. The relevant quantum information is contained in $\Delta(\hat h^{II})^2$ evaluated on the state in the extended Hilbert space that we have identified with $|k,\eta\rangle$. The semiclassical self consistent way to reinterpret this effect is to correct the Bogolyubov transformation by including the quantum correction of the Hamiltonian $\hat h^{II}$ induced by the non vanishing $\Delta(\hat h^{II})^2$. This correction is a quantum correction in $\hbar$ {\it but in the $\hbar$ entering into the reference frame algebra used to define the type $II$ factor}. But since the type $II$ factor is designed to correct the $\epsilon=0$ divergence of pure dS this {\it reference frame} effects appears as an $\epsilon$ effect that modifies the Bogolyubov transformation in the same way the factor $\frac{z^{''}}{z}$ modifies the key equation (\ref{Chibisov}) in the slow roll approximation \footnote{This is the standard self consistent approach to include quantum corrections. In symbolic form we evaluate the quantum correction determined by $\Delta(\hat h^{II})^2$ and we add this $\hbar$ effect as a correction to the equation defining the Bogolyubov transformation with (\ref{Chibisov}) the resulting equation. The type $II$ description reveals that the $\hbar$ correction of $\Delta(\hat h^{II})^2$ comes {\it from the $\hbar$ defining the reference frame algebra} i.e. is a quantum type $II$ effect in the weak gravity limit and therefore it encodes information about the type $II$ regularization used to tame the type $III$ divergence going like $\frac{1}{\epsilon}$.}

This last comment makes us confident on the possibility that the so defined type $II$ power spectrum accounts for both the correct amplitude as well as for the right spectral index $(1-n_s)$  \cite{Chibisov2, Hawking}\footnote{Recall that the non vanishing spectral index
comes from the deviation of $\frac{z^{''}}{z}$ in (\ref{Chibisov}) from the pure de Sitter value $\frac{2}{\eta^2}$.}. Thus if the type $II$ proposal is correct
we should expect that the former evaluation of ${\cal{C}}(k,\eta)$ also accounts for the value of $(1-n_s)$. In the quantitative analysis developed in \cite{Gomezcross} and \cite{GJ} for the squeezed state $|k,\eta\rangle$ this is indeed what happens with nice preliminary numerical predictions 
\begin{equation}
(1-n_s) \sim 0.0328
\end{equation}

and 
\begin{equation}
\epsilon\sim 0.0027
\end{equation}
( see \cite{GJ} and \cite{Gomezcross} for technical details ). Although these numbers should be taken with a grain of salt the main message is that the interpretation of the inflationary period as a type $II$ version of the type $III$ pure de Sitter case \footnote{Not to be confused with a type $II$ version of {\it classical} quasi de Sitter.}, opens the possibility of attempting a predictive approach to primordial Cosmology.

\subsection{Comments}

{\bf 1.-} A crucial aspect of our argument has been to use time dependent states $|\psi(t)\rangle = e^{i\hat h^{II} t} |\psi\rangle$ in the extended Hilbert space. To visualize the meaning of these states let us first consider the two sided $|TFD\rangle$ purification. In the type $III$ version this state is invariant under the modular transformation $e^{i\hat h t}$. However when we move into a type $II$ modification we can define $|TFD\rangle (t) = e^{i\hat h^{II}t} |TFD\rangle$ that formally accounts for the type $III$ ill defined time translation $e^{it(H_L+H_R)}$. When we work in the {\it planar purification} these states define what we have denoted a quantum foliation i.e. the family of states $|\psi(t)\rangle$ is interpreted as a foliation in $t$ with the quantum states $|\psi(t)\rangle$ describing the physics on each fixed $t$ hyper surface. We have used the simplest {\it pure dS} Bogolyubov transformation to define these states. However even in this simpler case we get a non vanishing entanglement capacity for these time dependent states. More precisely in the type $II$ version these states define a family of type $II$ density matrices ( each one purified by $|\psi(t)\rangle$ ) with the entanglement capacity equal to the corresponding quantum Fisher information ( relative to $t$ ). Even in the two sided version we get non vanishing entanglement capacity for the type $II$ states $|TFD\rangle (t) = e^{i\hat h^{II}t} |TFD\rangle$. In the simplest case this effect is simply related to the quantum uncertainty for the added RF Hamiltonian. Intuitively (\ref{Chibisov}) describes the quantum fluctuations of the full system with the added RF Hamiltonian characterized by the quantum uncertainty of $\hat h_{RF}$ \footnote{The former discussion is reminiscent of some recent approaches to wormhole traversability \cite{Jefferis, maldatrans}. In such a cue you manufacture a modification of the TFD state supporting a fluctuation of null energy that leads to the time shift needed to achieve the desired traversability. In the Cosmological setup we look for the time delay at horizon crossing.}.

{\bf 2.-} In the {\it planar representation} we can define the type $II$ factor associated to ${\cal{A}}_{dS}$ adding a RF to the static patch. In order to associate a type $II$ factor with the planar commutant ${\cal{A}}_h$ we need to add a RF in the region beyond the primordial cosmological horizon. Heuristically a RF for the static patch defines a physical way to identify the time at which modes created inside the static patch exit the horizon. Since, as already said, in the planar representation the type $II$ {\it dressing} contains the factor $e^{it(\hat h_{planar} +\beta \hat h_{RF})}$ this induces an asymmetry and consequently a potential {\it time shift} ( or time delay ) associated with the presence in the type $II$ dressing factor of $\hat h_{RF}$. This can be interpreted as the abstract algebraic counterpart of the standard time delay used in simple derivations of the power spectrum.

But what about the RF used in the region beyond the primordial cosmological horizon? The static patch observer cannot know what happens in this region, however a RF is needed to associate a type $II$ factor to this region. We will call this RF a "CMB" observer.

If for some reason inflation ends we can assume that the type $II$ factor associated with the region beyond the primordial cosmological horizon will describe the history of the Universe from the moment the primordial exponentially expanding phase ends until the present phase. If this type $II$ factor is $II_{\infty}$ we could expect to describe our present Universe as a slowly rolling {\it quintessence}. The reason we associate with this potential phase a type $II_{\infty}$ factor is because in this case the entropy can grow without any upper bound. However it can happens that the present Universe is dominated by a small but finite cosmological constant and in such a case the factor should be a type $II_1$. An appealing possibility is that we could encode the differences between the primordial phase
and the present as differences between the two RF's observers. Somehow the type $II_1$ primordial exponentially expanding phase could {\it know} about the present phase through the continuous Murray von Neumann dimension $d$ i.e. through the entropy deficit between the two phases. In this scenario  we could formally relate
$M_P^2( \frac{1}{H_{primordial}^2} - \frac{1}{H_{present}^2})$ as $\sim \log (\frac{1}{d})$. In a certain sense what in standard inflation we parametrize by a change of the effective value of $H$ could be thought  in terms of the change of the continuous dimension $d$ of the primordial type $II_1$ factor. Obviously this is just a qualitative and very speculative comment. 

\section{The algebra of large $N$ gauge theories}
\subsection{Preliminary discussion}
We will study, following the line of thought developed in the previous sections, the algebra of physical observables for $SU(N)$ gauge theory, at finite temperature $\beta$, defined on $S^3\times R$ with finite volume and in the  $N=\infty$
limit. More specifically we will address the problem of how to define in this large $N$ limit
thermal correlators for single trace operators in the high temperature regime i.e. for $\beta<\beta_H$ with $\beta_H$ the inverse Hagedorn temperature. 

The phase diagram of $SU(N)$ pure Yang Mills at finite volume and in the large $N=\infty$ limit has been extensively studied and it is reasonably well understood \cite{Minwalla1,Minwalla2,Minwalla3}. The partition function of the theory is
\begin{equation}\label{partition}
Z(\beta) =\int dE \rho(E)e^{-\beta E}
\end{equation}
for $\rho(E)$ the spectral energy density. In the large $N=\infty$ limit we know, based on reliable estimations of $\rho(E)$, the existence of a Hagedorn temperature $\beta_H$.

The simplest way to define the Hagedorn temperature is as the temperature $\beta_H$ for which the Boltzmann suppression $e^{-\beta E}$, is, for $\beta < \beta_H$ and in the limit of high energies, subdominant relative to the growth with energy of the spectral density $\rho(E)$. A typical example are those systems where the multiplicity of physical states at a given energy $n(E)$ grows exponentially as
\begin{equation}
n(E) \sim e^{\alpha E}
\end{equation}
with $\alpha$ some fixed physical scale with units of length and with $\rho(E)=\frac{dn(E)}{dE}$. This is, in particular, the case of strings where the number of states at level $n$ i.e. with energy of order $\sqrt{n}$, in string units, grows as $e^{\sqrt{n}}$. 

Obviously if the system has a Hagedorn temperature we will find that $Z(\beta)$ is divergent for $\beta<\beta_H$. In the holographic setup the gravitational dual of the Hagedorn transition, for $SU(N)$ Yang Mills, was put in correspondence, in \cite{Wittenconf}, with the Hawking-Page transition in the gravitational dual \footnote{The Hagedorn temperature $\beta_H$ defined at zero coupling is higher than the HP phase transition. Once we work with finite 't Hooft coupling $\lambda=g_{YM}^2N$ we discover a first order phase transition at some temperature $T_{HP} < T_H$ ( see \cite{Minwalla2} ). In the context of AdS/CFT this gap in temperatures can be large since we work at strong t'Hooft coupling.} . 

\subsection{Hagedorn and Bekenstein bounds}
As a parenthesis it could be worth to say few words on the relation between Hagedorn temperature and Bekenstein bound. Bekenstein bound \cite{Bek} is an absolute bound on the entropic capacity of a bounded region of space of finite volume and typical size $L$ as a function of the total energy $E$ enclosed in that region. This bound is independent of gravity and sets the upper bound on the entropy ( for an sphere of radius $L$ ) as
 \begin{equation}
 S \leq \frac{2\pi k LE}{\hbar}
 \end{equation}
 with $k$ the Boltzmann constant. Note that if the bounded system is a black hole with $E=M$ and $L=ML_P^2$ i.e. the corresponding gravitational size, then the Bekenstein upper bound $S= M^2L_P^2$ agrees with the Bekenstein Hawking black hole entropy. 
 
 Using Bekenstein formula we can define an upper bound on the multiplicity of physical states $n(E)$ we can enclose in a finite volume $S^3$ of radius $L$ as
 \begin{equation}
  n(E)\leq e^{ 2\pi LE}
 \end{equation}
 where we use natural units and we define the entropy $S$ as $\ln(n(E))$. Thus we observe that saturation of Bekenstein bound at high energies leads to a Hagedorn temperature $\beta_H$ of the order
 \begin{equation}
 \beta_H \sim 2\pi L
 \end{equation}
 In other words we can think of Hagedorn temperature, for a system defined in a finite volume, as reflecting {\it Bekenstein entropy saturation at high energies}. Thus the existence of a Hagedorn temperature for $SU(N)$ Yang Mills in the large $N$ limit and in finite volume strongly indicates that, in this limit, the gauge system enclosed in $S^3$, is close to Bekenstein saturation for $\beta<\beta_H$. How close to Bekenstein saturation will depend on the concrete value of $\beta_H$ that we will not discuss in this section.

\subsubsection{Order parameters}
 Once we give for granted the existence of a Hagedorn phase for the large $N$ limit of Yang Mills at finite volume, we should identify the {\it order parameter} distinguishing both phases. In the weak coupling limit these two phases are characterized as the confinement ( $\beta>\beta_H$ )  and deconfinement phase ($\beta<\beta_H$). The typical order parameter distinguishing both phases, in the large $N$ limit, is the rescaled free energy, namely
\begin{equation}\label{free}
\lim_{N=\infty} \frac{F(\beta)}{N^2}
\end{equation}
This quantity will be zero in the low temperature {\it confining} phase. This reflects the fact that in this phase the free energy $F$ scales with $N$ as $N^0$. Conversely in the high temperature phase ( deconfinement ) the free energy scales with $N$ as $N^2$ making the order parameter non vanishing.

In gauge theories the standard order parameter for confinement is the expectation value of the Polyakov loop i.e. the temporal Wilson loop
\begin{equation}
\hat{\cal{P}} = Tr P exp \int_C A
\end{equation}
with $C$ the circular loop in time that for finite temperature $\beta$ will have radius $\beta$. The order parameter, distinguishing both phases, analogous to (\ref{free}) is \footnote{The use of $|{\cal{P}}|^2$ as order parameter is due to the fact that for finite volume we cannot have spontaneous breaking of the center of the gauge group and therefore we should use as order parameter an operator invariant under the action of the center. Note that ${\cal{P}}$ transforms non trivially under the center.}
\begin{equation}
\langle |\hat{\cal{P}}|^2\rangle_{\beta}
\end{equation}

\subsection{The algebraic approach}
Let us define ${\cal{A}}_{YM}$  the Yang Mills algebra of local single trace operators which are not central. For a given and generic temperature $\beta$ we can define the von Neumann algebra ${\cal{A}}_{\beta}$ as the {\it GNS representation of ${\cal{A}}_{YM}$ in a GNS Hilbert space ${\cal{H}}_{\beta}$}. Recall, from the previous sections, that the GNS construction is done associating with any element $a\in {\cal{A}}_{YM}$ a state $|a\rangle$ and defining the scalar product by $\langle b|a\rangle = f_{\beta}(b^*a)$ for a particular linear form $f_{\beta}$ defined on 
${\cal{A}}_{YM}$. The completion of this set of states defines the Hilbert space ${\cal{H}}_{\beta}$. Using for $f_{\beta}$ the thermal expectation values, the von Neumann algebra ${\cal{A}}_{\beta}$ is defined by the bounded operators  in $B({\cal{H}}_{\beta})$ representing the algebra ${\cal{A}}_{\beta}$ as $\pi(b)|a\rangle =|ba\rangle$. 

In this GNS representation it exists in ${\cal{H}}_{\beta}$ a special state, that we will denote $|\beta\rangle$, namely the one associated with the identity of ${\cal{A}}_{YM}$. Moreover once we count with a GNS representation of ${\cal{A}}_{YM}$ i.e. once we have defined the von Neumann algebra ${\cal{A}}_{\beta}$ we can define the commutant ${\cal{A}}_{\beta}'$ and to check the basic property of vN algebras ${\cal{A}}_{\beta}={\cal{A}}_{\beta}^{''}$.

For finite $N$ the former construction is well understood using the TFD formalism. To do that we formally define two identical copies of our system, normally denoted by the labels $L$ and $R$, and we define the state
\begin{equation}\label{TFD}
|TFD\rangle = \frac{1}{\sqrt{Z(\beta)}} \sum_i e^{-\frac{E_i}{2} \beta} |E_i\rangle_L|E_i\rangle_R
\end{equation}
Now for any single trace operator $a$ in ${\cal{A}}^R_{\beta}$, representing the algebra ${\cal{A}}_{\beta}$ for the copy $R$, we define the corresponding linear form $f_{\beta}$ as
\begin{equation}
f_{\beta}(a) = \langle TFD|a|TFD\rangle
\end{equation}
Defining  $\hat \rho_{\beta}^R = tr_{L} |\beta\rangle\langle\beta|$ with the trace over the Hilbert space of copy $L$ (an equivalent representation for $L$ can be obtained defining the trace over the $R$ copy) we get
\begin{equation}
 \langle TFD|a|TFD\rangle= tr (\hat \rho_{\beta}^R a)
 \end{equation}
 The corresponding GNS Hilbert space ${\cal{H}}_{\beta}$ is defined as the completion of $\{{\cal{A}}^R_{\beta}|TFD\rangle \}$ with the commutant ${\cal{A}}^{R '}_{\beta}= {\cal{A}}^L_{\beta}$.
 
 Note that the TFD state describes an entangled state of the double system. The entanglement is given by the VN entropy of the density matrix $\hat\rho_{\beta}$, namely
\begin{equation}\label{entan}
S(\beta)= - tr \hat\rho_{\beta} \log \hat\rho_{\beta} = \beta\langle \hat H\rangle_{\beta} + \log Z(\beta)
\end{equation}
The entanglement capacity \cite{boer} defined in section 4  is given by the variance of $\hat H$ i.e.
\begin{equation}
{\cal{C}}(\beta) = \beta^2(\langle \hat H^2\rangle_{\beta} -\langle \hat H\rangle_{\beta}^2 )
\end{equation}
Defining the free energy $F(\beta) = -\frac{1}{\beta} \log Z(\beta)$ we get
\begin{equation}
{\cal{C}}(\beta) = -\beta^2(2 F'(\beta) + \beta F^{''}(\beta))
\end{equation}
for $F'= \frac{\partial F(\beta)}{\partial \beta}$. Finally the {\it quantum Fisher information} $I_F(\beta)$ for the variation of $\hat\rho_{\beta}$ with respect to $\beta$ is given by $\frac{{\cal{C}}(\beta)}{4}$. Thus in the TFD purification the quantum Fisher information or equivalently the entanglement capacity is represented by
\begin{equation}
\Delta(\hat H ^2) =\langle \beta|\hat H^2 |\beta\rangle -(\langle \beta|\hat H|\beta\rangle)^2
\end{equation}
with $\hat H$ the Hamiltonian of either the $L$ or the $R$ system.

Let us now consider the large $N$ limit. Following \cite{LL1} we will consider the renormalized single trace operators $t= T-\langle \beta|T|\beta\rangle$. The large $N$ limit of the vN algebra ${\cal{A}}_{\beta}$ will be defined by truncating the OPE of single trace operators to those elements surviving in the large $N$ limit. The GNS representation of this large $N$ algebra will be associated with a GNS cyclic state $|\hat\beta\rangle$ \footnote{We have denoted as $|\hat\beta\rangle$ the $N=\infty$ limit of the TFD state to distinguish it from the TFD state  defined at finite $N$.}. The obvious question we should face at that point is how the Hagedorn {\it phase transition} at $\beta= \beta_H$ is described in terms of the large $N$ algebra ${\cal{A}}_{\beta}$ and the cyclic state $|\hat \beta\rangle$.

The previous discussion contains several subtle aspects. A priori we could think the Yang Mills algebra as generated by arbitrary gauge invariant operators. In order to define the large $N$ limit we can reduce to the sub algebra, defined by the corresponding OPE, where we ignore sub-leading contributions, order $\frac{1}{N}$, to the OPE defining the algebra. This definition is however state dependent and consequently the definition of the large $N$ algebra is also state dependent. Once this state dependent algebra is defined the formal GNS representation selects a GNS vacuum state i.e. the one naturally associated in the GNS construction with the identity. The so defined GNS Hilbert space can be interpreted as describing the large $N$ small fluctuations around such a GNS vacuum. The different GNS vacuum states correspond to different states used to truncate the large $N$ algebra and they describe different semiclassical states\cite{LL3}.  

\subsection{Hagedorn as a type $I$ type $III$ transition}
In \cite{LL2} an interesting conjecture has been suggested on how the vN algebra ${\cal{A}}_{\beta}$ changes when we move from the low temperature phase $\beta>\beta_H$ into the high temperature phase $\beta<\beta_H$. The conjecture is that:
\begin{itemize}
\item For $\beta>\beta_H$ the vN algebra ${\cal{A}}_{\beta}$ is a type $I$ factor
\item For $\beta<\beta_H$ the vN algebra ${\cal{A}}_{\beta}$ is a type $III_1$ factor
\end{itemize}
This conjecture is partially supported by the explicit computations for the large $N$ limit of YM in finite volume and by the holographic dual description. Indeed we know that for $\beta<\beta_H$ $Z(\beta)$ is divergent as well as the entanglement (\ref{entan}) of the corresponding TFD state. From the holographic point of view we can replace $\beta_H$ by the Hawking-Page temperature $\beta_{HP}$ and the algebra ${\cal{A}}_{\beta}$ for $\beta<\beta_{HP}$ can be identified, in the gravity dual, with the algebra of QFT local observables on the $R$ or $L$ exterior region of the {\it two sided AdS eternal black hole}. From general arguments this algebra can be expected to be type $III$. 

Let us mention that we are effectively discussing the Hagedorn phase transition at zero coupling. As shown in \cite{Minwalla2} once we include finite coupling effects we discover a first order phase transition at a temperature smaller than $\beta_H$. The conjecture in \cite{LL2} is that, when we include finite $\lambda$ corrections, it is at this lower temperature where the transition from type $I$ into type $III$ should takes place. 

Irrespectively of these arguments let us first discuss, as we did for the de Sitter example, what the type $III$ nature of ${\cal{A}}_{\beta}$ implies for the physics describing the high temperature phase. 

The first implication of this conjecture is that {\it while} for $\beta>\beta_H$ the correlators
\begin{equation}\label{corre}
\langle \beta|t_1..t_n|\beta\rangle
\end{equation}
for any product of single trace operators\footnote{Assuming, on the basis of defining the algebra on an OPE, that the single trace operators generate the large $N$ algebra.}  in ${\cal{A}}_{\beta}$  can be represented as:
\begin{equation}\label{corre2}
 \langle \beta|t_1..t_n|\beta\rangle = tr(\hat \rho_{\beta} t_1..t_n )
 \end{equation}
 for $\beta<\beta_H$ it does not exist any density matrix $\hat \rho_{\beta}$ defining the representation (\ref{corre2}) of correlators (\ref{corre}). More specifically this means that while the linear form $f_{\beta}$ defining the GNS construction, at finite temperature, satisfies the trace property for $\beta>\beta_H$ it does not for $\beta<\beta_H$.  

The second consequence is more interesting and implies that {\it time evolution} works in a very different way in the low temperature and in the high temperature regime. Indeed in the high temperature regime, where ${\cal{A}}_{\beta}$ is type $III_1$, it exists a unique, up to inner automorphisms, outer automorphism implementing time translations in ${\cal{A}}_{\beta}$. The generator of this automorphism is the already familiar Tomita Takesaki (TT) state dependent modular Hamiltonian $\hat h_{\beta}$. In the low temperature regime $\beta>\beta_H$ time evolution is implemented using the standard Hamiltonian of our system. In the TFD formalism we have two identical copies with Hamiltonians $\hat H_L$ and $\hat H_R$ defining the time evolution for both systems independently. However in the high temperature regime {\it the only well defined time evolution is represented by the TT operator $\hat h_{\beta}$ while the operators $\hat H_L$ and $\hat H_R$ are ill defined}.

Why for $\beta<\beta_H$ the operators $\hat H_R$ and $\hat H_L$ are ill defined and what that means? The reason are ill defined can be easily explained. From explicit computations we know that the large $N$ limit of the TFD expectation value $\lim_{N=\infty} \langle\beta|\hat H_R|\beta\rangle = \langle \hat \beta|\hat H_R|\hat \beta \rangle$ goes as $N^2$. We can renormalize $\hat H_R$ and to define $h_{R}=\hat H_R- \langle \hat \beta|\hat H_R|\hat \beta \rangle$ \footnote{We can identify $\langle \hat \beta|\hat H_R|\hat \beta \rangle$ as the ground state energy $E_0(\beta)$. In the regime $\beta<\beta_H$ we can get an explicit expression for $E_0=N^2f(T)$ with $f(T)$ playing the role of the order parameter of the Hagedorn phase transition.}. However the {\it quantum fluctuations}, as measured by {\it the variance $\Delta(\hat h_R^2)$ evaluated on the large $N$ limit TFD state $|\hat \beta\rangle$, diverge with $N$ as $N^2$}. This means that when we act with $h_R$ we move out of the large $N$ GNS Hilbert space with cyclic state $|\hat \beta\rangle$ \footnote{Indeed the divergence of $\langle \hat \beta|h_R^2|\hat \beta\rangle$ implies that $h_R|\hat \beta\rangle$ is not a normalizable state and therefore is out of the GNS Hilbert space.}. 

This situation is reminiscent of what we have found in previous sections. More precisely, once we have identified the algebra of observables ${\cal{A}}_{\beta}$, in the large $N$ limit, we discover that for $\beta<\beta_H$ {\it quantum fluctuations} carry infinite energy and therefore they need to be {\it renormalized} \footnote{An intuitive way to visualize the problem is thinking that when we increase the temperature we need to use operators that cannot be represented in terms of the generators of the large $N$ algebra defined using a truncation of the OPE with respect to a given state. The type $III$ nature of the algebra indicates that in order to account for those operators we need to use the commutant. In some sense the high energy effective decoupling survives, in the type $III$ factor, in the form of the ( actually infinite ) entanglement between the algebra and its commutant.}.

The simplest possibility to define this {\it renormalization of the quantum fluctuations} will be to enlarge our original algebra ${\cal{A}}_{\beta}$ and the corresponding Hilbert space representation in such a way that it contains now a well defined bounded operator representing $h_R$ with the quantum fluctuation created by $h_R$ living now in an extended Hilbert space. 

What can we add to the large $N$ algebra ${\cal{A}}_{\beta}$ to achieve this goal ? After our former training the most natural answer will be to {\it add a formal RF algebra} generated, as usual, by the canonical commutation relation $[\hat h_{RF},\hat t]=-i$ for {\it some} RF Hamiltonian $\hat h_{RF}$ and to include, as a well defined operator of the extended algebra, the renormalized Hamiltonian $h_R^{II}$ defined now as
\begin{equation}\label{hamiltonian1}
h_R^{II}= \hat h_{\beta} +\beta \hat h_{RF}
\end{equation}
with $\hat h_{\beta}$ the modular Hamiltonian defined relative to the state $|\hat \beta\rangle$. 

Note that the so defined extended algebra is generated by $e^{i\hat t \hat h}ae^{-i \hat t \hat h}$ and $\hat h_{RF}$ and therefore is a crossed product of ${\cal{A}}_{\beta}$ by the action of the modular Hamiltonian $\hat h_{\beta}$. We will denote, following what we did in previous sections, this algebra ${\cal{A}}^{cr}_{YM}$. Now we can follow the same steps that in our analysis of the algebra of observables for de Sitter with some minor but important differences. 

Let us start defining in the extended Hilbert space a product RF-Yang Mills state
\begin{equation}
|\tilde \beta\rangle = \int d{\epsilon} f(\epsilon)|\hat \beta\rangle |\epsilon\rangle
\end{equation}
for $\epsilon$ the continuous spectrum of $\hat h_{RF}$. As before we can now define, using the KMS property, a trace over the crossed product algebra \cite{Witten1} as
\begin{equation}
tr(\hat a) = \langle \tilde \beta|\frac{\hat a e^{\beta \hat h_{RF}}}{|f|^2} |\tilde \beta\rangle
\end{equation}

Note again that this trace is {\it independent} of what function $f$ we choose to define the state $|\tilde \beta \rangle$
\begin{equation}\label{state1}
tr(\hat a) = \int d{\epsilon} e^{\beta \epsilon} \langle \hat \beta|\hat a|\hat \beta\rangle
\end{equation}
and is fully determined by the state $|\hat\beta\rangle$.

If we assume that the {\it spectrum of $\hat h_{RF}$ is the whole real line} as it would be  if we define $\hat h_{RF}$ by the commutation relation $[\hat h_{RF},\hat t]=-i$ with the spectrum of $\hat t$ the whole real line, then (\ref{state1}) implies that $tr(1)=\infty$ or equivalently a type $II_{\infty}$ factor. This means that we cannot find any normalizable state, let us say $|\hat \Psi_{\beta}\rangle$, such that $tr(\hat a)$, as defined above, could be represented as $\langle \hat \Psi_{\beta}|\hat a |\hat \Psi_{\beta}\rangle$. Recall the in the case of dS by imposing positivity of the spectrum of $\hat h_{RF}$ we were able to define such normalizable state that becomes the maximal entropy state. However as in the case of dS we can define for any product state $|\hat \Phi\rangle$ in the extended Hilbert space a density matrix $\hat \rho_{\hat \Phi}$ such that
\begin{equation}
tr(\hat a \hat \rho_{\hat \Phi}) = \langle \hat \Phi|\hat a|\hat \Phi \rangle
\end{equation}
We can easily prove that, contrary to the case of dS, does not exist any normalizable state $|\hat \Phi\rangle$ with density matrix $\hat \rho_{\hat \Phi} =1$. Indeed in such a case $\langle \hat \Phi|\hat \Phi\rangle = tr(1) = \infty$. Among other things this means that in the Hagedorn regime we have not any maximal entropy state.

However we can be interested in discovering the density matrix associated with the state $|\tilde \beta\rangle$. This can be achieved by solving the equation $\langle \tilde \beta|\hat a |\tilde \beta \rangle = tr(\hat \rho_{\tilde \beta} \hat a)$ with the trace defined by (\ref{state1}). The solution is given by \footnote{The key ingredient in this derivation \cite{Witten3} is the computation of the Tomita generator $\Delta_{\hat \Psi_{\beta}}$. Actually for generic $f$ satisfying that $\Delta(h_R^{II 2})$ is finite we observe that $\Delta_{\hat \Psi_{\beta}}$ splits.}
\begin{equation}
\hat\rho_{\tilde \beta}= e^{-\beta h_R^{II}}|f(h_R)|^2
\end{equation}
The state $|\tilde \beta \rangle$ for $\int d\epsilon|f|^2=1$ with the integral over the whole real line is what in \cite{Witten3} is defined as the {\it micro canonical version of the TFD}. Note that for the state $|\tilde \beta\rangle$ the variance of the type $II$ operator $h_R^{II}= \hat h_{\beta} +\beta\hat h_{RF}$
\begin{equation}
\langle \tilde \beta| ( h^{II}_R)^2|\tilde \beta\rangle 
\end{equation}
is finite. By finite variance we mean that it scale with $N$ as $N^0$ in the large $N$ limit. Note already that this variance will measure {\it entanglement capacity} of the ground state in the high temperature regime $\beta<\beta_H$. In summary what we have achieved adding the RF system and therefore working with the type $II_{\infty}$ factor is going from the type $III$ divergent fluctuations with 
\begin{equation}
\Delta (\beta h_R)^2 = O(N^2)
\end{equation} 
into the {\it renormalized} type $II$ version with
\begin{equation}\label{uncertainty}
\Delta ( h^{II}_R)^2 = O(1)
\end{equation}

Moreover the corresponding vN entropy will be given by
\begin{equation}\label{entropy}
\langle \tilde \beta| \ln \hat\rho_{\tilde \beta} |\hat \tilde \beta\rangle
\end{equation}
Since $\hat h |\hat \Psi_{\beta}\rangle =0$ this entropy is fully determined by the expectation value of the RF Hamiltonian $\hat h_{RF}$. However this entropy is ambiguous in an arbitrary constant reflecting the transformations in the fundamental group of the type $II_{\infty}$ factor. Nevertheless, as already discussed in previous sections, this ambiguity is not affecting the value of (\ref{uncertainty}).

\subsection{Statistical versus Quantum effects}
As already mentioned for pure Yang Mills in finite volume and at large $N$ we can evaluate, in the weak coupling limit, the value of $\frac{F(T)}{N^2}$ even for $T>T_H$ \cite{Minwalla1} and to use this quantity to define an order parameter for the Hagedorn phase transition. A natural question after the former algebraic discussion should be: What is the meaning of this quantity in the algebraic setup once we assume that the high temperature phase is described by a large $N$ algebra ${\cal{A}}_{\beta}$ that is a type $III$ factor? And even more sharply: What new physics practical information about the high temperature phase this algebraic setup provides?

The most direct answer to the first question would be that in this regime the large $N$ limit of $\frac{F(T)}{N^2}$ should be determined by the large $N$ limit of i.e.
\begin{equation}\label{largeN}
\lim_{N=\infty}\frac{\langle (H-\langle H \rangle)^2\rangle}{N^2}
\end{equation}
where the expectation values $\langle..\rangle$ are defined with respect to the TFD state at temperature $\beta$. Note that in the limit in (\ref{largeN}) also the TFD state should be pushed to its large $N$ limit defined by the GNS representation of the large $N$ algebra ${\cal{A}}_{\beta}$. Following \cite{Witten1} let us define the rescaled operator
\begin{equation}
{\cal{U}}= \lim_{ N=\infty}( \frac{(H-\langle H \rangle)}{N})
\end{equation}
This operator is central in the large $N$ limit i.e. $[{\cal{U}},{\cal{A}}_{\beta}] =0$. Thus formally adding ${\cal{U}}$ as an extra generator it will take us to an extended algebra $\tilde{\cal{A}}_{\beta}$ that although is still type $III$ will not be a factor since it has a non trivial center generated by ${\cal{U}}$. After introducing ${\cal{U}}$ we can represent
$\frac{F(T)}{N^2}$ as the correlator $\langle {\cal{U}}^2 \rangle$. Moreover in the large $N$ limit correlators for ${\cal{U}}$'s are gaussian and we can represent $\langle {\cal{U}}^2 \rangle$ as
\begin{equation}
\langle {\cal{U}}^2 \rangle = \int dU p_{\beta}(U) U^2
\end{equation}
for $U$ representing the continuous spectrum of ${\cal{U}}$ and for $p_{\beta}(U)$ the gaussian probability distribution
\begin{equation}
p_{\beta}(U) = (\frac{1}{\pi \sigma^2})^{1/2} e^{-\frac{U^2}{\sigma^2}}
\end{equation}
with the variance $\sigma^2= \frac{F(T)}{N^2}$. Defining $\tilde \rho_{\beta} = \int dU p_{\beta}(U)|U\rangle\langle U|$ for ${\cal{U}}|U\rangle = U |U\rangle$ we get
\begin{equation}
 \frac{F(T)}{N^2} = tr(\tilde \rho_{\beta}   {\cal{U}}^2)=  \int dU p_{\beta}(U) U^2
 \end{equation}
 
 Until this point what we get is that $ \frac{F(T)}{N^2}$ is the variance of a gaussian {\it statistical ensemble} defined on the spectrum of ${\cal{U}}$. 
 
 Now let us try to view the former statistical description with {\it quantum mechanical eyes}. In principle the way to get this quantum mechanical description will consist in defining quantum states $|\psi_{\beta}\rangle$ with wave functions $f_{\beta}(U)$ such that $|f_{\beta}(U)|^2 = p_{\beta}(U)$ i.e.
\begin{equation}\label{coherent}
|\psi_{\beta}\rangle= \int dU (\frac{1}{\pi \sigma^2})^{1/4} e^{-\frac{U^2}{2\sigma^2}}|U\rangle
\end{equation}
This wave function is nothing exotic but simply the typical wave function of a coherent state with vanishing phases.  However the change of view from the statistical gaussian distribution $p_{\beta}(U)$ into the full fledged quantum wave function $f_{\beta}(U)$ has dramatic implications. 

Indeed to give sense to the quantum wave function we need to define a Heisenberg algebra with two conjugated operators let us say $\hat U, \hat t$ satisfying $[\hat U,\hat t]=-i$ playing the role of position and momentum operator in the standard definition of coherent states. Let us denote $u$ the continuous spectrum of $\hat U$. In this case  a generic coherent state with variance $\sigma$ will be given, up to phases, by the wave function
\begin{equation}\label{coherentstate}
(\frac{1}{\pi \sigma^2})^{1/4} e^{-\frac{(u-u_0)^2}{2\sigma^2}}
\end{equation}
with $u_0$ representing the expectation value of $\hat U$. Moreover the quantum expectation value on this state of $(\hat U-u_0)^2$ will be given by the variance $\sigma^2$. Thus in this {\it quantum representation} $\frac{F(T)}{N^2}$ becomes the quantum expectation value of the operator $(\hat U-u_0)^2$. But what is $\hat U$ ? Obviously $\hat U$ cannot be identified with the rescaled operator ${\cal{U}}$ that in the large $N$ limit was a central term. Moreover $\hat U$ contrary to ${\cal{U}}$ can be associated with translations of the conjugated operator $\hat t$ and in that sense works as a Hamiltonian if we think of $t$ as a time. 

Thus in order to go from the statistical ensemble description of $\frac{F(T)}{N^2}$ into the quantum mechanical one we need to add the Heisenberg algebra defined by $\hat H$ and $\hat t$. Then $\frac{F(T)}{N^2}$ becomes, itself, the expectation value of $(\hat U-u_0)^2$ in a particular coherent state in the Hilbert space representation of the reference frame quantum algebra. 

 The reader will find obvious, at this point of the discussion, to identify the algebra $\hat U,\hat t$, needed for going from the statistical ensemble into the quantum description, with our old good friend {\it the reference frame algebra}.
 
Before proceeding further let us make some comments on the previous construction.

c-1) The operators $\hat U$ and $\hat t$ i.e. the reference frame algebra is defined in the $N=\infty$ limit.

c-2) The extended algebra obtained adding this reference frame algebra is the crossed product algebra defined in the previous section ${\cal{A}}^{cr}$. This becomes manifest once we represent $\hat U$ as a translation operator on $t$ and we identify $t$ with the "time" parameter defining the modular automorphism.

c-3) The $h^{II}$ defined above becomes simply $\hat h + \beta \hat U$ and consequently we represent $\frac{F(T)}{N^2}$ as the variance of $h^{II}$. 

c-4) The quantum state representing the ground state at $T>T_H$ can be identified, up to quantum phases, with the coherent state (\ref{coherentstate}) with $u_0$ the ground state energy $E_0$. This makes explicit that transformations in the fundamental group of ${\cal{A}}^{cr}$ correspond to redefinitions of $E_0$.

The type $II$ philosophy consists in thinking the Gaussian probability distribution with variance determined by the large $N$ limit of $\frac{F(T)}{N^2}$ in terms of {\it the quantum wave function} of a quantum state in the Hilbert space representation of the quantum reference frame algebra i.e. in terms of a quantum state in the type $II$ extended Hilbert space. A skeptical reader can wonder what is the real difference between an statistical ensemble characterized by a probability distribution $p(\epsilon)$ and a real quantum wave function, defining a state in the extended Hilbert space, with probability amplitude $f(\epsilon)$ satisfying $|f|^2=p$. The key difference, as stressed in former sections, is the quantumness of the reference frame algebra i.e. the existence of the conjugated operator $\hat t$ and the corresponding quantum uncertainties $\delta(\hat t)$ that are non vanishing even in the $N=\infty$ limit.

Intuitively we can parametrize different GNS representations by the eigenvalue $u$ of $\hat U$ and to think the quantum state in the extended Hilbert space with wave function $f(u)$ as a sort of GNS-Hartle-Hawking {\it wave function} on different GNS Hilbert spaces, each one representing, in the large $N$ limit, a semiclassical ground state. The crucial new feature of the type $II$ approach is that the quantumness of $f$, as it is the quantumness of the reference frame algebra, is  {\it not} a $\frac{1}{N}$ effect.

\subsection{The $N=\infty$ type $II$ limit}

Let us first briefly review the matrix model derivation of the partition function (\ref{partition}). While we can evaluate the spectral density $\rho(E)$ by direct counting of gauge singlets in the tensor product of different irreps of $U(N)$ \footnote{Recall that the reason of counting singlets is in order to implement the Gauss law constraint. For $U$ any element of $U(N)$ the number of singlets in, for instance, the tensor product of $n$ adjoint representations is $\int dU (tr(U)tr(U^{\dagger})^n$.}  we can also define the partition function using the path integral of Euclidean Yang Mills on $S^3\times S^1$ with the thermal circle of radius $\beta$. Using the standard decomposition of Yang Mills in Kaluza Klein modes we can define the partition function 
\begin{equation}
Z(\beta)=e^{-\beta F} = \int [dU] e^{-S_{eff}(U)}
\end{equation}
with $U$ defined in terms of the KK zero mode $\alpha$ as
\begin{equation}
U=e^{i\alpha\beta}
\end{equation}
with $\alpha = \frac{1}{vol}\int_{S^3} A_0$ satisfying the gauge constraint $\partial_t\alpha =0$ and with $S_{eff}(U)$ defined by integrating the higher KK modes \footnote{We will discuss the non trivial t'Hooft coupling $\lambda=0$ limit of $S_{eff}(U)$.}. The $U$ defined above is related to the temporal Polyakov loop by
\begin{equation}
{\cal{P}}=\langle \hat {\cal{P}} \rangle = \langle \frac{tr(U)}{N} \rangle
\end{equation}
The effective action $S_{eff}(U)$ can be represented in terms of the $N^2$ eigenvalues $\lambda_i$ of $U$. We can think of these eigenvalues as $N^2$ "particles" located on the unit circle that we will parametrize by the angle $\theta$. The dynamics of these eigenvalues contains two pieces. One is a repulsive interaction that is independent of $\beta$ and comes from the integration measure $[dU]$. The other piece is a temperature dependent attractive interaction induced by the integration of the higher KK modes. This attractive interaction increases with the temperature. In the low temperature regime the dominant interaction is the temperature independent repulsive force and the equilibrium eigenvalue distribution $\rho(\theta)$ defining the saddle point is the uniform distribution. In the high temperature regime $\beta<\beta_H$ the attractive interaction becomes dominant and the saddle point distribution of the eigenvalues correspond to "bound states" with all the eigenvalues located in a finite interval of the unit circle.

In order to make contact with the former algebraic construction we will associate the order parameter $\langle |(\hat{\cal{P}})|^2 \rangle$ with the thermal expectation value
\begin{equation}
\langle \frac{h_R^2}{N^2} \rangle
\end{equation}
with $h_R$ the {\it type $III$ operator} simply defined as $H_R-\langle H_R\rangle_{\beta}$. What is now the type $II$ version of the order parameter ? Obviously it should be defined as
\begin{equation}
\Delta(( h^{II}_R)^2) 
\end{equation}
with $h^{II}_R=\hat h+\beta\hat h_{RF}$ and to evaluate $\Delta(h^{II}_R)^2$ on the state, in the extended Hilbert space, $|\tilde \beta\rangle$. To distinguish this type $II$ order parameter from the type $III$ one we will denote the corresponding {\it type $II$ Polyakov operator ${\hat{\cal{P}}}^{II}$}. 

The state $|\tilde \beta\rangle$ is a {\it product state} 
\begin{equation}\label{product}
|\tilde \beta \rangle = |\hat \beta\rangle \otimes |\psi (\beta)\rangle
\end{equation}
where the state $|\psi(\beta\rangle$ was defined in (\ref{coherent}). The type $II$ von Neumann entropy associated to a generic product state in the extended Hilbert space reflects the product structure (\ref{product}) in the form of of a generalized entropy formally of the type $S(|\Phi\rangle |\psi\rangle) = S(\Phi) + S_{RF}(\psi)$ with $S(\Phi)$ measuring the distinguishability distance between the GNS states $\Phi$ and $\hat \beta$ and with $S_{RF}(\psi)$ roughly measuring the expectation value of the reference frame Hamiltonian $\hat h_{RF}$ on the state $|\psi\rangle$.

In the holographic setup this type $II$ QFT entropy has a beautiful geometrical meaning as a consequence of RT and QSE prescription. Indeed we expect that 
\begin{equation}
S(|\Phi\rangle |\psi\rangle) = S_{gen}(EW(cft))
\end{equation}
with $S_{gen}(EW(cft))$ the {\it extremal} Bekenstein generalized entropy defined for the bulk {\it entanglement wedge} of the full time evolution of the boundary CFT. In this case we expect
\begin{equation}
S_{RF} = S(\partial(EW(cft))) \sim \frac{A(\partial(EW(cft)))}{4G_N}
\end{equation} with $A(\partial(EW(cft)))$ the bulk area of the boundary of the entanglement wedge \footnote{This relation between the type $II$ reference frame algebra and the geometrical features of the bulk entanglement wedge is one of the key outputs of the type $II$ approach \cite{Witten2,Witten3}.}.

For the particular product state $|\tilde \beta\rangle$ the dominant contribution to $S(|\tilde \beta\rangle)$ comes from the RF since we have used the state $\hat \beta$ satisfying $\hat h |\hat \beta\rangle =0$ for $\hat h$ the modular Hamiltonian. Hence the corresponding type $II$ entanglement capacity ${\cal{C}}(\tilde \beta)$ can be related with the matrix model representation of the Polyakov loop as
\begin{equation}
{\cal{C}}(\tilde \beta) \sim |\rho_1|^2
\end{equation}
for $\rho_1$ the first moment of the eigenvalue distribution. Recall that at weak coupling $\rho_1$ determines the size of the interval where the eigenvalues are "bounded" \cite{Minwalla1}. In summary we conclude
\begin{equation}
\rho_1 \sim \langle \psi|\hat h_{RF}|\psi\rangle
\end{equation}
relating the size of the interval of the confined eigenvalues in the high temperature phase and the reference frame Hamiltonian of the type $II$ version. 

Note that the value of the first moment $\rho_1$ is also the solution of an {\it extremality} problem, namely extremality with respect to small perturbations corresponding to add one eigenvalue in the interval.

Now the quantumness of the reference frame algebra i.e. $[\hat h_{RF},\hat t]= -i\hbar$ becomes very important to distinguish a purely statistical ensemble approach to $\rho_1$ from a purely quantum approach based on the quantum state $|\psi\rangle$. In this second case we expect {\it quantum fluctuations} of $\rho_1$ or equivalently quantum fluctuations of the boundary of $EW(cft)$ {\it not} suppressed in the $N=\infty$ limit.

The way we understand physically what is going on is roughly as follows. In the high temperature phase we need to work with the large $N$ algebra {\it and with its commutant}. The interval of eigenvalues represent a semiclassical bulk geometry and the quantum uncertainty of $\hat t$, the conjugated in the quantum reference frame algebra of $\hat h_{RF}$, reflects non vanishing quantum fluctuations of this semiclassical background state even in the $N=\infty$ limit \footnote{These quantum effects are most likely the analog of the time shift discussed in \cite{Witten3} for pure Yang Mills in the Hagedorn phase.}.

\section{Acknowledgments}
 I thank Sumit Das for discussions. This work was supported by grants SEV-2016-0597, FPA2015-65480-P and PGC2018-095976-B-C21.


\begin{thebibliography}{99}
\bibitem{Schr} Schrodinger, E., 1935. "Discussion of Probability Relations Between Separated Systems," Proceedings of the Cambridge Philosophical Society, 31: 555?563; 32 (1936): 446?451.
\bibitem{EPR} Einstein, A., Podolsky, B., Rosen, N., 1935. "Can Quantum-Mechanical Description of Physical Reality be Considered Complete?," Physical Review, 47: 777?780.
\bibitem{vN}  J.V.Neumann "Mathematical Foundations of QuantumMechanics" Princeton University Press 1955
\bibitem{wittenrev1}E. Witten, "Some Entanglement Properties of Quantum Field Theory," Rev. Mod. Phys.90(2018) 045003, arXiv:1803.04993
  \bibitem{wittenrev2}E. Witten, "Why Does Quantum Field Theory In Curved Spacetime Make Sense?  And What Happens To The Algebra of Observables In The Thermodynamic Limit?"  arXiv:2112.11614.
\bibitem{Raam}
  M.~Van Raamsdonk,
  ``Building up spacetime with quantum entanglement,''
  Gen.\ Rel.\ Grav.\  {\bf 42} (2010) 2323
   [Int.\ J.\ Mod.\ Phys.\ D {\bf 19} (2010) 2429]
[arXiv:1005.3035 [hep-th]].
\bibitem{Malda}
  J. Maldacena and L. Susskind,"Cool horizons for entangled black holes",[arXiv:1306.0533[hep-th]].
  \bibitem{Ads1} J. M. Maldacena,The Large N limit of superconformal field theories and supergravity,Int. J.Theor. Phys.38(1999) 1113?1133, [hep-th/9711200].
  \bibitem{Ads2} S. S. Gubser, I. R. Klebanov and A. M. Polyakov,Gauge theory correlators from noncriticalstring theory,Phys. Lett. B428(1998) 105?114, [hep-th/9802109].
  \bibitem{Ads3} E. Witten,Anti-de Sitter space and holography,Adv. Theor. Math. Phys.2(1998) 253?291,[hep-th/9802150].
  \bibitem{sorkin}  R. Sorkin, "On The Entropy of the Vacuum Outside a Horizon" in B. Bertotti, F. de Feliceand A. Pascolini, eds.,Tenth International Conference on General Relativity and Gravitation(Padova, July 4-9, 1983), Contributed Papers, vol. II, pp. 734-736 (Roma, Consiglio Nazionale Delle Ricerche, 1983), available at arXiv:1402.3589
  \bibitem{unruh} 
  Unruh, W.G, "Notes on black-hole evaporation". Physical Review D. 14 (4): 870-892. 
  \bibitem{Bek} J. D. Bekenstein,Black Holes and Entropy,Phys. Rev.D7(1973) 2333?2346
 \bibitem{H}S. W. Hawking, "Particle Creation By Black Holes," Commun. Math. Phys.43(1975)199-220.

 \bibitem{GH}
  G.~W.~Gibbons and S.~W.~Hawking,
  ``Cosmological Event Horizons, Thermodynamics, and Particle Creation,''
  Phys.\ Rev.\ D {\bf 15} (1977) 2738.
  doi:10.1103/PhysRevD.15.2738
\bibitem{QEC}
E.~Gesteau,
``Large $N$ von Neumann algebras and the renormalization of Newton's constant,''
[arXiv:2302.01938 [hep-th]].

 \bibitem{Papa1}
K.~Papadodimas and S.~Raju,
``An Infalling Observer in AdS/CFT,''
JHEP \textbf{10}, 212 (2013)
doi:10.1007/JHEP10(2013)212
[arXiv:1211.6767 [hep-th]].
 \bibitem{Papa2}
K.~Papadodimas and S.~Raju,
``State-Dependent Bulk-Boundary Maps and Black Hole Complementarity,''
Phys. Rev. D \textbf{89}, no.8, 086010 (2014)
doi:10.1103/PhysRevD.89.086010
[arXiv:1310.6335 [hep-th]].
\bibitem{LL1}  S. Leutheusser and H. Liu, "Causal connectability between quantum systems and the black hole interior in holographic duality" 2110.05497.
\bibitem{LL2} S. Leutheusser and H. Liu, "Emergent times in holographic duality" 2112.12156.
\bibitem{LL3} 
S.~Leutheusser and H.~Liu,
``Subalgebra-subregion duality: emergence of space and time in holography,''
[arXiv:2212.13266 [hep-th]].
\bibitem{Page}  D. N. Page, "Information in black hole radiation" ,Phys. Rev. Lett.71(1993) 3743?3746,[hep-th/9306083].
\bibitem{Almieri}
A.~Almheiri, T.~Hartman, J.~Maldacena, E.~Shaghoulian and A.~Tajdini,
``The entropy of Hawking radiation,''
Rev. Mod. Phys. \textbf{93} (2021) no.3, 035002
[arXiv:2006.06872 [hep-th]].
\bibitem{class1} 
G.~Dvali and C.~Gomez,
``Self-Completeness of Einstein Gravity,''
[arXiv:1005.3497 [hep-th]].
\bibitem{class2}
G.~Dvali, G.~F.~Giudice, C.~Gomez and A.~Kehagias,
``UV-Completion by Classicalization,''
JHEP \textbf{08} (2011), 108
[arXiv:1010.1415 [hep-ph]].

\bibitem{Witten1}E.~Witten,
"Gravity and the Crossed Product''
[arXiv:2112.12828 [hep-th]].
\bibitem{Witten2} V.~Chandrasekaran, R.~Longo, G.~Penington and E.~Witten,
``An Algebra of Observables for de Sitter Space,''
[arXiv:2206.10780 [hep-th]].
\bibitem{Witten3}
V.~Chandrasekaran, G.~Penington and E.~Witten,
``Large N algebras and generalized entropy,''
[arXiv:2209.10454 [hep-th]].
\bibitem{Bek2}J. D. Bekenstein, "Black Holes and the Second Law" Lett. Nuovo Cim.4(1972) 737-740. 
\bibitem{Unglum}  L. Susskind and J. Uglum, "Black Hole Entropy In Canonical Quantum Gravity and Superstring Theory," Phys. Rev.D50(1994) 2700-11, arXiv:hep-th/9401070.
\bibitem{RT} S. Ryu and T. Takayanagi, "Holographic derivation of entanglement entropy from AdS/CFT" ,Phys. Rev. Lett.96(2006) 181602, [hep-th/0603001].
\bibitem{HRT} V.  E.  Hubeny,  M.  Rangamani  and  T.  Takayanagi,  JHEP07,  062  (2007) [arXiv:0705.0016 [hep-th]].
\bibitem{LM} A. Lewkowycz and J. Maldacena, "Generalized gravitational entropy" JHEP08(2013) 090,[1304.4926
\bibitem{QES1} N. Engelhardt and A. C. Wall, "Quantum Extremal Surfaces: Holographic Entanglement Entropy Beyond the Classical Regime",JHEP01(2015) 073, [1408.3203]
\bibitem{QES2} T. Faulkner, A. Lewkowycz and J. Maldacena,"Quantum Corrections to Holographic Entanglement Entropy" JHEP11(2013) 074, [1307.2892].
\bibitem{QES3} A. C. Wall, "Maximin Surfaces, and the Strong Subadditivity of the Covariant Holographic Entanglement Entropy" ,Class. Quant. Grav.31(2014) 225007, [1211.3494]
\bibitem{AS}
  Y. Aharonov and L. Susskind
  "Charge superselection rule"
Phys. Rev. 155, 1428 (1967)
 \bibitem{RF}
S. D. Bartlett, T. Rudolph, and R. W. Spekkens
"Reference frames, superselection rules, and quantum information"
Rev. Mod. Phys. 79, 555 (2007)
\bibitem{kitaev}
 A.Kitaev, D. Mayers, J. Preskill
 "Superselection rules and quantum protocols"
Phys.Rev. A69 (2004) 052326
\bibitem{Gomezcross}
C.~Gomez,
``Cosmology as a Crossed Product,''
[arXiv:2207.06704 [hep-th]].
\bibitem{corean}
M.~S.~Seo,
``von Neumann algebra description of inflationary cosmology,''
[arXiv:2212.05637 [hep-th]].
\bibitem{mukhanov0} 
V.~F.~Mukhanov,
``CMB-slow, or how to estimate cosmological parameters by hand,''
Int. J. Theor. Phys. \textbf{43} (2004), 623-668
[arXiv:astro-ph/0303072 [astro-ph]].
\bibitem{mukhanov1}
L.~M.~Wang, V.~F.~Mukhanov and P.~J.~Steinhardt,
``On the problem of predicting inflationary perturbations,''
Phys. Lett. B \textbf{414} (1997), 18-27
[arXiv:astro-ph/9709032 [astro-ph]].
\bibitem{GJ}
C.~G\'omez and R.~Jimenez,
``Quantum Fisher Cosmology: confronting observations and the trans-Planckian problem,''
JCAP \textbf{09} (2021), 016
[arXiv:2105.05251 [astro-ph.CO]].
\bibitem{Vafa} 
G.~Obied, H.~Ooguri, L.~Spodyneiko and C.~Vafa,
"De Sitter Space and the Swampland,''
[arXiv:1806.08362 [hep-th]].
\bibitem{Gia1} 
G.~Dvali and C.~Gomez,
"Quantum Compositeness of Gravity: Black Holes, AdS and Inflation,''
JCAP \textbf{01} (2014), 023
[arXiv:1312.4795 [hep-th]].
\bibitem{GW}
D.~J.~Gross and E.~Witten,
"Possible Third Order Phase Transition in the Large N Lattice Gauge Theory,''
Phys. Rev. D \textbf{21} (1980), 446-453
doi:10.1103/PhysRevD.21.446
\bibitem{Wadia} S. R. Wadia, "N = Infinity Phase Transition In A Class Of Exactly Soluble Model Lattice Gauge Theories", Phys. Lett. B93, 403 (1980).
\bibitem{Wittenconf} E. Witten, "Anti-de Sitter space and holography" Adv. Theor. Math. Phys.2, 253(1998) [arXiv:hep-th/9802150]
\bibitem{Sudorg} B. Sundborg, "The Hagedorn transition, deconfinement and N = 4 SYM theory" Nucl.Phys. B573, 349 (2000) [hep-th/9908001]; B. Sundborg, "Stringy gravity, interactingtensionless strings and massless higher spins," Nucl. Phys. Proc. Suppl.102, 113(2001) [hep-th/0103247].
\bibitem{Minwalla1}
O.~Aharony, J.~Marsano, S.~Minwalla, K.~Papadodimas and M.~Van Raamsdonk,
"The Hagedorn - deconfinement phase transition in weakly coupled large N gauge theories,''
Adv. Theor. Math. Phys. \textbf{8} (2004), 603-696
doi:10.4310/ATMP.2004.v8.n4.a1
[arXiv:hep-th/0310285 [hep-th]].
\bibitem{Minwalla2}
O.~Aharony, J.~Marsano, S.~Minwalla, K.~Papadodimas and M.~Van Raamsdonk,
"The deconfinement and Hagedorn phase transitions in weakly coupled large N gauge theories,''
Comptes Rendus Physique \textbf{5} (2004), 945-954
doi:10.1016/j.crhy.2004.09.012
\bibitem{Minwalla3} 
O.~Aharony, J.~Marsano, S.~Minwalla, K.~Papadodimas and M.~Van Raamsdonk,
"A First order deconfinement transition in large N Yang-Mills theory on a small S**3,''
Phys. Rev. D \textbf{71} (2005), 125018
doi:10.1103/PhysRevD.71.125018
[arXiv:hep-th/0502149 [hep-th]].
\bibitem{gomezliu}
L.~Alvarez-Gaume, C.~Gomez, H.~Liu and S.~Wadia,
"Finite temperature effective action, AdS(5) black holes, and 1/N expansion,''
Phys. Rev. D \textbf{71} (2005), 124023
[arXiv:hep-th/0502227 [hep-th]].
\bibitem{Papa3}
E.~Bahiru, A.~Belin, K.~Papadodimas, G.~Sarosi and N.~Vardian,
``Holography and Localization of Information in Quantum Gravity,''
[arXiv:2301.08753 [hep-th]].
\bibitem{Fermi1}  E. Fermi: "Quantum Theory of Radiation",Rev. Mod. Phys.4, 87?132 (1932)
\bibitem{Fermi2} G.C. Hegerfeldt: "Causality Problems in Fermi?s Two Atom System",Phys. Rev.Lett.,72, 596-599 (1994)
\bibitem{Fermi3} D. Buchholz, J. Yngvason: "There Are No Causality Problems in Fermi?s TwoAtom System",Phys. Rev. Lett.73, 613?613 (1994).12
\bibitem{Fermi4} Yngvason:"The role of type $III$ factors in Quantum Field Theory" 	arXiv:math-ph/0411058
\bibitem{split1}R. Werner: ?Local preparability of States and the Split Property in QuantumField Theory?,Lett. Math. Phys.13, 325?329 (1987).
\bibitem{split2}H. Roos: ?Independence of Local Algebras in Quantum Field Theory?,Comm.Math. Phys.16, 238?246 (1970).
\bibitem{split3}S. Doplicher, R. Longo: ?Standard and split inclusions of von Neumann algebras?,Invent. Math.73, 493?536 (1984).
\bibitem{split4} D. Buchholz, E. Wichmann: ?Causal Independence and the Energy level Density of States in Local Quantum Field theory?,Comm. Math. Phys.106, 321?344(1986).
\bibitem{Das} S. Das, P. Majumdar and R. K. Bhaduri, "General logarithmic corrections to black hole entropy" ,Class. Quant. Grav.19(2002) 2355?2368, [hep-th/0111001]
\bibitem{Connes} A. Connes
Journal of Operator Theory
Vol. 4, (1980) No. 1  pp. 151-153 
\bibitem{Banks1} R. Bousso, "Bekenstein Bounds in de Sitter and Flat Space," JHEP04(2001) 035,hep-th/0010252.
\bibitem{Banks2}T. Banks, "Cosmological Breaking of Supersymmetry?  Or Little Lambda Goes Back to theFuture, II,? Int. J. Mod. Phys.A16(2001) 910-921, hep-th/0007146.
\bibitem{Banks3}T. Banks, "More Thoughts on the Quantum Theory of Stable de Sitter Space" arXiv:hep-th/0503066.
\bibitem{Banks4} T. Banks, B. Fiol, and A. Morisse, "Towards a Quantum Theory of de Sitter Space" arXiv:hep-th/0609062.
\bibitem{Torroba}  X. Dong, E. Silverstein, and G. Torroba, "De Sitter Holography and Entanglement Entropy" arXiv:1804.08623.
\bibitem{susskinddS} L. Susskind, "De Sitter Holography:  Fluctuations, Anomalous Symmetry, and Wormholes" Universe7(2021) 464, arXiv:2106.03964.
\bibitem{Wigner}Wightman, A.S. "Superselection rules; old and new". Nuov Cim B 110, 751?769 (1995). 
\bibitem{Wigner2}G. -C. Wick, A. S. Wightman, and E. P. Wigner "Super-selection Rule for Charge" Phys. Rev. D 1, 3267 (1970)
\bibitem{Wigner3}A. S. Wightman "Proof of the charge superselection rulein local relativistic quantum field theory" Journal ofMathematical Physics 15, 2198 (1974);
\bibitem{crossed} C. Phillips: "An Introduction to Crossed Product C*-Algebras and Minimal Dynamics". 2017
\bibitem{Paris} M. Paris - International Journal of Quantum Information, Vol 7, 125-137, 2009
\bibitem{Gomez-s} 
C.~G\'omez,
``Gravity, Superselection Rules and Axions,''
Fortsch. Phys. \textbf{69} (2021) no.2, 2000095
[arXiv:1907.03619 [hep-th]].
\bibitem{Jakiw1}  C. G. Callan,  Jr., R. F. Dashen and D. J. Gross,"The Structure of the Gauge Theory Vacuum", Phys.Lett. B63(1976) 334.
\bibitem{Jakiw2} R. Jackiw and C. Rebbi, "Vacuum Periodicity in a Yang-Mills Quantum Theory", Phys. Rev. Lett.37(1976) 172.
\bibitem{tHooft}
G.~'t Hooft,
``Computation of the Quantum Effects Due to a Four-Dimensional Pseudoparticle,''
Phys. Rev. D \textbf{14} (1976), 3432-3450
[erratum: Phys. Rev. D \textbf{18} (1978), 2199]
\bibitem{Wittentop1}]  E. Witten, "Current Algebra Theorems for the U(1)Goldstone Boson,?  Nucl. Phys. B156(1979) 269.
\bibitem{Wittentop2} G. Veneziano, "U(1) Without Instantons" Nucl. Phys.B159(1979) 213.
\bibitem{Banks0}
T.~Banks and N.~Seiberg,
``Symmetries and Strings in Field Theory and Gravity,''
Phys. Rev. D \textbf{83} (2011), 084019
doi:10.1103/PhysRevD.83.084019
[arXiv:1011.5120 [hep-th]].
\bibitem{Bekenstein} Bekenstein,J "Universal upper bound on the entropy-to-energy ratio for bounded systems". Physical Review D. 23 (2): 287?298.
\bibitem{vafa2} 
N.~Arkani-Hamed, L.~Motl, A.~Nicolis and C.~Vafa,
``The String landscape, black holes and gravity as the weakest force,''
JHEP \textbf{06} (2007), 060
[arXiv:hep-th/0601001 [hep-th]].
\bibitem{species1}G. Dvali, Black Holes and Large N Species Solution to the Hierarchy Problem, Fortsch.Phys.58 (2010) 528?536, [arXiv:0706.2050]
\bibitem{species2}  G. Dvali and C. Gomez,Species and Strings,arXiv:1004.3744.
\bibitem{BD1} N. A. Chernikov and E. A. Tagirov, "Quantum theory of scalar field in de Sitter space-time" Annales de l'Institut Henri Poincare A IX (1968) 109.
\bibitem{BD2} C. Schomblond and P. Spindel, "Conditions d'unicite pour le propagateur du champ scalaire dans l'univers de de Sitter," Annales de l'Institut Henri Poincare A XXV (1976) 67.
\bibitem{BD3}T. S. Bunch and P. Davies, "Quantum Field Theory in de Sitter Space:  Renormalization by Point Splitting" Proc. Roy. Soc. LondonA360(1978) 117-34
\bibitem{BD4}E. Mottola, "Particle Creation in de Sitter Space", Phys. Rev.D31(1985) 754.
\bibitem{BD5} B. Allen, "Vacuum States in de Sitter Space", Phys. Rev.D32(1985) 3136
\bibitem{complementarity}  
L. Susskind and U.Thorlacius "The Stretched Horizon and Black Hole Complementarity". Physical Review D. 48 (8): 3743?3761. arXiv:hep-th/9306069. 
\bibitem{Giainfla1} G.~Dvali and C.~Gomez,
  ``Quantum Exclusion of Positive Cosmological Constant?,''
  Annalen Phys.\  {\bf 528} (2016) 68
 [arXiv:1412.8077 [hep-th]].
\bibitem{Giainfla2} G.~Dvali, C.~Gomez and S.~Zell,
 ``Quantum Break-Time of de Sitter,''
 JCAP {\bf 1706} (2017) 028
 \bibitem{Bauman}
D.~Baumann,``Inflation,''
[arXiv:0907.5424 [hep-th]].
 
 \bibitem{Chibisovphonon} 
G.~V.~Chibisov and V.~F.~Mukhanov,
``Galaxy formation and phonons,''
Mon. Not. Roy. Astron. Soc. \textbf{200} (1982), 535-550
\bibitem{Bell} 
J. S. Bell, Physics1, 195 (1964)
\bibitem{MaldaB}
J.~Maldacena,
``A model with cosmological Bell inequalities,''
Fortsch. Phys. \textbf{64} (2016), 10-23
doi:10.1002/prop.201500097
[arXiv:1508.01082 [hep-th]]
\bibitem{Martin}
J.~Martin, V.~Vennin and P.~Peter,
``Cosmological Inflation and the Quantum Measurement Problem,''
Phys. Rev. D \textbf{86} (2012), 103524
[arXiv:1207.2086 [hep-th]].
\bibitem{Chibisov2}
V.~F.~Mukhanov and G.~V.~Chibisov,
``Quantum Fluctuations and a Nonsingular Universe,''
JETP Lett. \textbf{33} (1981), 532-535
\bibitem{Hawking}
S.~W.~Hawking,
``The Development of Irregularities in a Single Bubble Inflationary Universe,''
Phys. Lett. B \textbf{115} (1982), 295
\bibitem{boer}
J.~De Boer, J.~J\"arvel\"a and E.~Keski-Vakkuri,
``Aspects of capacity of entanglement,''
Phys. Rev. D \textbf{99}, no.6, 066012 (2019)
[arXiv:1807.07357 [hep-th]].
\bibitem{Jefferis}
P.~Gao, D.~L.~Jafferis and A.~C.~Wall,
``Traversable Wormholes via a Double Trace Deformation,''
JHEP \textbf{12} (2017), 151
[arXiv:1608.05687 [hep-th]].
\bibitem{maldatrans}
J.~Maldacena, D.~Stanford and Z.~Yang,
``Diving into traversable wormholes,''
Fortsch. Phys. \textbf{65} (2017) no.5, 1700034
[arXiv:1704.05333 [hep-th]].
%
%
%
%
%
%
%
%
%
%
%
%
%
%
%
%
%

\end{thebibliography}
\end{document}